\documentclass[twocolumn]{aastex631}
\usepackage{graphbox}
\usepackage{threeparttablex}
\usepackage{subfigure}
\usepackage{lipsum}
\newcommand{\code}[0]{\texttt{Silkscreen}}
\newcommand{\artpop}[0]{\texttt{ArtPop}}
\newcommand{\sbi}[0]{\texttt{sbi}}

\graphicspath{{./}{figures/}{figures/color_comp/}{figures/corner_im_all}{figures/outliers/}}

\begin{document}

% Title
\title{\texttt{Silkscreen}: Direct Measurements of Galaxy Distances from Survey Image Cutouts}

\author[0000-0001-8367-6265]{Tim B. Miller}
\affiliation{Center for Interdisciplinary Exploration and Research in Astrophysics (CIERA), Northwestern University,1800 Sherman Ave, Evanston, IL 60201, USA}
\affiliation{Department of Astronomy, Yale University, New Haven, CT 06511}

\author[0000-0002-7075-9931]{Imad Pasha}
\affiliation{Department of Astronomy, Yale University, New Haven, CT 06511}

\author[0000-0002-5283-933X]{Ava Polzin}
\affiliation{Department of Astronomy and Astrophysics, The University of Chicago, Chicago, IL 60637, USA}
\affiliation{Department of Astronomy, Yale University, New Haven, CT 06511}

\author[0000-0002-8282-9888]{Pieter van Dokkum}
\affiliation{Department of Astronomy, Yale University, New Haven, CT 06511}

\shorttitle{\code{}: Distances from Survey Cutouts}
\shortauthors{Miller et al.}

\begin{abstract}
With upcoming wide field surveys from the ground and space the number of known dwarf galaxies at $\lesssim 25$ Mpc is expected to dramatically increase. Insight into their nature and analyses of these systems' intrinsic properties will rely on reliable distance estimates. Currently employed techniques 
%such as tip of the red giant branch (TRGB) or surface brightness fluctuations (SBF)
are limited in their widespread applicability, especially in the semi-resolved regime. In this work we turn to the rapidly growing field of simulation based inference to infer distances, and other physical properties, of dwarf galaxies directly from multi-band images. We introduce \code{}: a code leveraging neural posterior estimation to infer the posterior distribution of parameters while simultaneously training a convolutional neural network such that inference is performed directly on the images.
%to extract summary statistics from the images. 
Utilizing this combination of machine learning and Bayesian inference, we demonstrate the method's ability to recover accurate distances from ground-based survey images for a set of nearby galaxies ($2 < D ({\rm Mpc)} < 12$) with measured SBF or TRGB distances. We discuss caveats of the current implementation along with future prospects, focusing on the goal of applying \code{} to large upcoming surveys, like LSST. While the current implementation performs simulations and training on a per-galaxy basis, future implementations will aim to provide a broadly-trained model that can facilitate inference for new dwarf galaxies in a matter of seconds using only broadband cutouts. We focus here on dwarf galaxies, we note that this method can be generalized to more luminous systems as well.
\end{abstract}


\section{Introduction} \label{sec:intro}

Dwarf galaxies --- loosely defined here as galaxies with stellar mass $10^5<M_{*}<10^9$ $M_{\odot}$ --- represent an important population for studying the physics of galaxy formation and dark matter. Dwarf galaxies are typically more dominated by dark matter than higher stellar mass systems making them useful sites to distinguish different dark matter models. Due to their shallower potential well, they are more susceptible to feedback from supernovae and massive stars than $\sim L_*$ galaxies, enabling tests of models of star-formation. In current wide-field surveys, such as HSC-SSP \citep{aihara2018} and DECaLS \citep{dey2019}, it is possible to find samples of 100s or 1000s of dwarf galaxies, as demonstrated by, e.g., \cite{greco2018} and \citet{zaritsky2019}. With upcoming surveys, including the Legacy Survey of Space and Time (LSST) at Vera Rubin Observatory~\citep{ivezic2019} and those that will be carried out by the Roman ~\citep{Akeson2019} and Euclid Space Telescopes \citep{Euclid2022, Marleau2024} this number will grow dramatically as we push to lower masses and greater distances.

These discoveries will usher in a new era of dwarf galaxy studies; however, such studies will require accurate distance measurements to these systems in order to measure their intrinsic properties. This is especially crucial for galaxies at $\lesssim 30$ Mpc, where spectroscopic redshifts are not reliable distance indicators, as peculiar velocities of galaxies tend to dominate over the Hubble flow. These nearby galaxies represent an important discovery space. Follow-up can be carried out more effectively for closer galaxies as they are more likely to be bright and possibly resolved with, e.g., spectroscopy or higher resolution imaging with HST and JWST. Without accurate distance measurements it is nearly impossible to constrain the nature of individual dwarf galaxies or understand their population-level statistics.

\subsection{Distance Measurements to Dwarf Galaxies}
There are at present two predominant methods used to measure the distances to dwarf galaxies which are too nearby for Hubble-flow distances: the tip of the red giant branch (TRGB) and surface brightness fluctuations (SBF). 

TRGB, as the name suggests, aims to measure the peak brightness of the red-giant branch as a standard candle \citep[e.g.,][]{lee1993,sakai1996}. The primary strength of the TRGB method is the robust theoretical underpinnings; this phase of stellar evolution is well understood, and the method is well-calibrated to the level of a few percent ~\citep{Freedman2020, Newmann2024a}. The method relies explicitly on photometric measurements of individual, resolved stars, which means that the distance range over which the TRGB can be used is highly constrained. From the ground, it is possible to measure TRGB distances out to around 5 Mpc \citep{mutlupakdil2021} but crowding and blending can limit the accuracy of such photometry, depending on the galaxy's morphology and mass \citep{olsen2003,greco2021}. From space, the accessible distance is considerably greater; TRGB distance measurements have been made out to 20 Mpc with extremely deep Hubble Space Telescope (HST) observations \citep[e.g.][]{Danieli2020,Shen2021}. The James Webb space telescope (JWST) pushes the accessible distance even further, including  recent observations by \citet{Carleton2023} who identified a potentially-isolated dwarf at 30 Mpc with the potential of measurements out to 50 Mpc~\citep{Newmann2024b}. However, for large samples of galaxies, TRGB distances, especially when space based observatories are required, are simply observationally intractable.

Surface Brightness Fluctuations --- first presented in \citet{tonry1988} --- operate on unresolved stellar populations, and leverage the pixel-to-pixel variance in brightness resulting primarily from the stochastic distribution of the number of giants contained in each pixel. For a given stellar density, the average number of giants per pixel increases with distance, which in turn decreases the Poisson fluctuations, leading to lower variance. This method has been widely applied to both ground- and space-based data to measure distances to both early-type galaxies and dwarf galaxies ~\citep{tonry2001,jerjen2001,blakeslee2010,cohen2018,Jensen2021,foster2024}.

The SBF method requires calibration; the absolute fluctuation amplitude varies significantly with the system's stellar populations. Thus, to use SBF to measure distances, the absolute fluctuation magnitude needs to be calibrated based on other observable properties. The most commonly-used calibration takes the form of a color-fluctuation relation derived either from stellar population synthesis models or from galaxies with external distance measurements~\citep{tonry1997,blakeslee2010,cantiello2018,lee2018,carlsten2019}. This calibration introduces additional uncertainties into distance measurements. When applying SBF there is potentially-irreducible scatter as the color, or other observable properties, may not be perfectly correlated with absolute SBF magnitude. For dwarf galaxies there are a number of additional concerns, e.g., stochastic sampling of the IMF \citep{Raimondo2005,Raimondo2009,greco2021}. Despite these challenges, the applicability of SBF to larger distances, and its ability to operate on ground-based imaging, make it one of the primary paths toward assessing distances to galaxies discovered with, e.g., the LSST survey.

\begin{figure*}
    \centering
    \includegraphics[width=\linewidth]{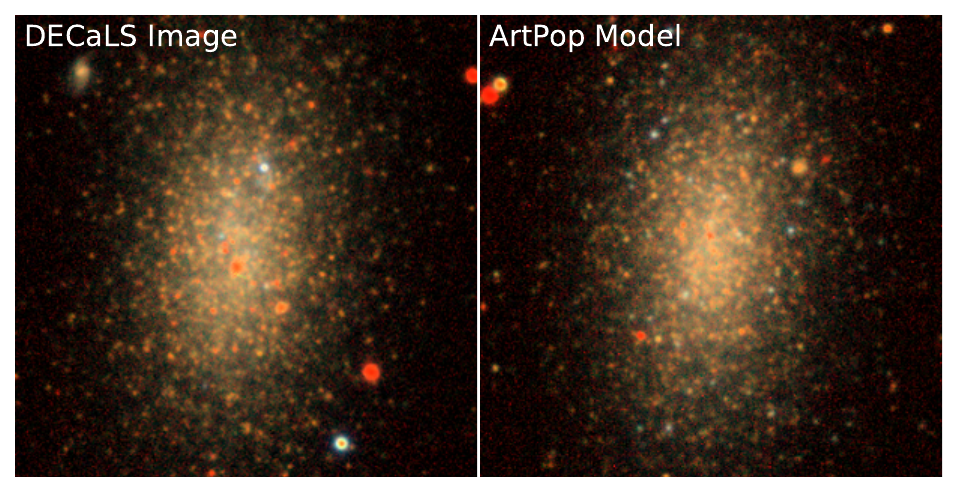}
    \caption{HSC three-color image of ESO 294-010 ($\sim$2 Mpc) compared with an \texttt{ArtPop} model output. The images generated by \texttt{ArtPop} achieve remarkable fidelity in the regime of semi-resolved galaxies, combining both the intrinsic properties of the stellar systems with the observational properties of a given input telescope and survey to produce images that are in many cases indistinguishable from the data. This ability to produce realistic ``simulations'' unlocks the use of simulation based inference as a technique for comparing \texttt{ArtPop} models to real images in a likelihood-free manner. In this example, the galaxy parameters for the \texttt{ArtPop} model shown were inferred using the framework presented in this paper.}
    \label{fig:artpop}
\end{figure*}

\subsection{Forward Modeling and Direct Inference of Semi-Resolved Stellar Populations}

Both the SBF and TRGB methods for distance determination utilize point-statistics which ultimately miss much of the potentially available information in galaxy images. This is especially true in the semi-resolved regime: TRGB ignores the unresolved stellar component, while SBF ignores the presence of any resolved stars. Furthermore, these methods tend to employ one or two bands of imaging, while modern surveys contain more color information across various bands. 

An additional possible avenue for constraining the properties of dwarf galaxy candidates is the use of forward modeling. Forward modeling is the process of generating realistic models of the data from a theoretical basis. These models often contain specific parameters of interest that can be tweaked to determine a ``best-fit''.
%Forward modeling, i.e., generating realistic models from the parameters of interest and comparing these models to the data to determine a best-fit (or posterior space), has been successfully employed in a variety of astrophysical contexts. 
Forward modeling is often paired with Bayesian frameworks to produce posterior distributions, that is, a probabilistic view of parameters given the data available, which is useful in helping to assess parameter uncertainties~\citep[see, e.g.,][]{sharma2017}. 

With modern computer infrastructure it is possible to forward model images of dwarf galaxies by simulating them ``from the ground up''~\citep{cook2019,mutlupakdil2021,greco2022}. These methods populate galaxies ``star by star,'' with masses drawn from an assumed initial mass function (IMF). Then, luminosities are assigned using isochrone libraries before the galaxies are constructed, generally assuming a parametric distribution of light (e.g., a Sérsic profile). These ``ideal'' model images can then be injected directly into images, adding in observational effects such as the point spread function (PSF) and the expected noise for a given instrument and exposure time. This procedure leads to a significant degree of flexibility, as almost anything can in principle be varied, e.g., star-formation history, metallicity, and total mass.

In Figure \ref{fig:artpop}, we show a real galaxy image of ESO 294-10 (at a distance of $\sim$2 Mpc) from the DECaLS survey imaging next to a model galaxy generated via \texttt{ArtPop} \citep{greco2022}. The same point spread function is applied to the model as the one in the data, and the model was injected into a blank patch of sky near the real galaxy. The model-generated image presents remarkable fidelity to the real image, capturing both the stellar populations of the galaxy in a star-by-star sense as well as the properties of the survey and telescope used to obtain the data. The ability for \texttt{ArtPop} to simulate galaxy images so realistically, given both intrinsic properties and observational ones, suggests that these models can be used in an inference framework to actually constrain the properties of galaxies, such as distance and mass, from imaging alone.\footnote{The galaxy parameters for the model shown in Figure \ref{fig:artpop} come from a posterior draw of \code{} after fitting this system; see Section 2.}

The challenge in performing traditional least-squares fitting or Bayesian sampling for such models is that they require a likelihood function with which to compare the models to the observed data. Traditionally this is done using a functional metric, e.g. $\chi^2$. In the case of simulating images, however, the simulator is \textit{stochastic} (as are the data), as the specific set of stars drawn from the IMF and their spatial locations will vary. Thus, the immediate quantitative nature of the data (i.e., the pixel values) cannot be compared one-to-one. While the naive residual obtained by, e.g., subtracting two similar galaxy images will be in principle smaller than those of two disparate images, a large amount of scatter is introduced in any such attempts.

Ultimately, the goal, given the ability to generate simulations of inherently stochastic systems, is not to ``match" exactly the observed image, rather, it is to marginalize over the features of the image that are stochastic (such as the precise locations of individual stars) and to fit to the higher order features that are shared in common between two galaxies sharing the same properties (such as, e.g., 2-point correlations in the distribution of resolved star locations). This often involves compressing the images into a set of summary statistics which aim to provide a low dimensional description of the data. However, this risks the loss of information if the summary statistics chosen do not fully capture the information present in the original image. We can see when a model fits well ``by eye'' but an analytic comparison requires dimensionality reduction or use of imperfect summary statistics.

One such forward modeling approach to studying semi-resolved populations was introduced in~\citet{conroy2016}, utilizing the pixel-CMD --- the distribution of pixels in luminosity-color space~\citep{bouthon1986,lanyonfoster2007,lee2018}. Unlike a traditional color-magnitude diagram, this method forgoes the need to fully resolve and measure photometry of individual stars, while capturing the distribution of pixels in the image. The pixelized-CMDs are thus reduced to a space for which a likelihood can be constructed, making it a viable approach for forward modeling and standard inference techniques \citep{cook2019}. Pixel-CMD's have been used to infer the distances and star-formation histories of the bulge of M31~\citep{conroy2016} and nearby elliptical galaxies~\citep{cook2020}. The limitation in this method is that by reducing the data space to a pixel-CMD, valuable spatial information is lost. In particular, the information that groups of adjacent pixels make up individual resolved stars is lost.

\subsection{Simulation Based Inference}
\label{sec:ml}
In this paper we introduce a new approach for utilizing realistic dwarf galaxy simulations to fit distances to real systems, leveraging recent advancements in the field of simulation-based inference (SBI; also known as likelihood-free inference -- see \citet{Cranmer2020} for a recent review). SBI represents a collection of methods aimed at performing inference when a method to simulate  observations exist, but direct comparison to data is challenging or intractable. Unlike traditional Bayesian methods which compute the likelihood with a metric such as $\chi^2$, simulation based inference forgoes this calculation using other techniques to compare model to data. An example of simulation based inference is Approximate Bayesian Computation, where a parameter value is added to the posterior if its simulated observation is ``close'' to the real observation using a self-defined distance metric, often computed by the euclidean distance between a set of tractable summary statistics. In these cases, the choice of summary statistic becomes of key importance, driving whether or not the right information is available when assessing model quality. 

In the past decade the field of simulation based inference has seen a rise in the use of probabilistic machine learning methods, specifically normalizing flows \citep{kobyzev2020,papamakarios2021}. Normalizing flows begin with a simple distribution --- often a standard normal --- before applying a series of bijective transforms to create a more complex distribution. The parameters of these transformations are then optimized to match a target distribution. 
An additional benefit is that these transformations can be made dependent on context; a normalizing flow can be trained to produce a different output distribution based on differing inputs. The specific type of transformation varies depending on the method used, but they are often designed to be easily invertible to reduce the computational demand while training and performing inference, yet remain expressive enough to match nearly any probability distribution \citep[e.g.][]{papamakarios2016, durkan2019}.

In simulation based inference, normalizing flows are used in neural posterior estimation~\citep{papamakarios2017,greenberg2019}. This method directly learns the posterior distribution by training a neural flow to estimate the posterior given the context of the observed data. The network is trained directly on parameters sampled from a proposal distribution (often the prior) and mock observations produced by the simulator. When trained solely on samples from the prior, the inference can be amortized. The computational burdens of simulating observations and training the network are placed up front, and as a result, performing inference on new observations takes very little time once training is complete. Within astronomy this is being utilized to quickly infer physical properties of stars, galaxies or transients from modern surveys where the vast number of sources means traditional inference techniques are intractable~\citep{hahn2022,Khullar2022, villar2022,wang2023, Zhang2023}.

Alternatively, the network can be trained in successive rounds focusing on a single observation, where in each round proposals are drawn from the posterior derived in previous rounds, known as sequential~\citep{greenberg2019}, or truncated sequential neural posterior estimation~\citep{deistler2022}. This approach is much more simulation efficient, as the simulations from successive rounds are targeted towards the region of parameter space that matches the given observation, but the trained network is no longer universally applicable.

A final crucial benefit to the machine learning based approach is that for high dimensional data, such as images, embedding networks can be trained alongside the flow \citep{greenberg2019}. An embedding network is an additional component that is needed to distill the original, high dimensional data, into a smaller set of summary statistics which are then passed to the normalizing flow. Due to the architecture of normalizing flows it is preferable to use a relatively small number of summary statistics, of order tens, rather than the potentially thousands of pixels in an image. For images this is most often a convolutional neural network (CNN) and because they are trained simultaneously in theory it will learn the ``best'' summary statistics which maximize the ability for the network to discern how parameters affect the data. As a result, these methods utilize the entirety of the information content, rather than relying on manually extracted summary statistics that are likely imperfect descriptors of the data.

In this paper we introduce the use of simulation based inference to directly infer the distances of dwarf galaxies from multi-band imaging data. We present this method in the Python package \texttt{Silkscreen}\footnote{This name is a reference to the silkscreen process used by Andy Warhol to define the Pop Art movement.} that utilizes the \artpop{}~\citep{greco2022} package to simulate realistic dwarf galaxies and the \texttt{sbi} \citep{tejero-cantero2020} package to train the model and perform inference. The code is available open source on github. \footnote{At this URL:\url{https://github.com/tbmiller-astro/silkscreen}}  We show inference applied to several real-world examples and highlight that \code{} is applicable to a wide range of stellar systems, from fully resolved to unresolved. We note that while our focus in this work is on measuring distances to dwarf galaxies, the methods developed herein are generalizable. They could be applied to other systems e.g., elliptical galaxies and the haloes of spirals or to infer stellar population properties alongside the distance. 

This paper is organized as follows: in Section ~\ref{sec:silkscreen} we introduce the concept of \code{} and the details of the simulation procedure, network architecture and training procedure. \code{} is applied to a mock galaxy in Section~\ref{sec:self-test} and then to a sample of real galaxies in Section~\ref{sec:real_gal_res}. We discuss limitations with the current implementation in Section~\ref{sec:issues} and provide a future outlook and development plans in Section \ref{sec:future}.

\begin{figure*}
    \centering
    \includegraphics[width=\linewidth]{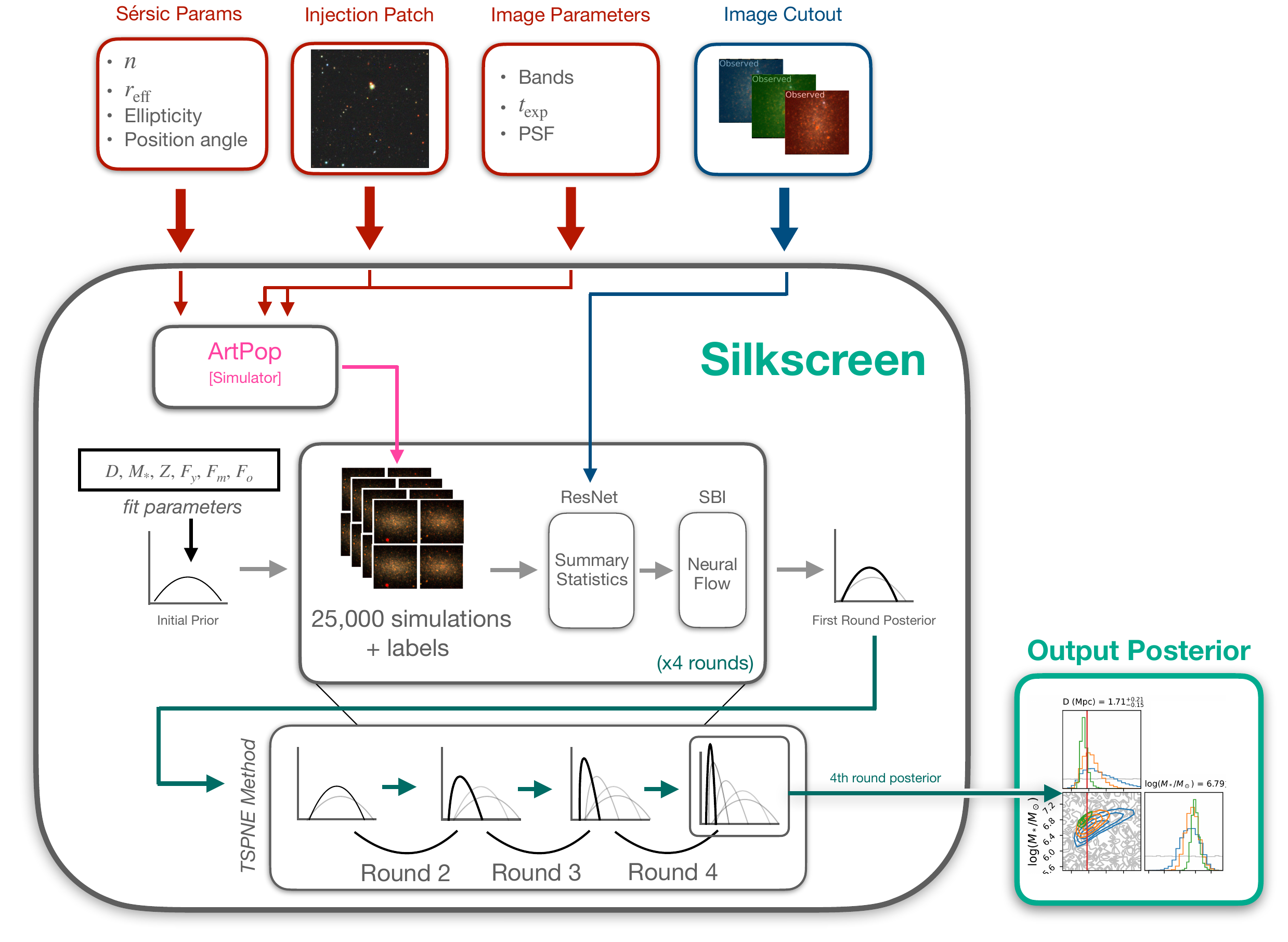}
    \caption{Schematic overview of the \code{} architecture. The external inputs to \code{} are shown at the top, colored by where in the code they are used; in our \textit{bespoke} fitting paradigm, easily-measurable parameters of input galaxies, including the Sérsic parameters, are fixed, along with the parameters describing the cutout image (i.e., which survey bands the imaging was obtained in, the exposure times therein, and the empirical PSF in the vicinity of the target). These inputs are passed to \texttt{ArtPop}, and all generated simulations will share the values set by these inputs. Additionally, we pass a large ``injection patch'' of sky from the survey being used in the vicinity of the galaxy cutout to \texttt{ArtPop}. During training, generated model galaxies are placed in randomly chosen locations in this large patch. The final input to \code{} is the multi-band image cutout of the real galaxy, which is utilized in the learning stage. In four rounds, \texttt{ArtPop} generates 25,000 simulations, drawn (at first) from highly uninformative priors. These simulations and labels are then passed to the ResNet for summary statistic extraction; then, SBI is used to infer via neural flows the posterior on the fit parameters given the span of generated models and the true input image. The posterior inferred in each round is then adopted as the prior for the subsequent round, and the process is repeated. As illustrated in the bottom row of the schematic, each sequential round (tends to) tighten the posterior, as progressively smaller regions of parameter space (at progressively higher likelihoods) are sampled. The final round's posterior is adopted as the output posterior distribution on the parameters. }
    \label{fig:ss-network}
\end{figure*}

\section{The \texttt{Silkscreen} Framework}
\label{sec:silkscreen} 
\subsection{Simulating Images of Dwarf Galaxies}
\label{sec:sfh}
To simulate dwarf galaxies we utilize the \artpop{} package developed in \citet{greco2022}. \artpop{} generates artificial images of dwarf galaxies ``from the ground up'' by drawing stellar masses from an IMF, calculating luminosities using isochrones, drawing positions for each start using a predefined brightness profile and then directly injecting stars into an image before convolving with a provided PSF.

\artpop{} is designed to be highly modular and flexible, meaning many types of systems can be generated, but also many choices must be made. For all the simulations shown here we use a Kroupa IMF \citep{kroupa2001} and MIST isochrones with stellar rotation $v/v_{\rm crit} = 0.4$~\citep{choi2016, dotter2016}. In \artpop{} we specify a magnitude limit (using the \texttt{mag\_limit} keyword) for which to resolve individual stars, below which stars are modelled as a smooth background. This ensures that stars far below the detection thresholds are not individually simulated, greatly reducing computational cost. For this study we set the $r$ band computation limit at $28\ mag$, well below ($\gtrsim 2 mag$) the detection limit in the HSC and DECaLS surveys, to ensure realistic simulated images~\citep{aihara2018, dey2019}. Note that while this limit is specified using $r$-band magnitudes, the resolved stars brighter than this limit are simulated in each of $g,r$ and $z$ or $g,r$ and $i$ bands for DECaLS and HSC respectively.

\begin{table*}
    \centering
    \caption{The parameters and priors for the distance, mass, stellar populations, and metallicity parameterization used in this study. Note that parameters marked as not free \textit{do} vary, but are explicitly dependent on other parameters. }
    \small
    \begin{tabular}{c c l l} 
        Parameter & Free? & Description & Prior \\  \hline \hline
         D & Y &Distance to galaxy (Mpc) & Uniform: $D_{\rm min}$ -  $D_{\rm max}$\\
         $\log (M_*/M_\odot)$ & Y &Total Stellar Mass & Log-uniform: $M_{\rm*, min}$ -  $M_{\rm* , max}$\\

         $F_{y}$& Y &Mass frac. in the young population (50 Myr - 250 Myr) & Trun. Norm. from $0-0.1$, $\mu = 0, \sigma = 0.025$\\
         $F_{m}$& Y & Mass frac. in the medium-age population (1.5 Gyr) & Uniform from $0-0.2$\\
         $F_{o}$& N &Mass frac. in the old population (10 Gyr) & Set to 1 - $F_{y}$ - $F_{m}$\\ 
         $Z_o$ & Y &Metallicity of old population, [Fe/H] & Gaussian with $\mu$ from MZR\tablenotemark{a}, $\sigma = 0.255$ \\
         $Z_{m,y}$ &N& Metallicity of medium and young population &  Set to $Z_o + 0.5$ \\         
         \hline  
    \end{tabular}
    \tablenotetext{a}{The prior for $Z_o$ depends on the mass, using a Gaussian with a center defined by the $Z$ associated with that mass in the \cite{kirby2013} MZR, and the $\sigma$ chosen being 1.5$\times$ the scatter in the measured MZR. }
    \label{tab:dwarf_model}
\end{table*}

The next important set of choices is how to parameterize the star formation and metallicity history. For this initial exploration we opt for a relatively simple three population parameterization with five free parameters. These decisions were informed by known star-formation histories of local group dwarf galaxies \citep{weisz2011} and the variation of SBF magnitude with stellar population parameters \citep{Blakeslee2001, Jensen2003, Raimondo2005, greco2021}. 

The parameters and their priors are summarized in Table~\ref{tab:dwarf_model}. 
The youngest population $F_y$ is modeled with a continuous star-formation ranging from 50\,Myr\,--\,250\,Myr. The prior of $F_y$ is a truncated normal, centered at 0 with a maximum of 10\%. The medium aged population has a fixed age of 1.5 Gyr, with a uniform prior of $F_m$ ranging from 0\%-20\%. The rest of the mass is modeled as an old stellar population with an age of 10 Gyr. For the metallicity of the old population we follow the stellar mass-metallicity relation of local-dwarf galaxies presented in~\citet{kirby2013}. Specifically we use a Gaussian prior in [Fe/H] for $Z_o$ with a width of 0.255, $1.5\times$ the scatter reported in~\citet{kirby2013}. To roughly mimic the expected metallicity evolution of chemical enrichment increasing with time, we set the metallicity for the middle and young populations as $Z_{m,y} = Z_o + 0.5$. The parameterizations of both the star-formation and metallicity histories are simplified assumptions of what we would expect for realistic dwarf galaxies. For this initial study these choices represent a reasonable starting place without adding too much complexity.

In order to produce mock images with realistic noise profiles we inject them into real data. Using \artpop{} we first render pristine, i.e. noise free, images that have been convolved with the PSF and have the pixel scale of the specified survey. Then we add Poisson noise before injecting these images into patches of real data. We do not pre-select ``empty'' regions as we want to pass the full distribution of data quality to the network such that it can produce realistic uncertainties. For each example shown we utilize a large (roughly $2000$ pixels $\times 2000$ pixels) region offset by about 12 arcminutes from the main target, see section~\ref{sec:data} for additional details on where the survey data is downloaded from . When a new simulation is performed we then simply select a random position within this patch to inject the simulated galaxy into. Given the field of view of DECam and HSC of 2.2 deg and 1.8 deg respectively, using a patch of sky close to the target galaxy should ensure the observations were taken at the same time to most closely match the noise properties and PSF.

\subsection{Network Architecture}
\label{sec:NN}
To perform simulation based inference we rely on the \sbi{} python package \citep{tejero-cantero2020}. Specifically we use the truncated sequential neural posterior estimation technique (TSNPE) to train the network in multiple rounds focusing on a single observable~\citep{greenberg2019,deistler2022}. The sequential nature of this method means the trained network is only applicable to a single galaxy, but provides better simulation efficiency as each successive round of simulations is drawn from the posterior derived in the previous round. The advantage of the TSNPE technique, compared to other sequential methods, is the ease of training, as the proposal distributions are truncated versions of the prior leading to a simpler loss function and more efficient optimization. Further details on the training procedure are discussed below in section~\ref{sec:walkthrough}.

The network contains two main components, the embedding network which distills the multiband images into a set of summary statistics (default of 16) which describe the input images. These are then passed to the normalizing flow,the second part of the network, which takes these summary statistics as input and predicts the posterior distribution. As discussed in section~\ref{sec:ml} these two parts of the network are trained simultaneously so that the embedding networks learn to extract optimal summary statistics to differentiate images produced with different input parameters. 

For our embedding network we use an instance of a Residual Neural network \citep[ResNets;][]{he2016}. Compared to traditional CNNs, ResNets include skip connections where the output of a given layer is added to the input following the calculation. They have been shown to be easier to optimize, and allow for deeper networks without degradation of generalizability. Our implementation is based on that in \texttt{torchvision} \citep{torchvision2016}. It contains an initial convolution layer which maps the given number of input filters to 16 channels, each representing an abstract learned representation of the data. The bulk of the network consists of four segments. Each segment contains the two bottleneck blocks, where the number of channels is first reduced using a 1$\times$1 convolution followed by a 3$\times$3 convolution before the number of channels is expanded again to the original number. At this point the input is added directly to the result completing the skip connection. The four segments increase the number of channels from 16 to 32, 64, and 128 respectively. This is followed by a spatial averaging to reduce the total number of features to 512. The final layer is a fully connected layer which further reduces the number of features to the desired number of summary statistics, 16 for this study. This represents a slight modification of the original ResNet implementation presented in \citet{he2016} that is reduced in complexity by lowering the number of channels in each of the four segments.

For the normalizing flow part of the network we use a Neural Spline Flow  \citep[NSF;][]{durkan2019} and its implementation in \texttt{sbi}. This architecture parametrizes each transformation using a monotonic quadratic spline, which is parameterized by the location and derivative of the transformation at each knot. Our implementation uses a NSF consisting of five transformations with eight spline bins each. Each transformation is parameterized using a fully connected layer with 50 features and a dropout layer~\citep{durkan2019}. A dropout layer randomly masks some subset of the features during each step of training, we use a probability of 0.2, to help with regularization of the network. This implementation of a NSF is close to the default included in the \texttt{sbi} package, with the only addition being a non-zero dropout probability.

%For our embedding network we use an instance of a Residual Neural network \citep[ResNets;][]{he2016}. Compared to traditional CNNs, ResNets include skip connections where the output of  a layer is added to the input. They have been shown to be easier to optimize and allow for deeper networks without degradation of generalizability. Our implementation is based on that in \texttt{torchvision} \citep{torchvision2016}. It contains an initial convolution layer which maps any number of input filters to 16 channels proceeded by four segments. Each segments contains two bottleneck residual blocks with an increasing number of channels, 16, 32, 64, and 128 respectively. This is followed by an average pooling to 512 features and a fully connected layer resulting in the desired number of summary statistics, which defaults to 16.

\subsection{Current Pipeline: Bespoke Simulations and Training}
\label{sec:walkthrough}

The current pipeline for inference with \code{} is tailored to each specific galaxy of interest; thus, a number of observations need to be assembled. An overview of the software architecture is provided in Figure \ref{fig:ss-network}. First, cutouts of the galaxy in a set of filters, along with their PSFs, must be provided. Additionally, details about the telescope and survey, including the average sky surface-brightness and exposure time are necessary to accurately calculate the noise properties. Finally a parameterized estimate of the observed light distribution is required; the current supported models are Plummer \citep{Plummer1911} and S\'{e}rsic \citep{sersic1968} profiles. These are all collected into a \texttt{SilkscreenObservation} class which can then be used for inference. We note that additional profiles can easily be implemented; here, the sample of observed galaxies used are well described by the two profiles above. 

Next, one must choose a parameterization for the star-formation history, and priors for the parameters. We provide routines to generate the default options discussed in section~\ref{sec:sfh} in the \texttt{silkscreen.simmers} and \texttt{silkscreen.priors} modules respectively. The neural network is also available in the \texttt{silkscreen.neural\_nets} module. The modularity and flexibility of parameterization with both \artpop{} and \code{} means further extensions can be easily made; here we focus on the default set. 

Training the posterior estimation network is performed using the \sbi{} python package. This methodology uses the posterior of each successive round to truncate the prior and achieve more targeted simulations for a given observation. %Through experimentation, we have found that including half the total simulation budget in the first round, drawn from the prior, and then splitting the other half among three successive rounds, leads to successful inference.
Unless otherwise specified, in this paper we train the posterior estimator in four rounds with 25,000 simulations each, restricting the prior each time following the TSNPE method. We tested the effect of using more simulations, specifically four rounds of 50,000 simulations each. One of the three galaxies that we used for testing showed marginally tighter constraints, with the other two showing no improvements. If the computing resources are available, more simulations should in general improve constraints but we have found four rounds of 25,000 simulations to be a reasonable starting place for most problems. Similarly, we find that at least four rounds are required to appropriately narrow the final posteriors to the type of constraints being sought (e.g., distances to $\sim$10\% or less), while further rounds would enforce too-tight of a parameter space. 

For training, we use the AdamW \citep{loshchilov2018} optimizer and apply a learning rate of $2\times 10^{-4}$ for the flow part of the network and $1\times 10^{-6}$ to the embedding part of the posterior estimator network. At each training step we use a batch size of 256 simulation and parameter pairs. After the initial round of training we lower each learning rate by a factor of 10. The weight decay, which describes the normalization applied to parameters during training , is set to $10^{-4}$. Finally the number of learned summary statistics is set to 16; during hyperparameter tuning (see below), we found that the exact number of summary statistics had little impact on the results. Each round of training continues until the loss of a validation set (10\% of total data) fails to improve for 15 successive epochs. Before training, the images are passed through an arc-sinh function, to normalize them and limit the effects of contaminants such as bad pixels or bright stars, and are then z-scored: the mean is subtracted from each sample, which is then divided by the standard deviation. Z-scoring is conducted across all filters simultaneously, to preserve color information. Similarly, the parameters used to generate the simulations are z-scored. These processes are primarily used because neural networks perform more efficiently when the dynamic range of the data is uniform and of order unity.

To find the best training hyper-parameters, i.e. all of the parameters described above, we use the \texttt{optuna} package \citep{akiba2019},  applied to the test case shown in Section~\ref{sec:self-test}. Optimizing these parameters is crucial for successful training. As with many different applications, we find the batch size and learning rates to be the most impactful hyper-parameters. For batch size we explore the range of 8--256 and find the optimal range to be at the larger end, $>200$. For learning rate we tested a wide range and found we only achieved consistent results when we used different learning rates for the flow and embedding components of the network. We explore both learning rates independently with values ranging from $10^{-7}$ to $10^{-2}$, sampled logarithmically. We find the optimal learning rate for the embedding network to be lower, around $10^{-6}$, and the learning rate for the flow section of the network at least an order of magnitude higher. We also varied the strength of the weight decay, which affects attempts to regularize the weights and biases back to zero, between $10^{-6}$ to $10^{-2}$ and the number of summary statistics between 8 and 64, but found these had only marginal effects on the network training.

A crucial final step is the validation of the trained model. The training loss curves are unfortunately not very useful for this style of sequential neural posterior estimation because the expected loss will depend on each specific observation and exactly how the prior is restricted at each step. However, it is crucial to assess if the posterior, and therefore uncertainties, are well calibrated. That is to say, does the true value lie within the 90\% confidence interval 90\% of the time~\citep{Lueckmann2021,Hermans2022}. To assess this we employ the technique known as simulation-based calibration~\citep[SBC][]{Cook2006,talts2018}. In short, SBC involves performing inference on a set of simulated observations where the underlying parameters are known. From this inference one can obtain a set of posterior draws. The ``rank statistic'' is calculated as the position of the true parameter value in an ordered list of the posterior draws. Once repeated for a large number of simulated observations, the distribution of the rank statistics is analyzed. For a well-calibrated posterior the distribution of rank statistics should be uniform, suggesting the true values are uniformly sampled from the posterior indicating it is well calibrated. The benefits of SBC are that it is universally applicable and deviations from a well calibrated posterior can be easily seen by analyzing the distribution of the rank statistic~\citep{talts2018}.

Another important step is performing posterior predictive checks; for~\code{} this involves analyzing images simulated using parameters from the posterior. This helps to assess any model mis-specification; if \artpop{} is not able to realistically simulate the galaxy of interest the results should be less trustworthy. These procedures are not directly integrated into the silkscreen pipeline but they are needed to better interpret the results model. We explain our workflow for these validation techniques in Section~\ref{sec:examp_gal}.

\section{Inference on Mock Galaxies}
\label{sec:self-test}

Before applying \code{} to real galaxies, we perform several injection-recovery tests to assess the framework's ability to recover distances in a controlled manner. First, we perform a simple test to confirm that the framework can distinguish between a galaxy placed at different distances for which the apparent size is held fixed. For this test, we create a mock galaxy using \texttt{ArtPop} with a stellar mass of log M$_{*}=7.4\; M_{\odot}$ and a metallicity of $-1.4$, with $F_y$ set to 0.02 (2\% by mass) and $F_m$ set to 0.17. We fix the Sérsic index of this model to 1.1, and also fix the angular size to 15$''$, the position angle to 45 degrees, and the ellipticity to 0.2. We then simulate the galaxy at a variety of distances ranging from 5--19 Mpc, and produce mock images at the HSC-SSP Wide depth, injecting into patches of blank sky near one of the real galaxies fitted later in this work and convolving with the HSC-SSP Wide PSF for that location.

\begin{figure}
    \centering 
    \includegraphics[width=\linewidth]{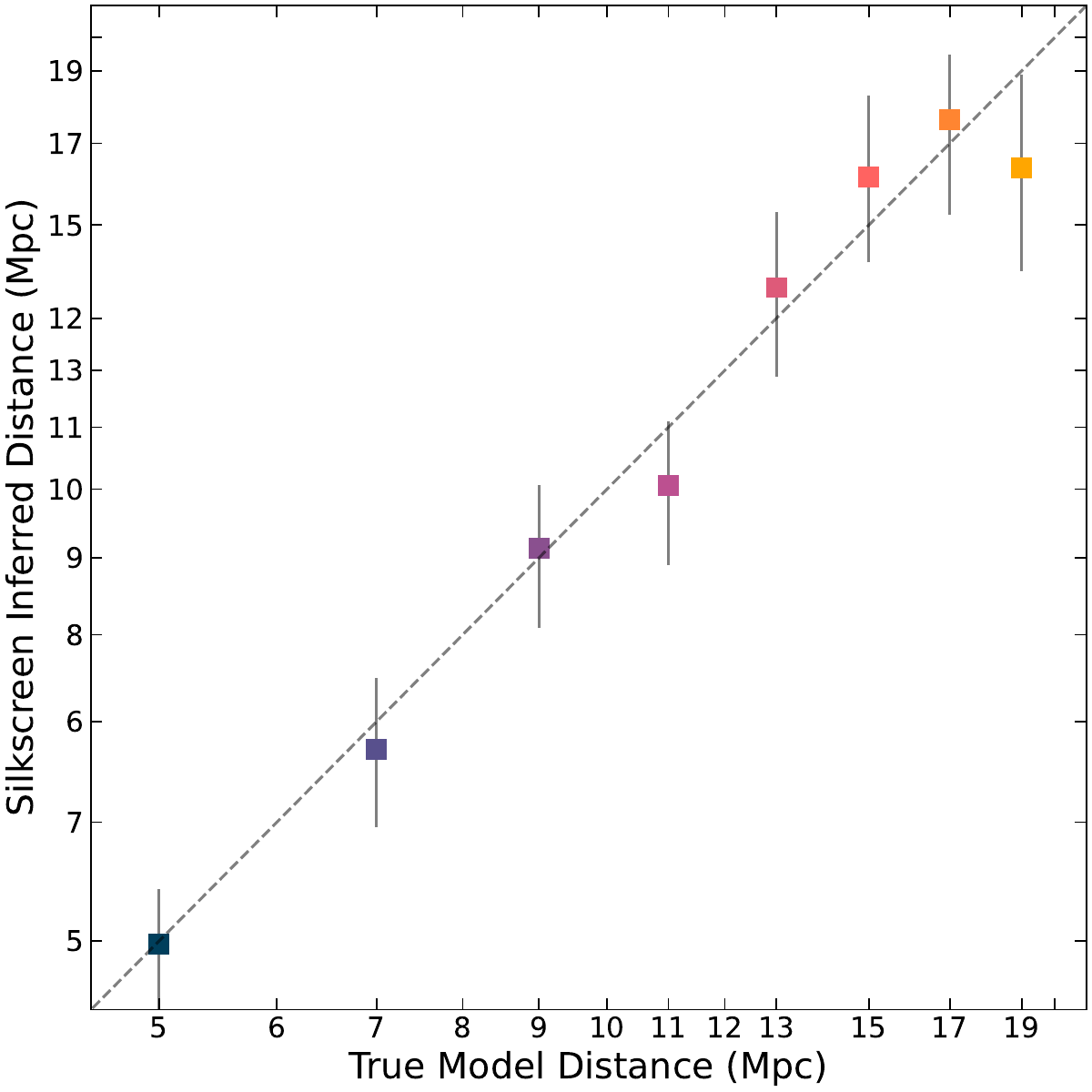}
    \caption{Inferred vs.\ actual distance for a simulated model galaxy of log M$_{*}=7.4\; M_{\odot}$, with a metallicity of $-1.4$, $F_y$ set to 0.02 (2\% by mass) and $F_m$ set to 0.17. We inject this galaxy at different distances and fit it with \code. We keep a fixed Sérsic index of 1.1 for this test, and fix the effective radius to 15$''$. This means that the more distant galaxies are intrinsically larger, and so allows for the demonstration that \code{} is not using the apparent size as a marker for distance.  We find that \code{} is able to recover the distances to the injected systems well, with all galaxies falling within 1-$\sigma$ of the true values and with no systematic bias in the recovered distances.}
    \label{distance-test}
\end{figure}

\begin{figure*}[htb!]
    \centering
    \includegraphics[width=\linewidth]{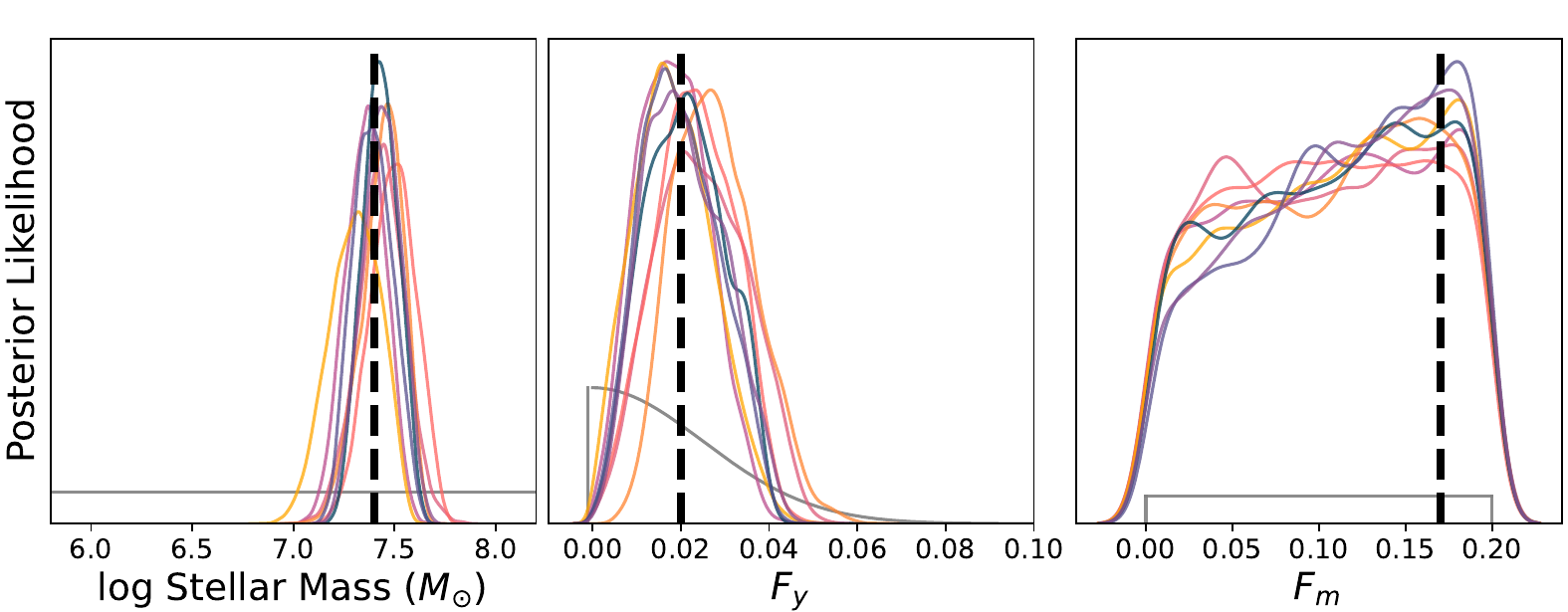}
    \caption{Kernel Density Estimates (KDEs) of posterior stellar mass (left), young fraction by mass (middle) and medium-aged fraction by mass (right) for the set of mock galaxies simulated at distances from 5 to 19 Mpc (see Fig.\;\ref{distance-test}); colors of these curves match the points in that figure. Priors are shown in light gray, and the true values are shown as vertical dashed lines. We note that the apparent leaking of the posterior distributions beyond the prior bounds is a visual artifact from the KDE smoothing. We find that in addition to the distances being recovered for these mocks, the stellar population parameters are also reasonably well-fit. We do not show the KDEs for metallicity as the mass-metallicity relation prior influences the derived metallicity (but they are consistent with the model values as well). The fraction of the mass in young ($50-250$ Myr old) stars has a prior which preferences an $F_y$ of 0, so the ability for the inference to constrain it to 0.02 (the value for this model) with no weight at 0.0 implies the value is not prior dominated. In contrast, the intermediate age population (a single-age burst of 1.5 Gyr in age) is relatively unconstrained from its uniform prior between 0 and 0.2, though there is a recovered increased likelihood toward the correct value for this model.  }
    \label{fig:injection-other}
\end{figure*}

We apply \code{} to this faux-galaxy as if it were a real input. The result of this injection test is shown in Figure \ref{distance-test}. We find that \code{} can successfully recover the true distances in this idealized case to within 1-$\sigma$ of the posterior-derived uncertainties for all but the simulated galaxy at 20 Mpc, which lies just outside the 1-$\sigma$ bounds. We see no major trends in accuracy or precision with distance across the ranges tested, other than uncertainties increasing moderately toward larger distance. We note the axes in Figure \ref{distance-test} are shown on a logarithmic scale. The fact that the uncertainties appear constant with distance using this scaling implies the fractional uncertainty, i.e. $\sigma(D)/D$, is constant. In Figure \ref{fig:injection-other}, we show Kernel Density Estimate (KDE) distributions for the other fit parameters (excluding metallicity, as it is influenced by the MZR prior). We find that across the set of simulated mocks at different distances, the stellar mass and fractional young population are well recovered, centering on the correct value and with no considerable bias. In particular, $F_y$ appears to be constrained by the fit; its prior (a truncated normal) has maximal weight at a value of 0. Stellar mass in contrast has an uninformative prior, so the constraints there are quite clearly not prior dominated. The intermediate-age population (the modeled 1.5 Gyr population) is relatively challenging to constrain; we find it roughly traces the uniform prior it is initialized with, though it does show increased likelihood at the upper end (where the correct value for the tested mock lies). Ultimately we are not interested in constraining $F_m$ so much as marginalizing over it.

The fact that this recovery test was successful, despite all simulated galaxies having the same apparent size, suggests that \code{} does not use (or need to use) the size of the galaxy within the cutout as a proxy for distance, and that it is indeed using pixel-to-pixel variations in some way in order to infer the distance. As we will investigate more fully, the primary benefits of ``learning'' this posterior directly via SBI and ResNets is the leveraging of multi-color information, and a more explicit modeling and marginalization over stellar population parameters, which removes the need for an empirical (e.g., color-based) calibration.

\begin{figure*}
    \centering
    \includegraphics[width=0.74\linewidth]{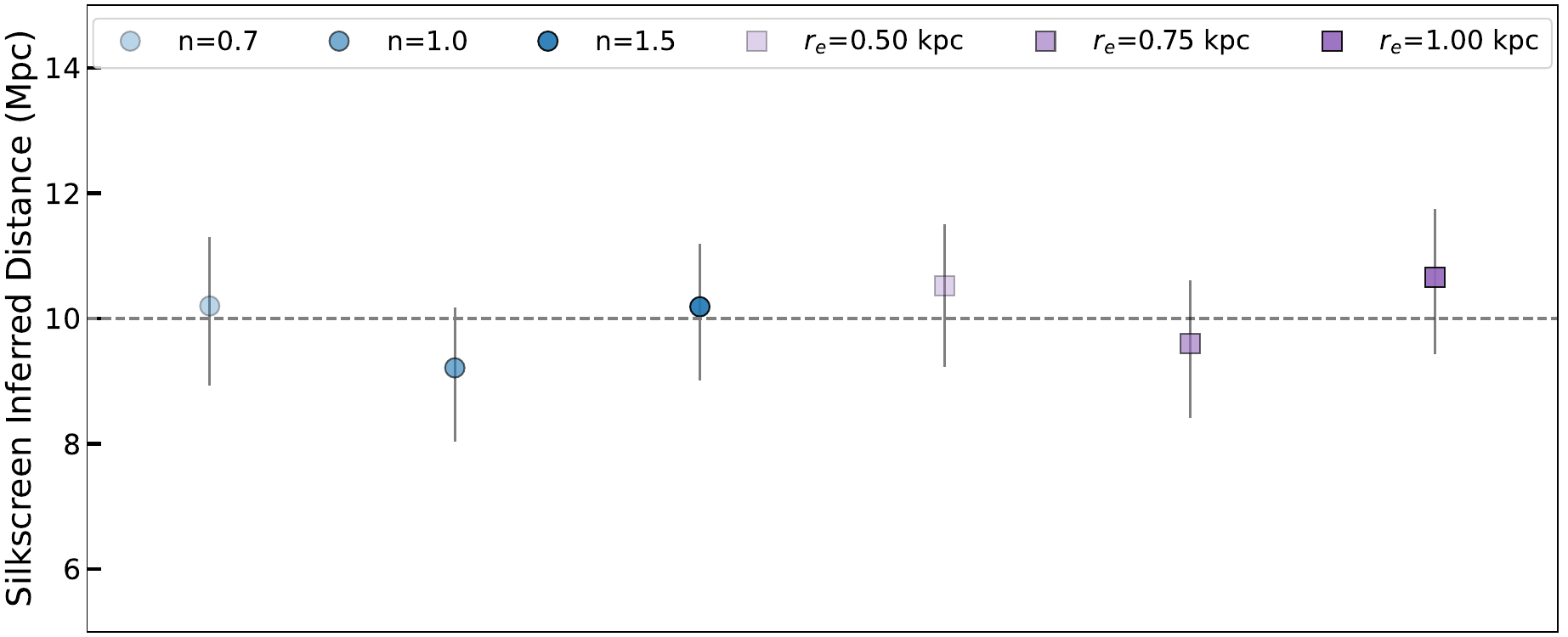}
    \caption{Distance recoveries for a set of injections all at 10 Mpc, varying the model intrinsic size between 0.5, 0.75, and 1 kpc with a fixed $n=1$, and the Sérsic index between 0.7, 1.0, and 1.5 with a fixed $r_e=0.75$ kpc. We find that \code{} recovers the correct distance within 1-$\sigma$ in all cases, with no apparent systematic biases in recovered distance based on these parameters (albeit with a small test sample). The scatter in the distance estimates is $\sim$0.5 Mpc, corresponding to 5\% at a distance of 10 Mpc.}
    \label{injection-recovery}
\end{figure*}

We additionally explore how, if at all, variations in galaxy morphology affect the inference by performing a set of fits using injected models all placed at the same distance (10 Mpc) but with varying intrinsic sizes and Sérsic indices. We create three mock galaxies with Sérsic indices of 0.7, 1.0, and 1.5, holding the effective radius fixed at 0.75 kpc, and three mock galaxies with sizes of 0.5 kpc, 0.75 kpc, and 1 kpc, holding the Sérsic index fixed at $n=1$. We then fit them using \code. The results of this test are shown in Figure \ref{injection-recovery} We find that \code{} successfully recovers the true distance (with a scatter around 10 Mpc of 5\%), with no immediately apparent systematic biases that trend with size or Sérsic index.

\section{Application to a Sample of Real Galaxies}
\label{sec:real_gal_res}

Here we describe the results of running \code{} on a sample of nearby galaxies with distances previously measured using TRGB or SBF. First, in Section~\ref{sec:data} we describe the selection of the sample and the data used, then analyze  the inference results for a representative galaxy (Section \ref{sec:examp_gal}), discussing the quality of the inference in detail. Then, in Section \ref{sec:allgals}, we assess the robustness of \code{} using the full sample, comparing our inferred distances to the literature values.

\subsection{Data and Galaxy Sample}
\label{sec:data}

We assess \code's ability to estimate distances for real galaxies by assembling and fitting a sample of 20 galaxies with distances measured using either HST-based TRGB or ground-based SBF. The galaxies span a range of distances from 2 - 12 Mpc. Imaging for these galaxies was retrieved from either the HSC-SSP DR2 \citep{aihara2018} ($gri$ bands) when available, or from DECaLS-DR10~\citep{dey2019} ($grz$ bands). 

The galaxy sample is largely based off of the Local Volume Galaxy Catalog (LVGC) presented in \citet{karachentsev2013,Anand2021}. Our selections fall into three main categories. This first is six dwarf galaxies in the direction of the Sculptor galaxy group. This is a loose group collection which stretches from approximately 1.5 Mpc to 5 Mpc in distance~\citep{karachentsev2003}. The second category is selected from the LVGC to lie within the legacy survey southern footprint, have measured SBF or TRGB distances of $5 < D ({\rm Mpc}) < 8$ and have a KK designation (i.e.presented in \citet{Karachentseva1998}) with observed $m_K < 14$, resulting in 5 galaxies. For these first two groups inference will be performed on imaging data from the DECaLS survey. All but one of these galaxies have HST measured TRGB distances. The final category is selected from the HSC-SSP wide footprint. This data is significantly deeper than that of DECaLS, so we expand our selection to larger distance. We select everything in the local catalog with $2 < D < 12$ Mpc and $m_K < 19$, again with TRGB or SBF measured distance. After cutting three galaxies for poor data quality or proximity to a bright star, this results in a sample of nine galaxies, of which three have distances measured from TRGB, with the remaining six having distances measured via SBF. We note these distance measurements come from numerous different studies which employ differing data, methods and calibrations (See Table~\ref{tab:dist}) so the reported distances may not be fully compatible with each other.

This is not meant to be an exhaustive list of galaxies in the local neighborhood, but was chosen to be representative selection while creating a reasonable number of galaxies to perform inference on. This sample spans a large range of distances across two different surveys, at $D\lesssim 4$ Mpc, the galaxies should be fully resolved while at $D \gtrsim 8$, they will be mostly unresolved. We discuss the literature TRGB and SBF measurements for these galaxies to provide context for the comparison not because they are directly simulated or used in the inference procedure of \code{}. \code{} performs inference directly on the images provided.

To begin, we download a large cutout of 500$\times$500 pixels surrounding each galaxy and PSF for each band directly from the Legacy Survey website for DECaLS data\footnote{Following the URL patterns described at this URL: https://www.legacysurvey.org/viewer/urls} or using the \texttt{unagi}\footnote{https://github.com/dr-guangtou/unagi} python package for HSC data. To determine structural parameters we smooth the $r$ band image with a Gaussian kernel with $\sigma = 3$ pixels before fitting a single Sersic profile~\citep{sersic1968} with \texttt{pysersic} \citep{Pasha2023}. The angular size, Sersic index, position angle and axis ratio are then used in the \artpop{} simulator as described in Sec.~\ref{sec:walkthrough}. Even though we select the original cutout at the location of the galaxy it may not be perfectly centered. To refine the location we  use the central location found in the Sersic fitting process and select a smaller cutout of 160$\times$160 pixels to be used for inference. The injection patch was then chosen as a 2000$\times$2000 pixel slice of imaging, offset by 12 arcminutes from each galaxy's location.

We run \code{} inference on each galaxy closely following the procedure outlined in Sec.~\ref{sec:walkthrough}; the same neural network is used along with four rounds of simulations and training consisting of 25,000 simulations each round. All galaxies use the same priors for the stellar population parameters described in Table~\ref{tab:dwarf_model}. For the DECaLS and HSC galaxies we use $1.5< D ({\rm Mpc}) < 9$ and $1.5 < D ({\rm Mpc}) < 15$ respectively as the limits on the distance prior. For all galaxies we use a log-uniform prior in stellar mass spanning $5.8 < \log\, M_*/M_\odot < 8.2$. The distance limits were chosen based on our selection from the catalog described above. The stellar mass prior was chosen to encompass the stellar mass of each galaxy implied by the K band magnitude measured from the LVGC, assuming a mass to light ratio near unity. While the exact bounds of the prior do not have large effects on the results, if a large fraction of galaxies simulated from the prior are too faint and undetected in the resulting image, training the network can struggle, leading to poor results. 

\subsection{Example galaxy: ESO 410-005}
\label{sec:examp_gal}
\begin{figure*}
    \centering
    \includegraphics[width = 0.9\textwidth]{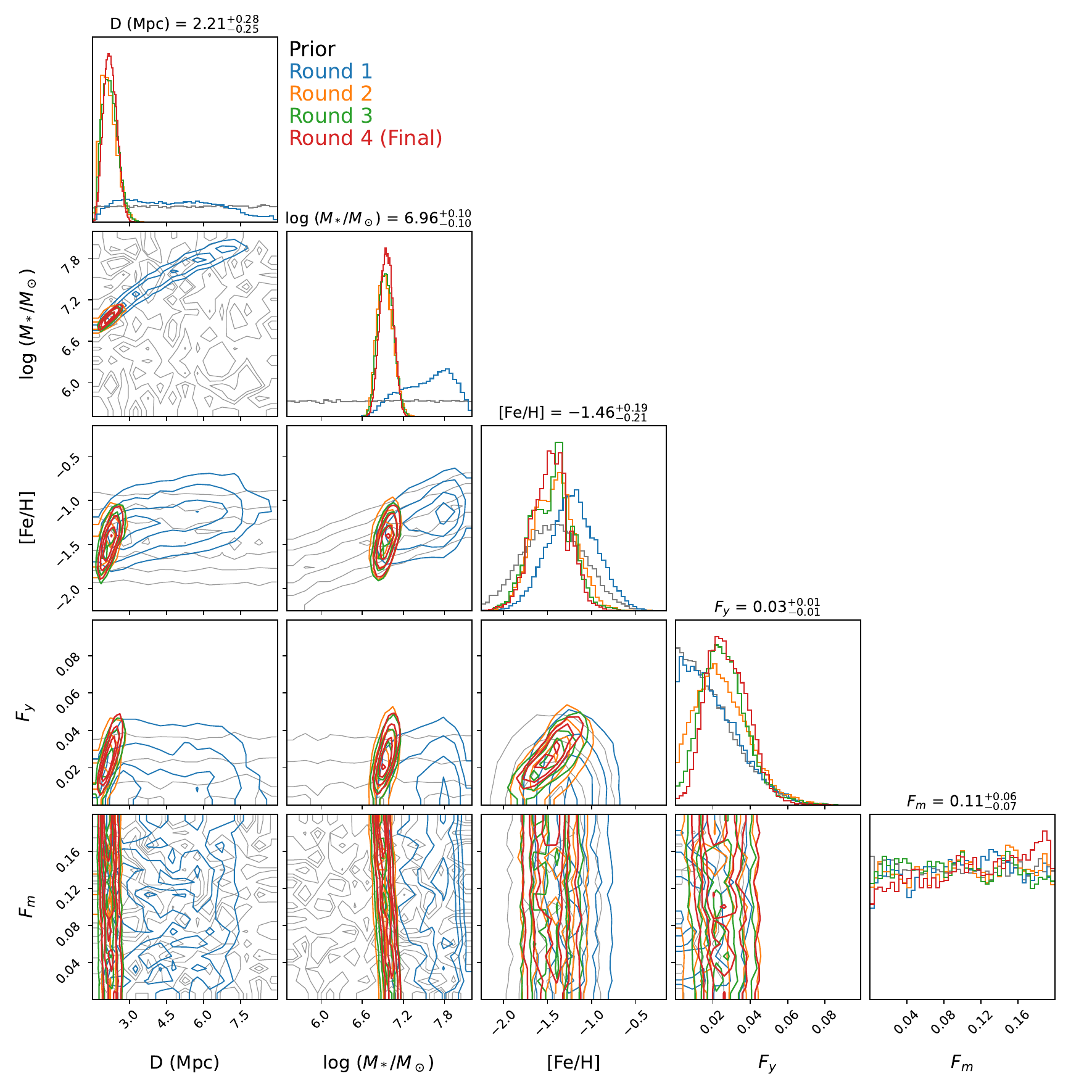}
    \caption{Showcasing \code{} inference performed on the galaxy ESO 410-005 at a known distance of 1.9 Mpc \citep{dacosta2010,Tully2013}. This corner plot displays the prior distribution along with the posterior produced from each successive round of training. The constraints from the posterior become more precise after the initial round of training, where targeted simulations are used and appear to converge between the third and fourth round. The distance is very well determined. The posteriors of the stellar population parameters $F_m$ and [Fe/H] follow the prior, but the stellar mass and $F_y$ are also reasonably well constrained.}
    \label{fig:eso410_05_corner}
\end{figure*}

Figure~\ref{fig:eso410_05_corner} displays joint posterior distribution produced by \code{}~for the galaxy ESO 410-005 as a representative example. The posterior distribution after each round of training is shown alongside the prior distribution. Investigating the initial rounds of the posterior, we find that it produces large ranges of possible distances and stellar masses along a tight relation. We suspect that this represents an ``iso-brightness'' surface; the most obvious characteristic for the network to learn is likely the total brightness, which produces this relation in the $\log\ M_* - D$ plane. It is only in the later rounds that the posterior tightens around the correct distance of 1.93 Mpc \citep{dacosta2010,Tully2013}. We also note the similarity between the posteriors of Round 3 and Round 4 suggesting that the training has converged; more rounds of simulations and training will likely only yield marginal gains.

The stellar mass is well-constrained as well, with an uncertainty of 0.10 dex. Even after the four rounds of training, we find that other stellar population parameters remain largely unconstrained. Specifically, the metallicity follows the input MZR with similar width, $\sigma$ = 0.2 in the posterior compared to 0.22 from the prior, and the fraction of mass in the medium population $F_m$ is wholly unconstrained. $F_y$ is different from the prior with a value of $0.03 \pm 0.01$, while the prior is centered at 0. Physically, varying $F_y$ impacts the image the most with the presence of bright blue main sequence stars compared to the more subtle effects of varying metallicity or changing the contribution of 1.5 Gyr compared to a 10 Gyr old population.

We emphasize that the inability to constrain the metallicity and star formation history is not a failure of the method and is indeed the expected outcome; the changes in measurable imaging properties produced by varying $F_m$ or $Z$ in ground-based images are subtle. Generally these parameters are measured using deep space-based CMDs which reach down to the main-sequence \citep[e.g.][]{weisz2011,mcquinn2024}. The function of these parameters is to allow their intrinsic uncertainty to be marginalized over in our estimates of the distance. In some cases where the galaxy is bright and well resolved, or using space-based data, \code{} may be able to learn about these stellar population properties. In most cases, however, we expect their inference to be prior dominated, which for $F_m$ simply means that the uninformative prior results in an uninformative posterior.

We do note that because the posteriors on several parameters tend to be prior dominated, care must be taken not to impose strong, incorrect shapes to the prior distributions, as this can lead to systematic biases in the retrieved distances. We attempt to mitigate this via the use of a flat prior on $F_m$. We found, however, that $F_y$ was too dominant given its lack of constraint when sampled by a flat prior. The choice of a truncated normal with a few percent (by mass) in scale effectively keeps the resulting $F_y$ close to 0 (as expected in reality) and requires significant evidence to be pushed to higher fraction; this works well for the sample presented here but may not generalize to all galaxies.

\begin{figure}
    \centering
    \includegraphics[width = \columnwidth]{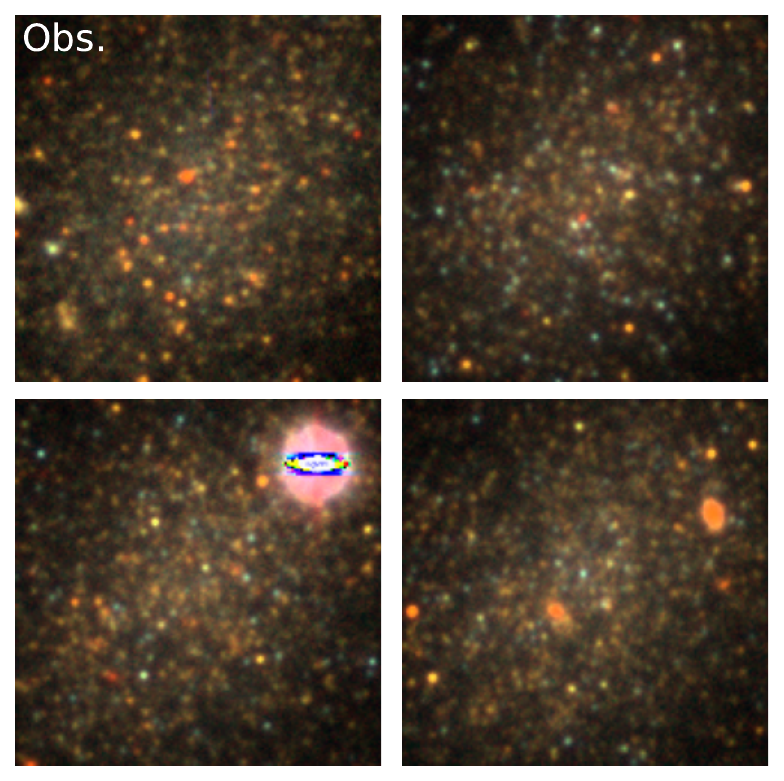}
    \caption{The observed image DECaLS image (top left) for the galaxy ESO 410-005 alongside posterior predictive images produced by \code{}. The simulated images match the overall characteristics of the observed image well ``by eye", reinforcing the idea that \artpop{}, and the parameterization chosen here, can produce realistic images of dwarf galaxies.}
    \label{fig:eso410_05_ims}
\end{figure}

In Figure~\ref{fig:eso410_05_ims} we investigate the posterior predictive distribution. The image of ESO 410-005 from DECaLS is displayed alongside simulated images using parameters drawn from the posterior. Visually, we find that the simulated images match the observed image well, indicating that the simulator we are using and the parameterization we have chosen are sufficient to model realistic data. In detail there are subtle differences, e.g., there appear to be more bright blue sources in the simulated images which could indicate an issue with the parameterization of the youngest population (see Section~\ref{sec:issues} for further discussion).

\begin{figure}
    \centering
    \includegraphics[width=0.95\columnwidth]{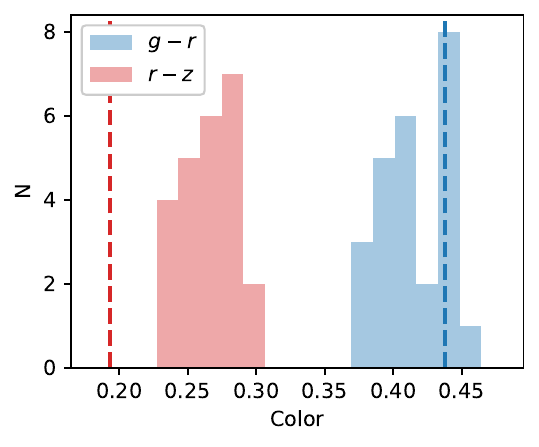}
    \caption{Comparing the observed colors of ESO 410-05 to those measured from the posterior predictive images produced by \code{}. The red and blue histograms show the spread of $g-r$ and $r-z$ colors measured directly for 25 posterior predictive images. The lines show the observed color. We find the observed $g-r$ color matches that of the posterior predictive images while the observed $r-z$ color is slightly bluer by 0.075 mag}
    \label{fig:eso410_05_col_comp}
\end{figure}

In a more quantitative comparison we compare the observed colors of the ESO 410-05 to those measured from the posterior predictive images, displayed in Figure~\ref{fig:eso410_05_col_comp}. For each image we measure the color within the effective radius in an elliptical aperture which follows the ellipticity and position angle found in the Sersic fitting process. We note that we make no attempt to mask interloping sources so we expect there to be some scatter in the measured color. For ESO 410-05 we find the simulated and observed $g-r$ colors agree well. The observed $r-z$ color is bluer than than the median of the posterior predictive images by a small amount, approximately 0.075 mag. This agreement, at least in $g-r$, appears somewhat at odds with the observation of an excess of bright blue sources. However we suspect this is due to the spatial distribution. In the simulated images the blue sources are more extended, as they are prescribed to follow the same spatial distribution as the older stars. In the observed image the blue light appears more compact leading to a similar total color while visually appearing different.

When comparing the observed and posterior predictive colors for the entire sample of galaxies in Appendix~\ref{sec:samp_res_appendix} we find similar results to ESO 410-05. The observed $g-r$ colors generally match those from the posterior predictive images but the observed $r-z$ (or $r-i$) colors tend to bluer by about 0.05-0.1 mag on average. This may be indicative of an issue with the isochrones as there can be systematic uncertainties in the near-infrared ~\citep{Raimondo2009, greco2022}.

\begin{figure}
    \centering
    \includegraphics[width = 0.95\columnwidth]{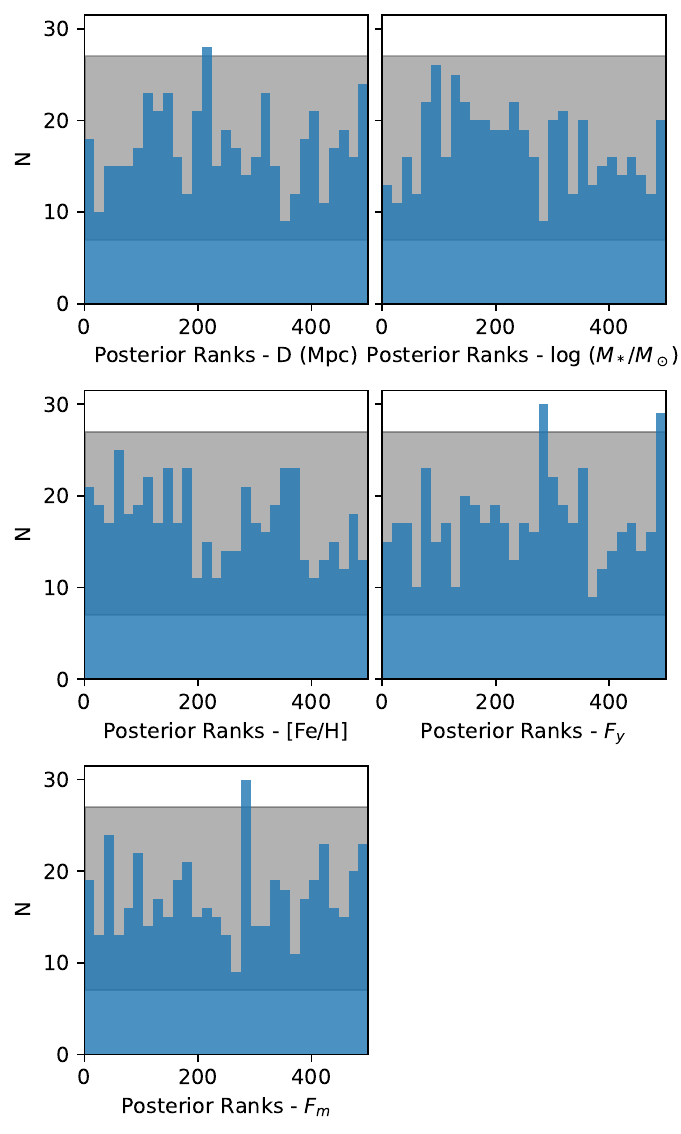}
    \caption{Simulation based calibration (SBC) tests applied to the initial round of training for the galaxy ESO 410-005 using 500 prior samples. The distribution of ranks for an unbiased posterior is uniformly distributed; for our network, we find no statistical departure from uniformity across the posterior ranks. The grey band shows the expected range of values calculated using the .5\%- 99.5\% percentile of the binomial distribution for a given number of bins ~\citep{talts2018}}
    \label{fig:sbc}
\end{figure}

While the posterior predictive checks discussed above are important for assessing if the inference algorithm is producing seemingly realistic results, we still must ensure that the posterior is well calibrated; i.e. that 90\% of the time the true value lies within the 90\% confidence interval. This is crucial for determining if the inference uncertainties are significantly over- or underestimated. For this we turn to simulation based calibration \citep[SBC;][]{Cook2006, talts2018} as described above in Sec~\ref{sec:walkthrough}. Using the implementation of SBC within the \texttt{sbi} package we calculate the rank statistics for the Round 1 posterior using 500 simulated observations and 500 posterior draws each. Figure~\ref{fig:sbc} displays the distributions of rank statistics for each variable. For each variable there is no deviation from a uniform distribution, indicating a well-calibrated posterior.

\subsection{The Full Sample}\label{sec:allgals}
\begin{figure*}[htp]
    \centering
    \includegraphics[width =\textwidth]{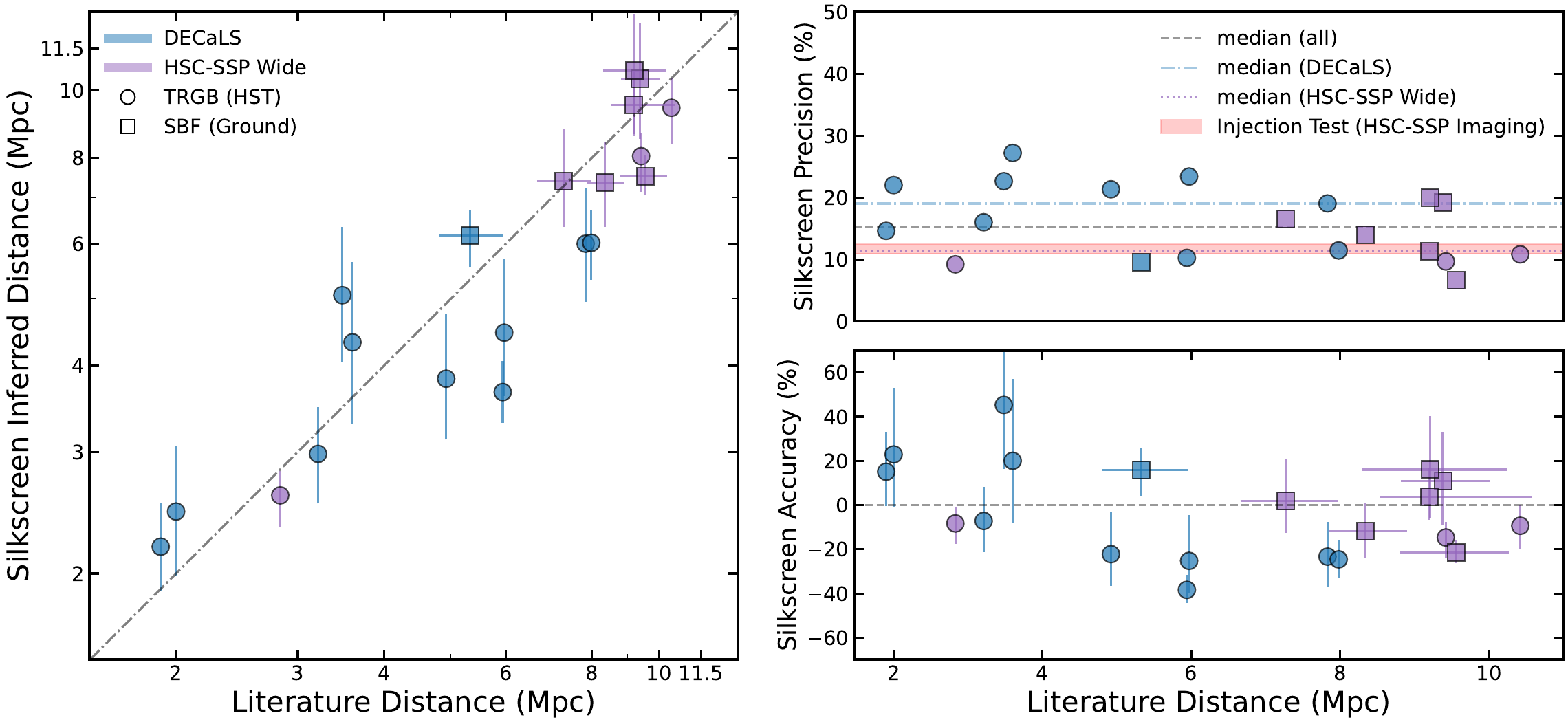}
    \caption{\code{} inferred distances compared to previously measured values in the literature. \textit{Left}: 1:1 plot comparing the literature distance to \code, with \code{} 1-$\sigma$ uncertainties generated from the learned posteriors. Though in this work we are considering literature distances as a ground-truth, the SBF measurements do have formal uncertainties which are in some cases similar to those from \code; thus, for reference we show the SBF uncertainties as x-errors. Reported uncertainties for TRGB distances measured with HST are typically much smaller ($\lesssim 3\%$); we do not plot these here. The color of the points denotes which imaging survey is used for \code{} inference: blue for DECaLS \citep{dey2019} and purple for HSC-SSP \citep{aihara2018}. The shape of the marker denotes how the literature value is measured, with a circle for TRGB measurements using HST imaging, and square for SBF measurements from ground based-imaging. Note that the distance on each axis is on a log scale. We find generally good agreement; For 15 of the 20 galaxies the \code{} measured distance matches that from the literature within uncertainties. For the other five, the literature distance lies outside the 5\%-95\% percentile of the posterior predicted by \code{}. \textit{Top Right:} Precision of \code, measured as the ratio of the standard deviation of the distance estimate (calculated as half of the difference between the 16th - 84th percentile) and the median inferred distance. We find the median precision to be $\sim$15\%, but we note that the precision is better for the HSC-SSP Wide sample than the overall average, while the shallower DECaLS is worse. Also shown is the precision of the injection test presented in Figure \ref{distance-test}, which used HSC-SSP survey properties. \textit{Bottom Right:} We assess the accuracy by plotting the ratio of \code{} inferred distance to literature distance, normalized by the literature distance. In both precision and accuracy, we find no notable trends with galaxy distance.}
    \label{fig:D_lit_comp}
\end{figure*}

To assess the accuracy and reliability of \code{} we compare the distances that it infers to the literature TRGB or SBF distances in Figure~\ref{fig:D_lit_comp} for the sample of 20 galaxies described above. The distances inferred from \code{} for all of the galaxies are displayed in Table~\ref{tab:dist}. We find good agreement between \code-inferred distances compared to the literature distances where most of the galaxies lie along the one to one line within uncertainties. This extends over the entire distance range of 2 Mpc to 12 Mpc and across both surveys used.

Of the 20 galaxies fit, we find that for five, the literature distances lie outside the 5\% to 95\% percentile of the distance posterior produced by \code{}. These galaxies are marked by an asterisk in Table~\ref{tab:dist}. Assuming the literature distances are not biased, the naive expectation is that 10\% of our sample should lie outside this range, so two from our sample. For four of these five galaxies, the \code{} inferred distance underestimates the literature distance,  which is at $D > 6$ Mpc. The larger number we observe, especially given that four are underestimated, could suggest a systematic bias. We investigate this further in Appendix~\ref{sec:outliers}; in short we find no obvious systematic issues when analyzing the posterior distribution or posterior predictive images for these systems. Another possibility is that the true star-formation history of these galaxies is much different than how we have parameterized it, e.g. $>95$\% of the stars formed 2-4 Gyr ago, which would lead to anomalous results.

The precision and accuracy of \code{} inference are further investigated using the 20 galaxy sample in Figure~\ref{fig:D_lit_comp}. To assess the precision, we show the ratio of the 1-$\sigma$ errors from \code{}~(calculated as half the 16\% -84\% percentile width) compared to the median distance from the posterior. We find the median precision of the sample to be 15\%, which is comparable to the scatter of the color-SBF magnitude relation for dwarf galaxies~\citep{carlsten2019, Moresco2022}. We note that there is no discernible trend with distance but there is a difference between the two surveys used, galaxies in the deeper HSC-SSP survey have a median precision of 11\% compared to those in DECaLS at 19\%. Compared to the precision achieved in the injection-recovery tests, which focused on HSC-SSP imaging we find a very similar median value of 11\%. We note that the perceived trend of lower precision at larger distances is likely due to the fact that at $D> 6$ Mpc, most of the galaxy's distances are based on the deeper HSC-SSP survey. The dominant factor in determining this value is likely the data quality but it is further suggestive that our parameterization is not hindering the inference in any way when applied to real galaxies. 

The residuals compared to the literature distances, a projection of the data presented in Figure~\ref{fig:D_lit_comp}, are plotted as a function of distance. Again, we do not see any discernible trends with distance apart from the four outlier galaxies discussed above at $D\gtrsim6$ Mpc that underestimate the literature distance. We have additionally compared the distance residuals to the integrated $g-r$ color and observed $r$-band magnitude and found no discernible correlation. 

The full posterior distributions for all parameters along with posterior predictive images for every galaxy in the sample are included in the appendix, in sections~\ref{sec:samp_res_appendix} and~\ref{sec:outliers}, with the latter focusing on the five galaxies where the \code{} distance is discrepant from the literature value.

We note that the stellar masses appear to be well-constrained by the inference, with tight posteriors predicted for the majority of galaxies in this study. We do not attempt to compare the derived masses to the literature in this work, primarily because there does not exist a robust, uniform method for measuring stellar mass for these systems. Even observationally-driven comparisons, e.g., $K$-band magnitude, which roughly traces stellar mass, are challenging for this sample, for which many $K$-band estimates are derived from bluer bands and uncertain color correction\citep{karachentsev2013}. Indeed, it is possible that \code{} is providing competitively accurate and precise stellar mass estimates (as suggested by the mass recovery shown in Figure~\ref{fig:injection-other}), a possibility that will be investigated in future work.  Crucial to this will be a quantitative comparison to assess not only the precision but the accuracy when compared to established methods.

\section{Current Limitations}
\label{sec:issues}

The largest issue encountered during development --- and the most time spent --- was ensuring proper training of the network. Initially we had difficulty obtaining well-calibrated posteriors, like the one shown in Fig.~\ref{fig:sbc}. The SBC ranks plot often showed `U' shaped or tilted distributions, indicating that the network was producing over-confident posteriors, i.e. too small error bars~\citep{Cook2006,talts2018}. Even after hyper parameter tuning and testing different architectures for the embedding network or the flow, these problems persisted. The key insight was to separate the learning rates of the embedding network and the neural flow sections of the posterior estimator. In particular, we found that the learning rate for the embedding network had to be much lower, at least an order of magnitude, compared to the learning rate for the neural flow. With this decoupling the training was successful but there likely remains architectural and training procedure improvements for this multi-component network consisting of the embedding network plus the normalizing flow.

A major limitation of the current pipeline is the computational resources required. The bespoke process is designed for tailored simulations to match the observations but necessitates the running of new simulations and training for each galaxy. For the examples in this paper it took an average of 75 CPU hours to run the simulations, and a combined 6 hours using an NVIDIA A100 GPU to complete training of the network. While this cost is reasonable for a small number of galaxies, it may become prohibitive to scale up to hundreds or thousands of galaxies. This can be overcome by training a contextually broad model that is applicable to any galaxy, which we discuss below in Section ~\ref{sec:future}.

In simulation based inference, the results are only as trustworthy as the simulator. A strength of \artpop{} is the flexibility in the simulation procedure, but in this work we have made simplifying choices that may have limited the accuracy of the simulator. One major area for improvement is the parameterization of the star-formation and metallicity history of dwarf galaxies. In this initial exploration we have opted for simplicity while still attempting to capture the diversity of dwarf galaxy formation histories~\citep[see e.g.][]{weisz2011}. During our testing we found the parameterization and the priors placed on the youngest stellar population had the largest effect on the resulting distance. Even though the young population is typically a small percentage ($\lesssim$ 5\%) of the overall mass, its high stellar light-to-mass ratio and ability to produce bright individual stars means it has an outsized impact on the appearance of galaxies. The properties of stellar populations with ages $< 300$ Myr change drastically over a short period of time, increasing the importance of modelling them correctly. We note that similar issues also affect SBF measurements.

\begin{figure}
    \centering
    \includegraphics[width=\columnwidth]{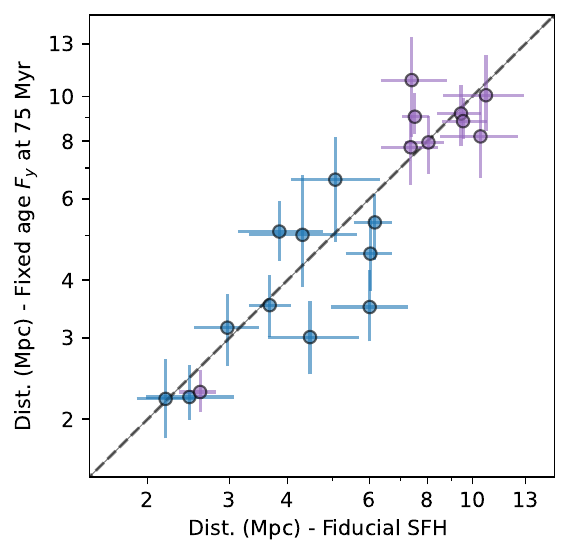}
    \caption{Comparison of the distances measured by \code{} using two different parameterizations of the youngest population of stars: On the horizontal axis the fiducial continuous SFH between 50-250 Myr and on the vertical axis assuming a fixed age of this population of 75 Myr. The color scheme is the same as Figure ~\ref{fig:D_lit_comp}, where blue denotes when DECaLs data is used and purple for HSC data. For most of the galaxies the distances agree between the two parameterizations however there are some galaxies where the fixed age model produces a lower distance, by up to a factor of two.}
    \label{fig:D_comp_sfh}
\end{figure}

We tested numerous variations including a single aged population with the age as a free parameter and a number of single population fixed age models. We also varied the prior on $F_y$, testing uniform and truncated normal distributions. To illustrate the effect these choices can have on the inferred distance in Figure ~\ref{fig:D_comp_sfh} compares the distance measurements from \code{} using two different parameterizations for the youngest population. We compare the fiducial continuous model described in Sec.~\ref{sec:sfh} to that with a fixed age of 75 Myr. For the majority of the sample the distances agree between the two measurements however for the galaxies at $D > 4$  where DECaLS data is used the fixed age model tends to produce lower distance estimates, by up to a factor of 2. For these galaxies the continuous model is closer to the literature distances. We prefer the continuous star-formation history for the young populations presented here because it limits the number of catastrophic outliers in the 20 galaxy sample and the continuous model avoids any systemic issues caused by transient populations when using a single aged population. 

A further complicating factor for star-forming galaxies is that we have neglected dust attenuation internal to the dwarf galaxies and clustering of young stars. Dust attenuation could play a significant role for young and gas rich systems. It is often seen that young stars appear spatially clustered, while we have assumed that they follow the same broad distribution as the older stars. This difference can be seen when comparing the observed to posterior-predictive images in Appendix~\ref{sec:samp_res_appendix}, for galaxies such as PCG1099440 and dw1236-0025 the young blue stars appear more centrally concentrated compared to older, redder stars. Given that these choices can systematically bias the inferred properties, more work is needed to find physically realistic parameterizations for the star formation history along with reasonable priors. We note that a simple solution is to not to galaxies with highly asymmetric morphologies or large clusters of blue stars.

There are assumptions inherent to our simulation procedure that could bias outcomes. For example, we have used the commonly assumed \citet{kroupa2001} IMF for this work but there is some evidence that the IMF differs in dwarf galaxies \citep{geha2013,mentz2016,yan2020}. As a key ingredient to the \artpop{} simulation procedure, systematic changes could alter the appearance of simulated galaxies and therefore the distances inferred by \code{}. Finally, the entire \artpop{} simulation procedure is built upon the use of precomputed stellar isochrones to predict the luminosity of a star in different bandpasses given its initial mass, age and metallicity. As discussed in \citet{greco2021} there are systematic variations in the predicted SBF magnitude between different libraries. This problem is particularly acute in the near infrared where predicted SBF magnitudes from MIST~\citep{choi2016}, Basti~\citep{Hidalgo2018} and Parsec~\citep{Margio2017} differ by up to 1.5 magnitudes \citep[see Figure 6 in][]{greco2021}. This is likely a result of the treatment of AGB stars in each of these different isochrone libraries~\citep{Raimondo2009,choi2016}. As a bedrock of our inference procedure it will be important to assess the effect of different isochrone libraries on the inferred properties of galaxies as it remains one of the largest systematic uncertainties.

An inherent downside to machine learning techniques, like those employed in this study, is the lack of interpretability. Unlike physically motivated methods it is difficult to understand what information the network is using to estimate the distance or other parameters. In regards to \code{}, it would be interesting to know if the network is simply an efficient interpolator between the resolved and unresolved regimes and measuring known point statistics such as the TRGB or SBF signal. Or if it is instead learning a wholly different signature of distance which could possibly be more universal and/or accurate. Unfortunately, properly answering this question and interpreting the behavior of the network is difficult and beyond the scope of this initial work, however we speculate on two possible avenues that could be useful for future study. 

The first is using techniques like GRAD-CAM~\citep{selvaraju2020} which aim to assess what pixels in the image have the biggest effect on the outputs of CNNs by tracing the strength of gradients through the network. In this scenario it could be used to test whether the resolved or unresolved components of a galaxy have more of an effect on the summary statistics measured from the CNN portion of our network. Another, more physically motivated, approach would be to take advantage of the flexibility of the ArtPop simulator and remove or change different aspects of the simulated stellar population and test what the recovered properties are. For example one could artificially move or smooth the TRGB for a given galaxy and test if \code{} would still be able to recover the distance. In this manner it would be possible to systematically test which parts of the stellar population have the largest effect of the estimated distance. In practice some combination of these two approaches will be needed to better understand what physical properties are the physical properties of galaxies that~\code{} is using to estimate the distance and other parameters.

To this end it is particularly stunning that for some of the galaxies in our sample, such as PGC1099440, ESO349-31 or DDO190, the simulated images and integrated colors vary greatly from those observed yet the inferred distances match the literature values. Alternatively galaxies like KK182 appear to be a good match in color and visual morphology yet the inferred distance does not agree with the literature value. Naively this suggests that whatever information the network is using to estimate the distance is independent of these characteristics. For example the brightness distribution of red stars, i.e. the TRGB, would not be captured by these metrics. A better understanding of the type of information being used by the network can lead us to where improvements to the simulation procedure are needed. By connecting the information back to well understood phases of stellar evolution or physical processes of stars we can improve the trustworthiness of the results obtained with \code{}.

\begin{table*}
    
\begin{threeparttable}
    \centering
    \caption{Distances inferred by \code{}~compared to those measured in the literature for the 20 galaxies used in this study}

    \begin{tabular}{c| c c |c c c}
         Galaxy & Data & \code{} D (Mpc)  &Lit. D (Mpc) & Method  & Ref. \\ \hline
        ESO 540 32 & DECaLS & $4.30_{ 0.99 }^{ 1.33 } $ & $3.60\pm 0.08$ & TRGB & \citet {Tully2013}\\
        ESO 540 30$^*$ & DECaLS & $5.07_{ 1.02 }^{ 1.26 } $ & $3.48\pm 0.08$ & TRGB & \citet {Tully2013}\\
        ESO 410 05 & DECaLS & $2.19_{ 0.30 }^{ 0.34 } $ & $1.90\pm0.08$ & TRGB & \citet {Tully2013}\\
        ESO 294 10 & DECaLS & $2.46_{ 0.48 }^{ 0.60 } $ & $2.00\pm0.08$ & TRGB & \citet {Tully2013}\\
        ESO 349 31 & DECaLS & $2.98_{ 0.45 }^{ 0.50 } $ & $3.21\pm0.1$ & TRGB & \citet {Tully2013}\\
        IC 1574 & DECaLS & $3.82_{ 0.70 }^{ 0.94 } $ &$4.92\pm0.1$ & TRGB & \citet {Tully2013}\\
        KK13 & DECaLS & $6.01_{ 1.05 }^{ 1.23 } $ & $7.83\pm 0.21$ & TRGB & \citet {McQuinn2014}\\
        KK16 & DECaLS & $4.46_{ 0.85 }^{ 1.26 } $ &$ 5.97^{+0.11}_{-0.09}$ & TRGB & \citet {McQuinn2014}\\
        KK65$^*$ & DECaLS & $6.04_{ 0.69 }^{ 0.68 } $ & $7.98\pm0.1$ & TRGB & \citet {Tully2013}\\
        KK182$^*$ & DECaLS & $3.67_{ 0.36 }^{ 0.39 } $ & $5.94\pm0.1$ & TRGB & \citet {Tully2013}\\
        KK2000 71 & DECaLS & $6.17_{ 0.63 }^{ 0.55 } $ & ${5.33}_{-0.53}^{+0.63}$ &  SBF & \citet {Carlsten2022}\\
        PGC1099440 & HSC-SSP & $9.44_{ 1.07 }^{ 0.99 } $ & $10.42\pm0.75$ & TRGB & \citet {Karachentsev2022}\\
        CGCG 014-054$^*$ & HSC-SSP & $8.03_{ 0.89 }^{ 0.67 } $ & $9.42\pm 0.6$ & TRGB & \citet {Anand2021}\\
        dw1232+0015$^*$ & HSC-SSP & $7.52_{ 0.46 }^{ 0.54 } $ & ${9.56}_{-0.76}^{+0.71}$ &  SBF & \citet {Carlsten2022}\\
        dw1236-0025 & HSC-SSP & $10.43_{ 1.88 }^{ 2.09 } $ & ${9.38}_{-0.56}^{+0.64}$ &  SBF & \citet {Carlsten2022}\\
        dw1238+0028 & HSC-SSP & $10.69_{ 2.03 }^{ 2.25 } $ & ${9.21}_{-0.91}^{+1.03}$ &  SBF & \citet {Carlsten2022}\\
        dw1238-0035 & HSC-SSP & $9.54_{ 0.95 }^{ 1.23 } $ & ${9.2}_{-0.66}^{+1.37}$ &  SBF & \citet {Carlsten2022}\\
        dw1238-0105 & HSC-SSP & $7.40_{ 1.05 }^{ 1.40 }  $ & ${7.27}_{-0.61}^{+0.70}$ &  SBF & \citet {Carlsten2022}\\
        KDG171 & HSC-SSP & $7.35_{ 1.01 }^{ 1.06 } $ & ${8.34}_{-0.49}^{+0.56}$ &  SBF & \citet {Carlsten2022}\\
        DDO190 & HSC-SSP & $2.60_{ 0.26 }^{ 0.21 } $ & $2.83\pm0.08$ & TRGB & \citet {Tully2013}\\
        \hline
    \end{tabular}
    \begin{tablenotes}
    \item[$^*$] Galaxies where the literature distance lies outside the 5\% - 95\% posterior range produced by \code; these cases are further discussed in Appendix~\ref{sec:outliers}.
    \end{tablenotes}
    \label{tab:dist}
\end{threeparttable}
\end{table*}

\section{Summary and Future Outlook}
\label{sec:future}
In this paper we apply simulation based inference techniques to infer distances to dwarf galaxies directly from their images. Simulations of dwarf galaxies are produced using the \artpop{} package, and are used in neural posterior estimation to estimate posterior distributions of distance, stellar mass and other stellar population parameters. By training a CNN simultaneously with the normalizing flow, \code{} uses the full information present in multi-band images to estimate the distance. We describe the details of our implementation, which is available open source\footnote{https://github.com/tbmiller-astro/silkscreen}. We apply \code{} to an injected mock test, as well as a sample of 20 real galaxies in DECaLS and HSC imaging, with distances ranging from 2 Mpc to 12 Mpc.

One of the main advantages of \code{} is its flexibility. With this procedure one can perform inference on any stellar system that can be simulated with \artpop{}. Much like other forward modelling techniques, e.g. pCMD \citep{conroy2016, cook2019}, it is able to bridge the gap between the resolved and unresolved regimes while marginalizing over uncertainties in the stellar populations of dwarf galaxies and not sacrificing any information in the image. Based on our initial exploration, \code{} appears to be reasonably well calibrated and produces robust measures of distance for most galaxies; however for a fraction of galaxies (5 of 20) tested in this study, this method does not accurately recover the distance. Further work can both improve the ability for the network to recover stellar populations, along with handling galaxies with properties more distinctly different than the models that are currently implemented. Additionally by analyzing the posterior predictive images and integrated colors, as shown for all galaxies in Appendix~\ref{sec:samp_res_appendix}, we find differences in the visual morphology and colors (specifically $r-z$ or $r-i$) even for galaxies where \code{} recovers the literature distances. This implies improvements to the simulator itself can be made in order to produce more realistic simulated images.

Looking ahead to wide field surveys performed from the ground, e.g., the Legacy Survey of Space and Time (LSST) with the Vera Rubin Observatory, or from space with the Roman and Euclid space telescopes, we expect the number of dwarf galaxies to exponentially grow. If we wish to perform inference with \code{} in its current state, the computational cost will rapidly become prohibitively expensive for increasingly large samples. The current bespoke pipeline requires simulations and training for each individual galaxy of interest. This need not be the case. In future work we plan to train a contextually-broad model that would be trained once, and then could be applied to any input galaxy. This is also known as amortized inference, where the computational cost is front-loaded in the initial training of the network and is a key benefit of Neural Posterior estimation. It is becoming increasingly popular in astronomy for accelerated inference as the amount of incoming data continues to grow.\citep[e.g.][]{hahn2022,villar2022, Zhang2023,iglesias-navarro2024}.

For \code{}, this process would involve training a single network on data representative of an entire survey rather than having priors progressively restricted to the single galaxy being fit. It would then be capable of performing posterior estimation for any new galaxy of interest on the order of seconds. The initial cost would be the necessary creation of and training on many more simulated galaxies spanning the targeted parameter space (and for a given survey), compared to the one hundred thousand used for the bespoke pipeline now. In the context of current and upcoming wide field surveys, such investment will be well-warranted.

Our next goal is to work toward this style of contextually broad (or amortized) model for the DECam legacy imaging survey ~\citep{dey2019}. This will allow us to benchmark \code{} against a much larger sample of galaxies with known distances such as those presented in ELVES~\citep{Carlsten2022} and the Local Volume catalog~\citep{karachentsev2013, Anand2021}. This will allow us to more robustly test various parameterization choices for the star-formation and metallicity history and their priors. Careful planning and preparation is required for this application, it will be necessary to accurately capture the variations in noise, PSFs, surface brightness profiles and other observational systematics across the entire survey. The application to DECam imaging is an important next step in preparation for the ultimate goal of applying \code{} to LSST imaging when the multi-year stacks are available later this decade.

\facilities{Blanco (DECam),Subaru (HSC)}

\section*{Acknowledgements}
The authors would like to thank Aritra Ghosh for valuable discussions on the design, implementation and training of the Neural Posterior Estimator, and Yasmeen Asali for helpful discussions regarding the description of the network, as well as Charlie Conroy for discussions about the parameterization of the star-formation history. TBM was supported by a CIERA Postdoctoral Fellowship. We thank the anonymous referee whose comments improved the manuscript.

We thank the Yale Center for Research Computing, for guidance and assistance in computation run on the Grace cluster.

The Legacy Surveys consist of three individual and complementary projects: the Dark Energy Camera Legacy Survey (DECaLS; Proposal ID \#2014B-0404; PIs: David Schlegel and Arjun Dey), the Beijing-Arizona Sky Survey (BASS; NOAO Prop. ID \#2015A-0801; PIs: Zhou Xu and Xiaohui Fan), and the Mayall z-band Legacy Survey (MzLS; Prop. ID \#2016A-0453; PI: Arjun Dey). DECaLS, BASS and MzLS together include data obtained, respectively, at the Blanco telescope, Cerro Tololo Inter-American Observatory, NSF’s NOIRLab; the Bok telescope, Steward Observatory, University of Arizona; and the Mayall telescope, Kitt Peak National Observatory, NOIRLab. Pipeline processing and analyses of the data were supported by NOIRLab and the Lawrence Berkeley National Laboratory (LBNL). The Legacy Surveys project is honored to be permitted to conduct astronomical research on Iolkam Du’ag (Kitt Peak), a mountain with particular significance to the Tohono O’odham Nation.

NOIRLab is operated by the Association of Universities for Research in Astronomy (AURA) under a cooperative agreement with the National Science Foundation. LBNL is managed by the Regents of the University of California under contract to the U.S. Department of Energy.

This project used data obtained with the Dark Energy Camera (DECam), which was constructed by the Dark Energy Survey (DES) collaboration. Funding for the DES Projects has been provided by the U.S. Department of Energy, the U.S. National Science Foundation, the Ministry of Science and Education of Spain, the Science and Technology Facilities Council of the United Kingdom, the Higher Education Funding Council for England, the National Center for Supercomputing Applications at the University of Illinois at Urbana-Champaign, the Kavli Institute of Cosmological Physics at the University of Chicago, Center for Cosmology and Astro-Particle Physics at the Ohio State University, the Mitchell Institute for Fundamental Physics and Astronomy at Texas A\&M University, Financiadora de Estudos e Projetos, Fundacao Carlos Chagas Filho de Amparo, Financiadora de Estudos e Projetos, Fundacao Carlos Chagas Filho de Amparo a Pesquisa do Estado do Rio de Janeiro, Conselho Nacional de Desenvolvimento Cientifico e Tecnologico and the Ministerio da Ciencia, Tecnologia e Inovacao, the Deutsche Forschungsgemeinschaft and the Collaborating Institutions in the Dark Energy Survey. The Collaborating Institutions are Argonne National Laboratory, the University of California at Santa Cruz, the University of Cambridge, Centro de Investigaciones Energeticas, Medioambientales y Tecnologicas-Madrid, the University of Chicago, University College London, the DES-Brazil Consortium, the University of Edinburgh, the Eidgenossische Technische Hochschule (ETH) Zurich, Fermi National Accelerator Laboratory, the University of Illinois at Urbana-Champaign, the Institut de Ciencies de l’Espai (IEEC/CSIC), the Institut de Fisica d’Altes Energies, Lawrence Berkeley National Laboratory, the Ludwig Maximilians Universitat Munchen and the associated Excellence Cluster Universe, the University of Michigan, NSF’s NOIRLab, the University of Nottingham, the Ohio State University, the University of Pennsylvania, the University of Portsmouth, SLAC National Accelerator Laboratory, Stanford University, the University of Sussex, and Texas A\&M University.

BASS is a key project of the Telescope Access Program (TAP), which has been funded by the National Astronomical Observatories of China, the Chinese Academy of Sciences (the Strategic Priority Research Program “The Emergence of Cosmological Structures” Grant \# XDB09000000), and the Special Fund for Astronomy from the Ministry of Finance. The BASS is also supported by the External Cooperation Program of Chinese Academy of Sciences (Grant \# 114A11KYSB20160057), and Chinese National Natural Science Foundation (Grant \# 12120101003, \# 11433005).

The Legacy Survey team makes use of data products from the Near-Earth Object Wide-field Infrared Survey Explorer (NEOWISE), which is a project of the Jet Propulsion Laboratory/California Institute of Technology. NEOWISE is funded by the National Aeronautics and Space Administration.

The Legacy Surveys imaging of the DESI footprint is supported by the Director, Office of Science, Office of High Energy Physics of the U.S. Department of Energy under Contract No. DE-AC02-05CH1123, by the National Energy Research Scientific Computing Center, a DOE Office of Science User Facility under the same contract; and by the U.S. National Science Foundation, Division of Astronomical Sciences under Contract No. AST-0950945 to NOAO.

\appendix

\section{Full Results for the Galaxy Sample}
\label{sec:samp_res_appendix}
In this section we show the full results from \code{} applied to each galaxy in the sample. The posterior distributions and posterior predictive images for the majority of the sample are shown in Figures ~\ref{fig:corner_im_all_0},~\ref{fig:corner_im_all_1},~\ref{fig:corner_im_all_2},~\ref{fig:corner_im_all_3}, and~\ref{fig:corner_im_all_4}. The five galaxies where the literature distance is discrepant with that inferred from \code{} are discussed in more detail below in Section~\ref{sec:outliers}.

\begin{figure*}
\centering
\includegraphics[width = 0.75\textwidth]{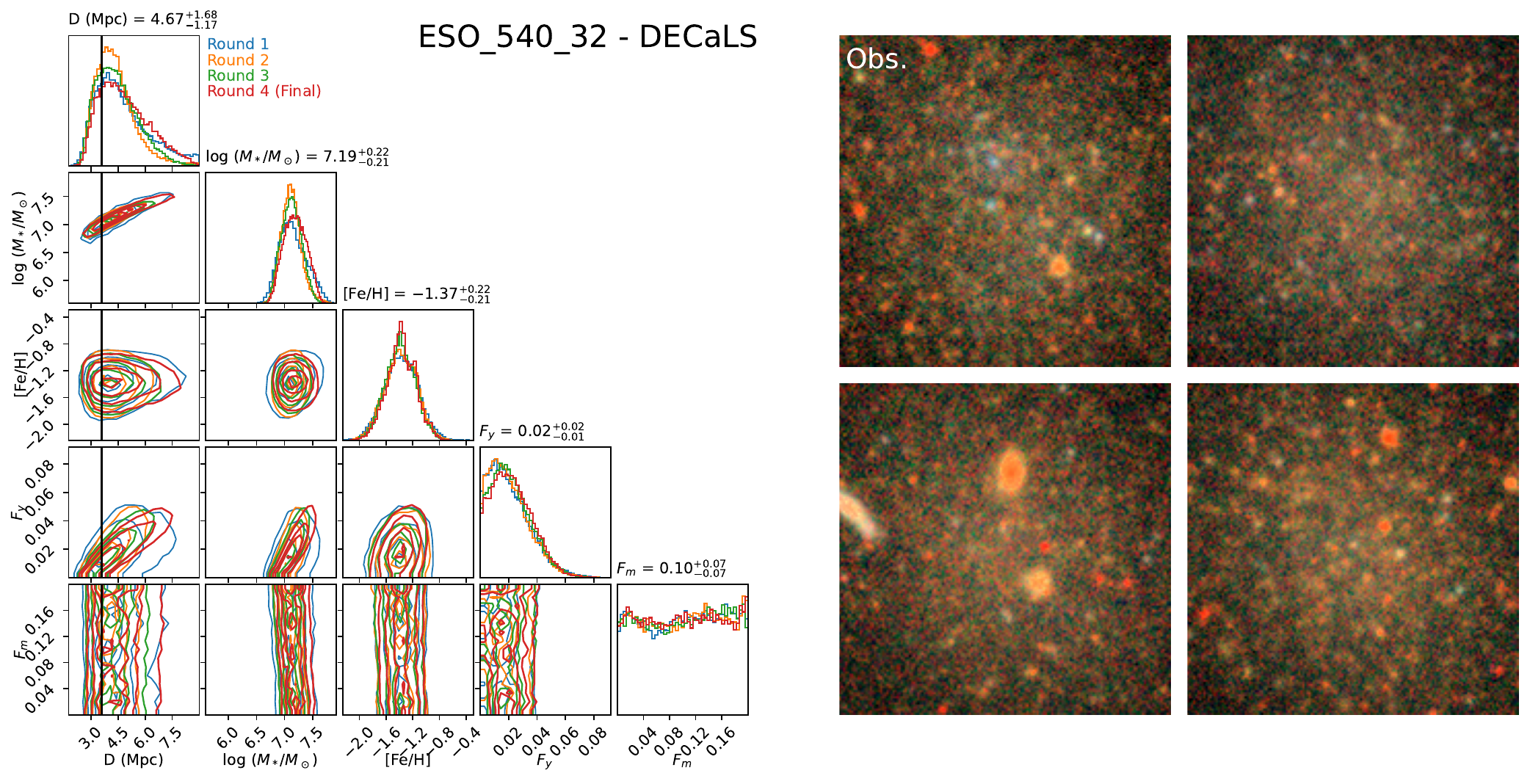}
\includegraphics[width = 0.2\textwidth]{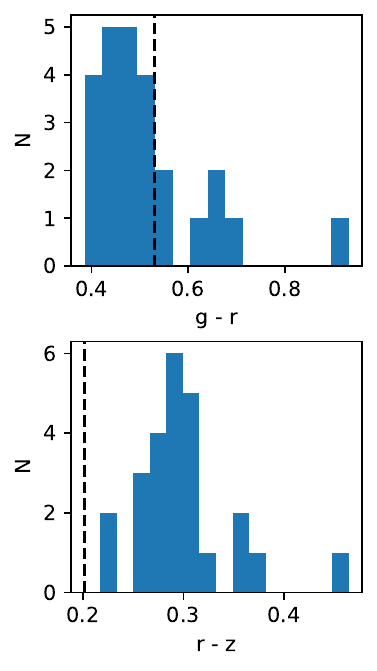}
\includegraphics[width = 0.75\textwidth]{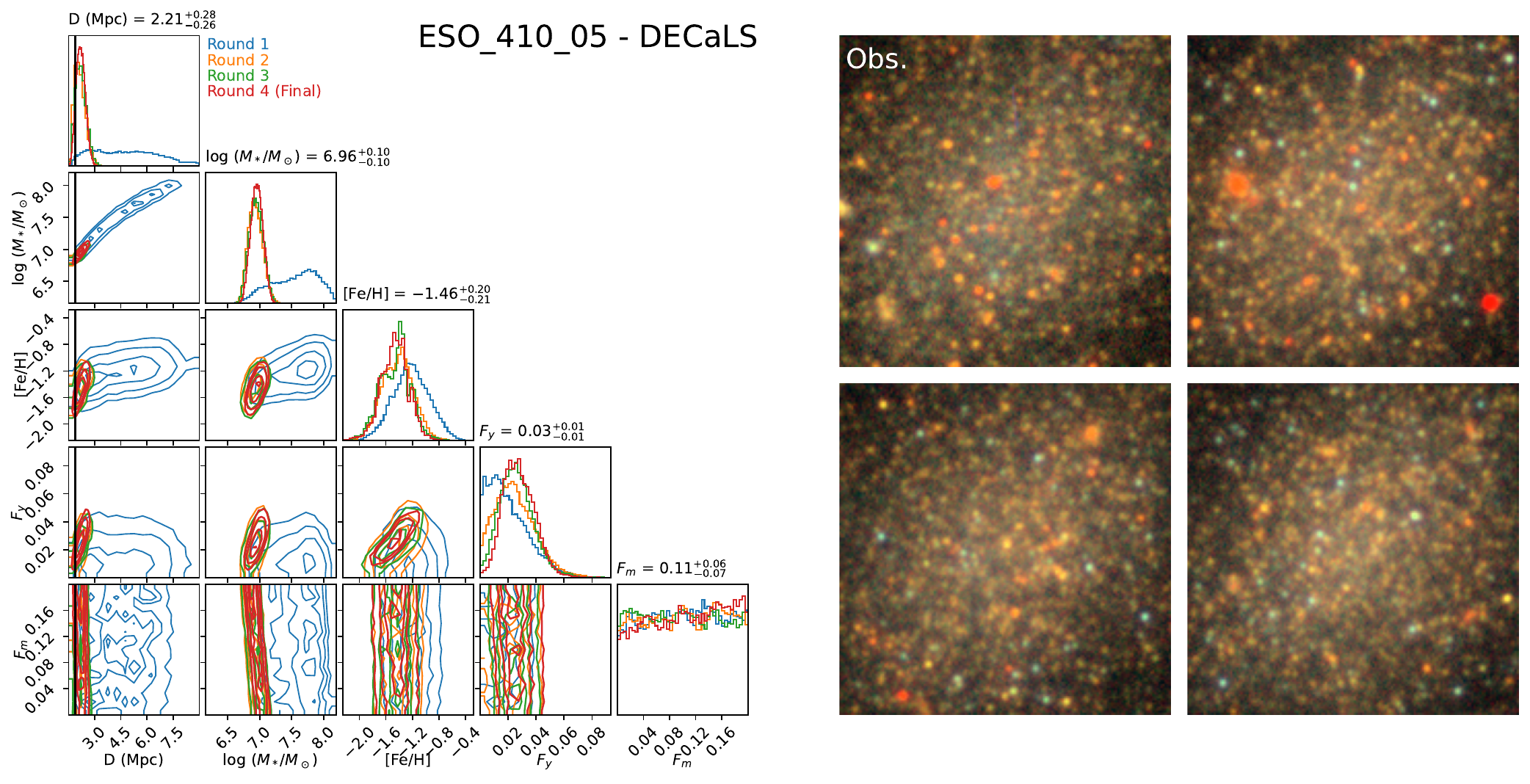}
\includegraphics[width = 0.2\textwidth]{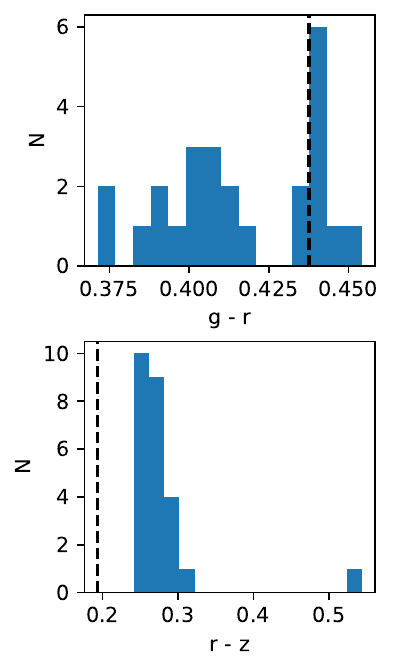}
\includegraphics[width = 0.75\textwidth]{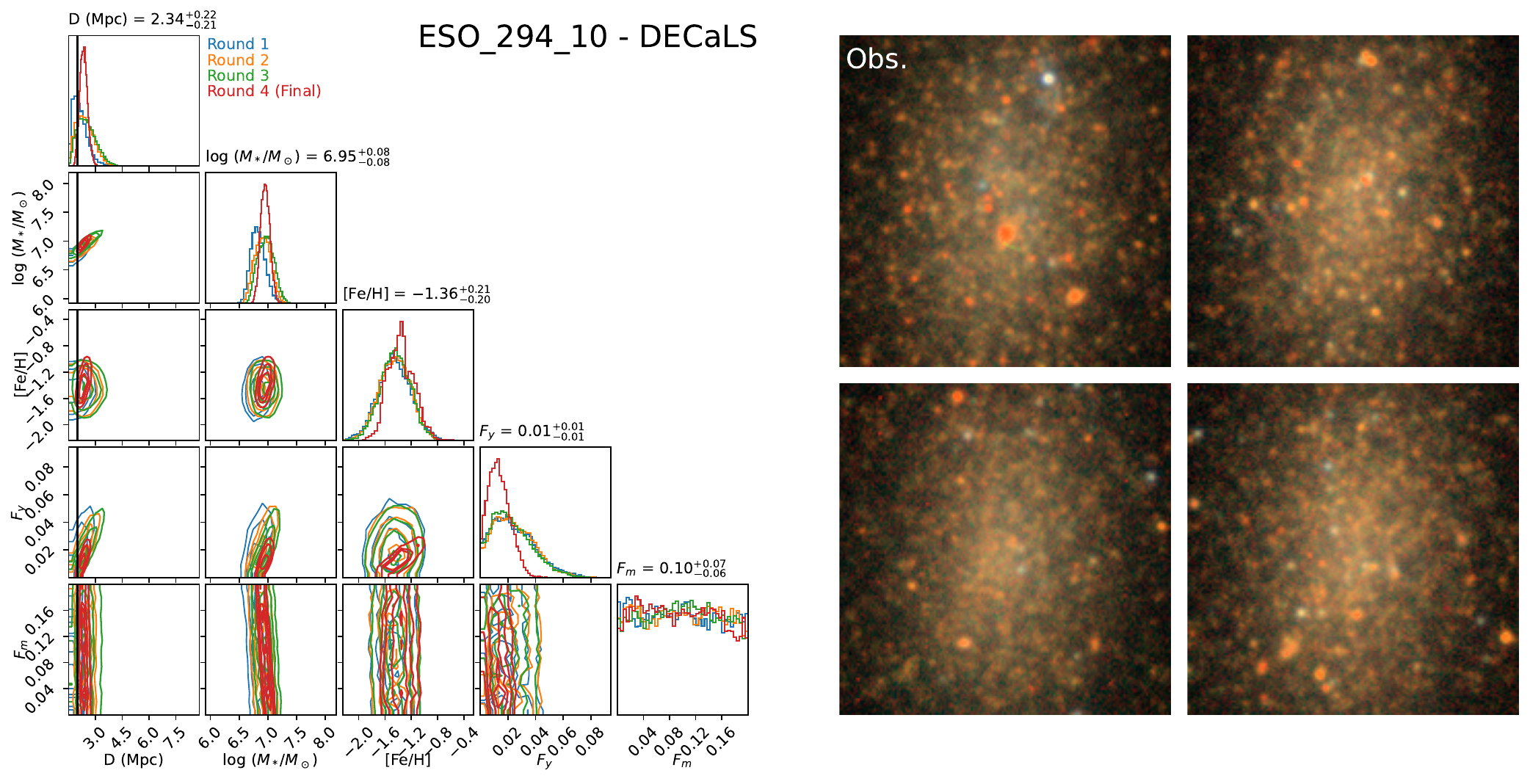}
\includegraphics[width = 0.2\textwidth]{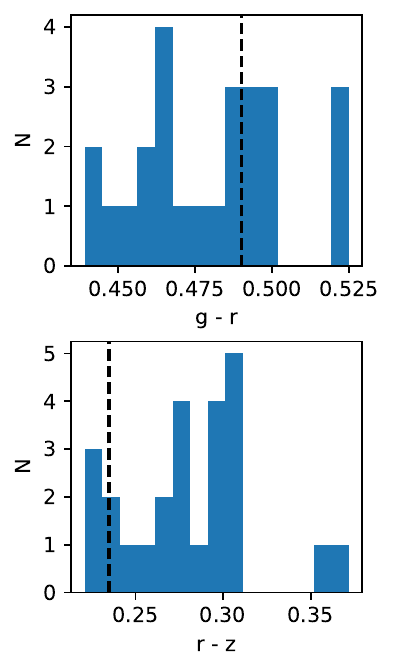}
    \caption{The joint posterior distributions and posterior predictive images produced by \code{}. Similar to Sec.~\ref{sec:examp_gal} in the main text, we show the posterior distribution for the four successive rounds of training and the observed image alongside posterior predictive images. Literature distances are denoted with a solid black line in the corner plots}
\label{fig:corner_im_all_0}
\end{figure*}
\begin{figure*}
\centering
\includegraphics[width = 0.75\textwidth]{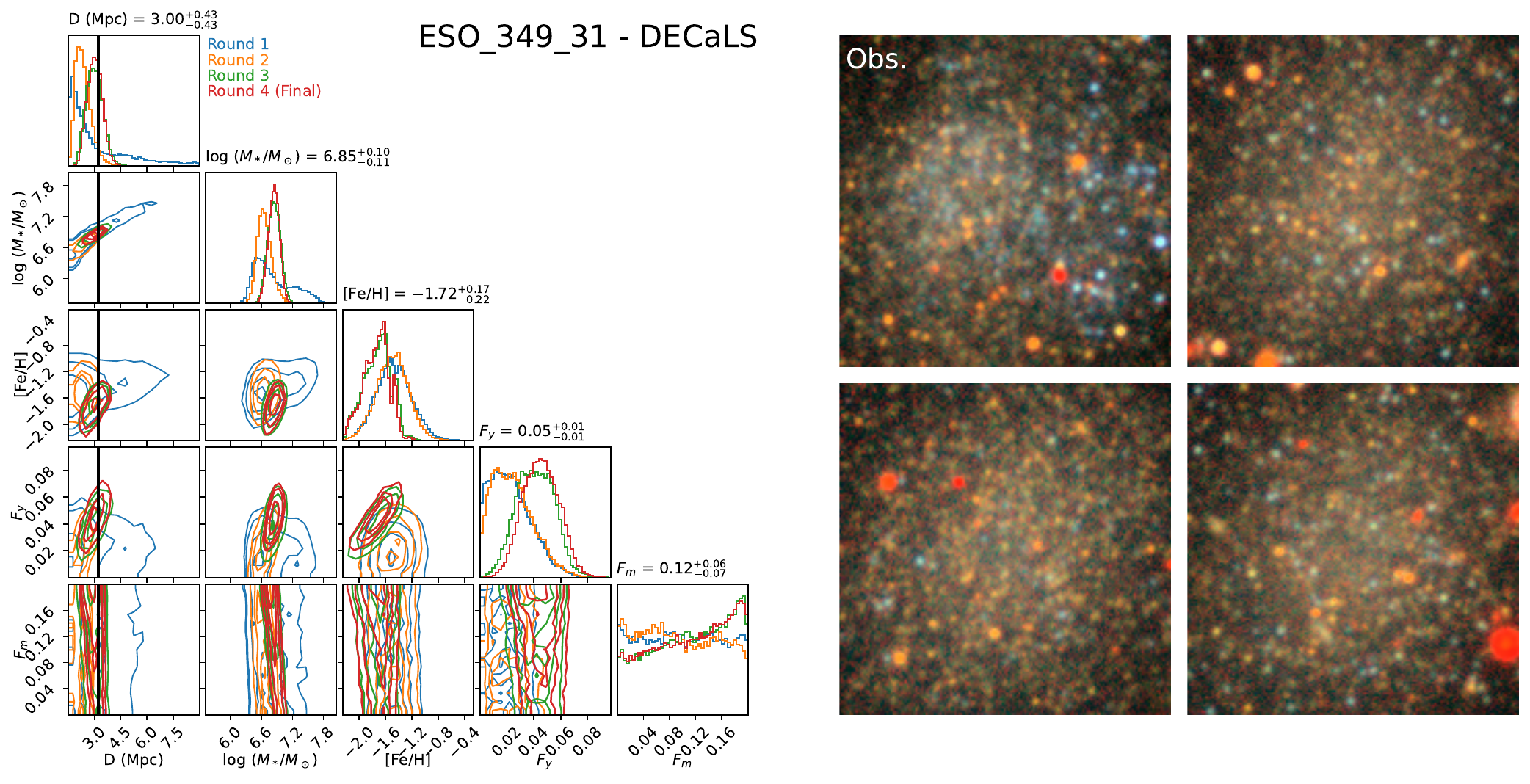}
\includegraphics[width = 0.2\textwidth]{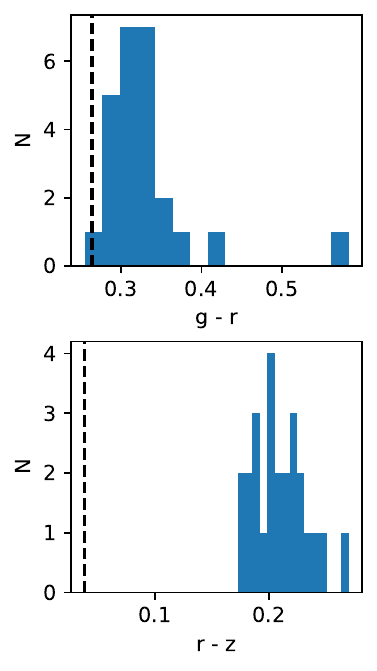}
\includegraphics[width = 0.75\textwidth]{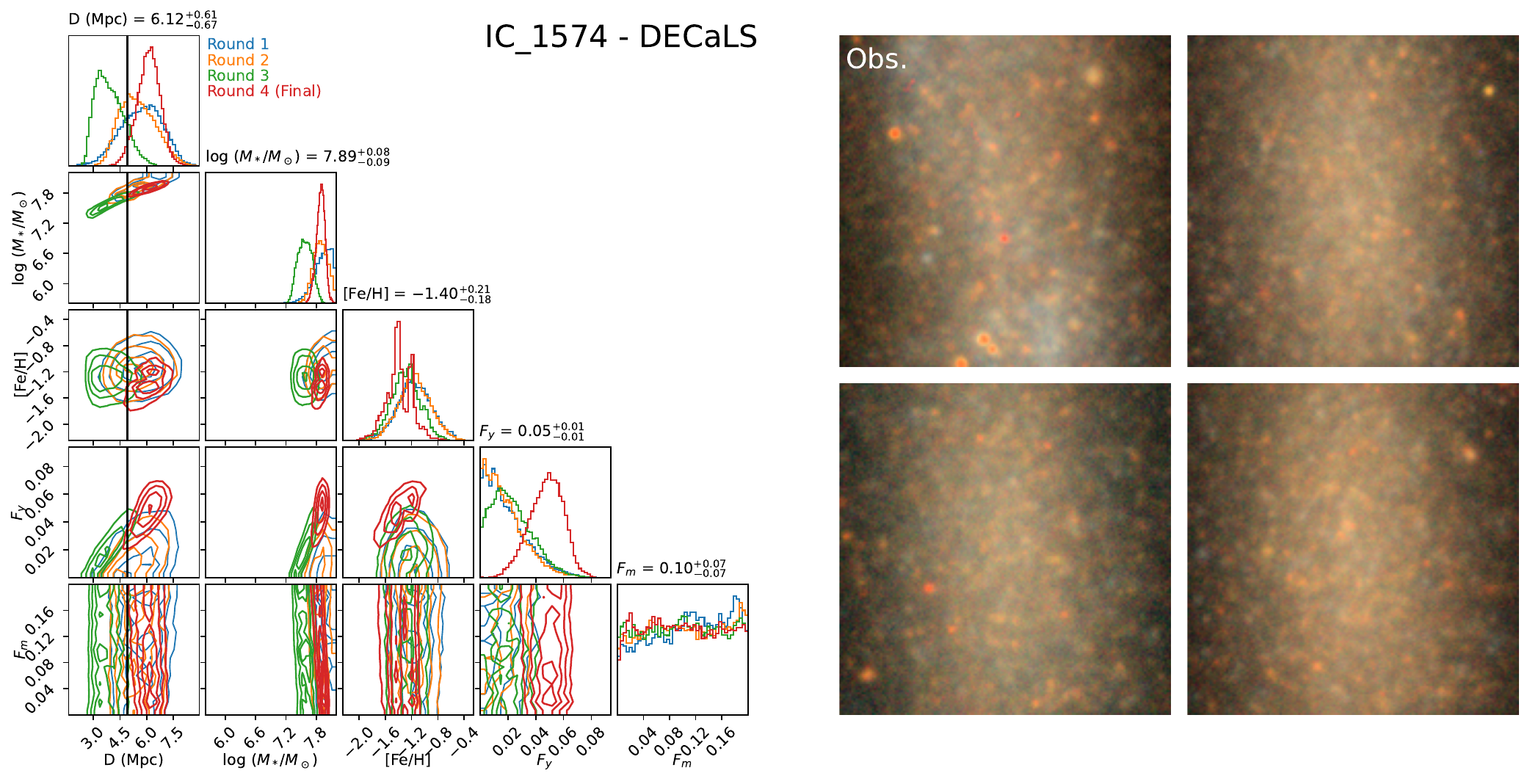}
\includegraphics[width = 0.2\textwidth]{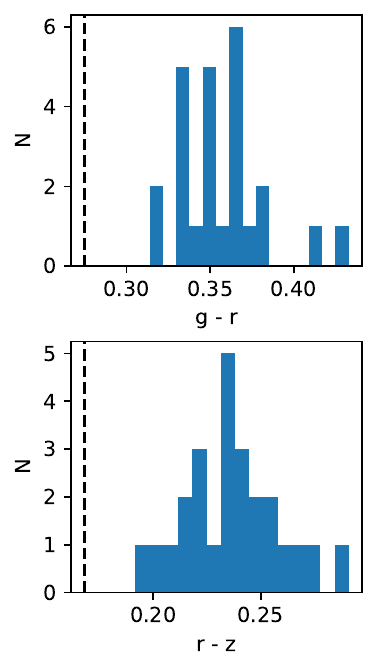}
\includegraphics[width = 0.75\textwidth]{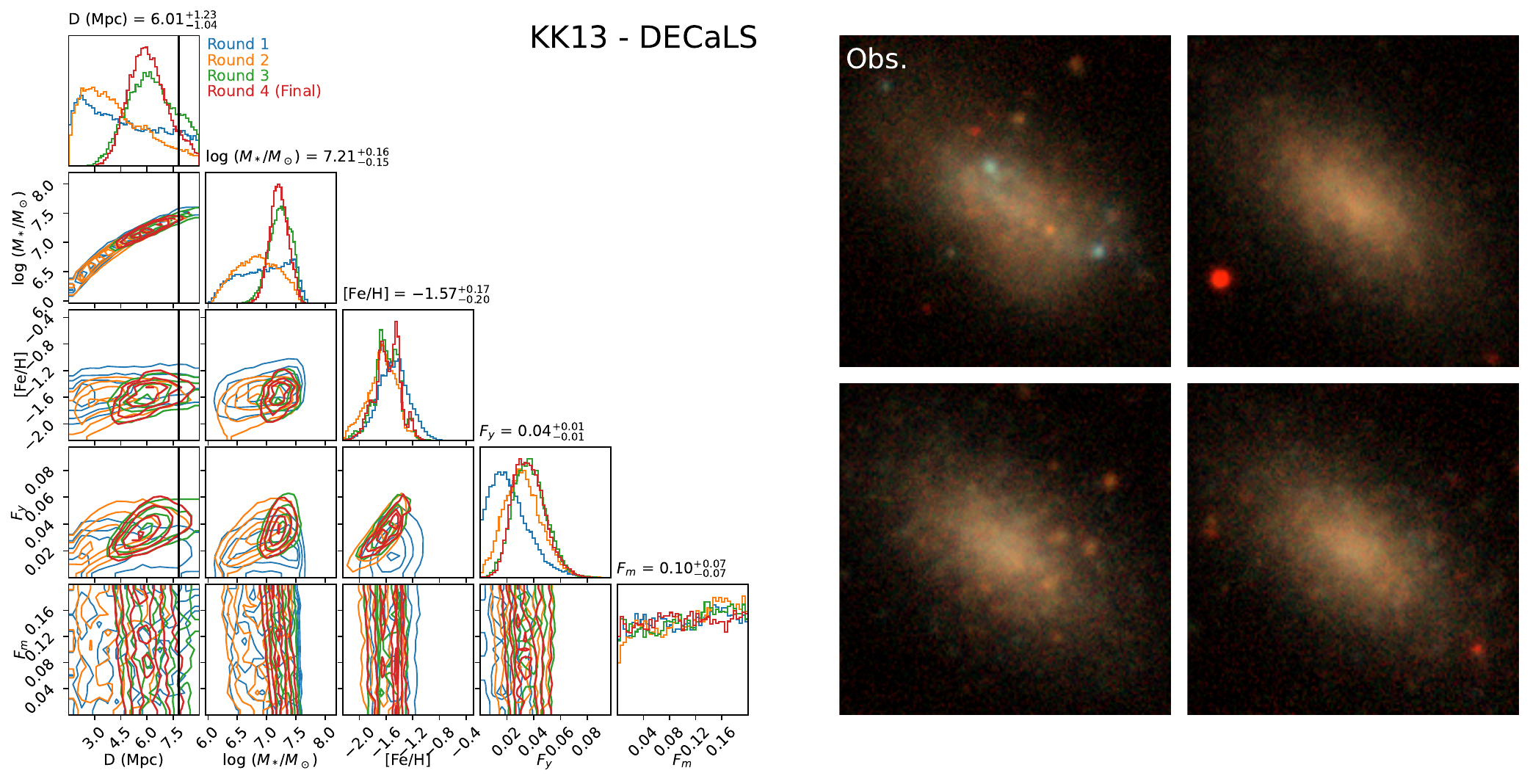}
\includegraphics[width = 0.2\textwidth]{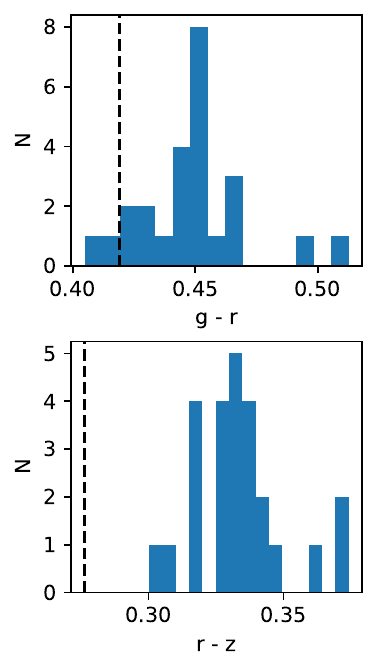}
\caption{Continuation of Figure~\ref{fig:corner_im_all_0}}
\label{fig:corner_im_all_1}
\end{figure*}
\begin{figure*}
\centering
\includegraphics[width = 0.75\textwidth]{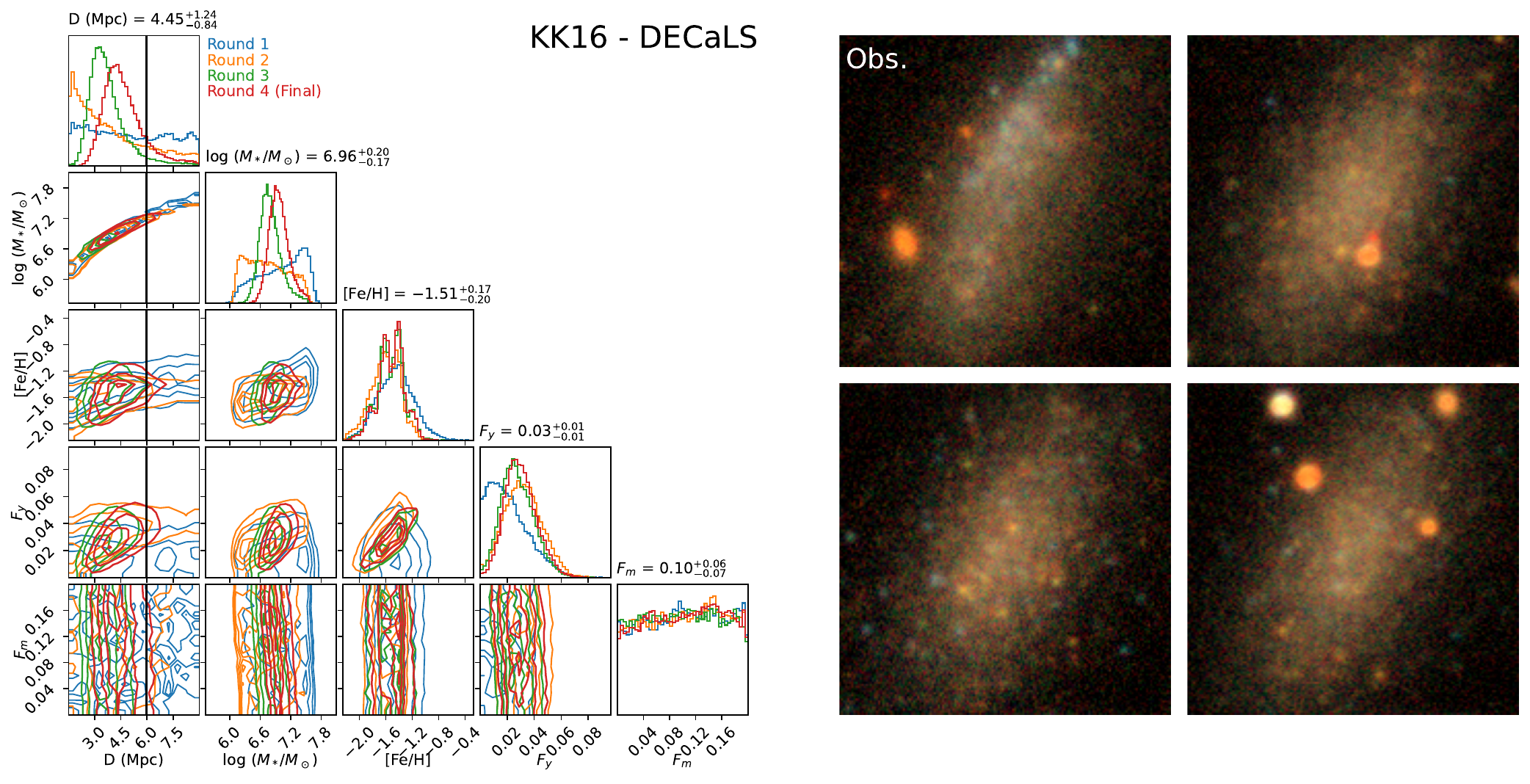}
\includegraphics[width = 0.2\textwidth]{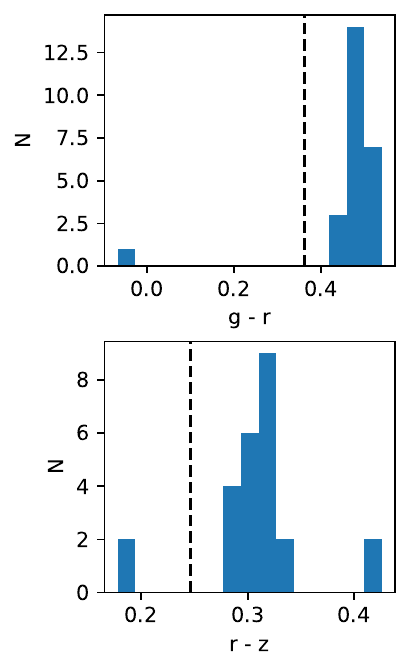}
\includegraphics[width = 0.75\textwidth]{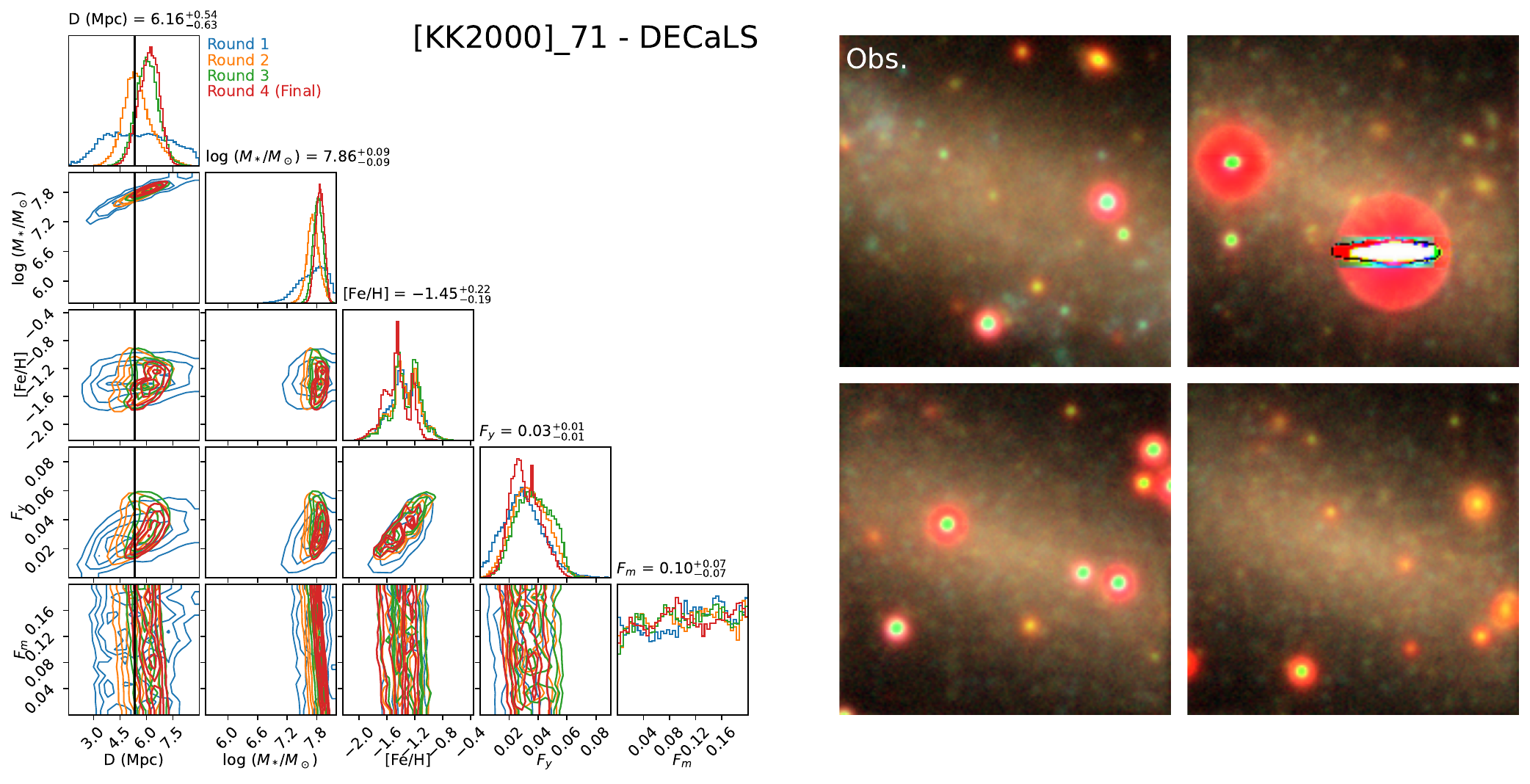}
\includegraphics[width = 0.2\textwidth]{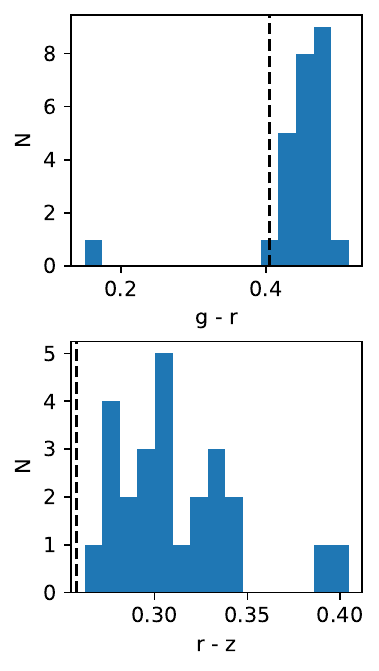}
\includegraphics[width = 0.75\textwidth]{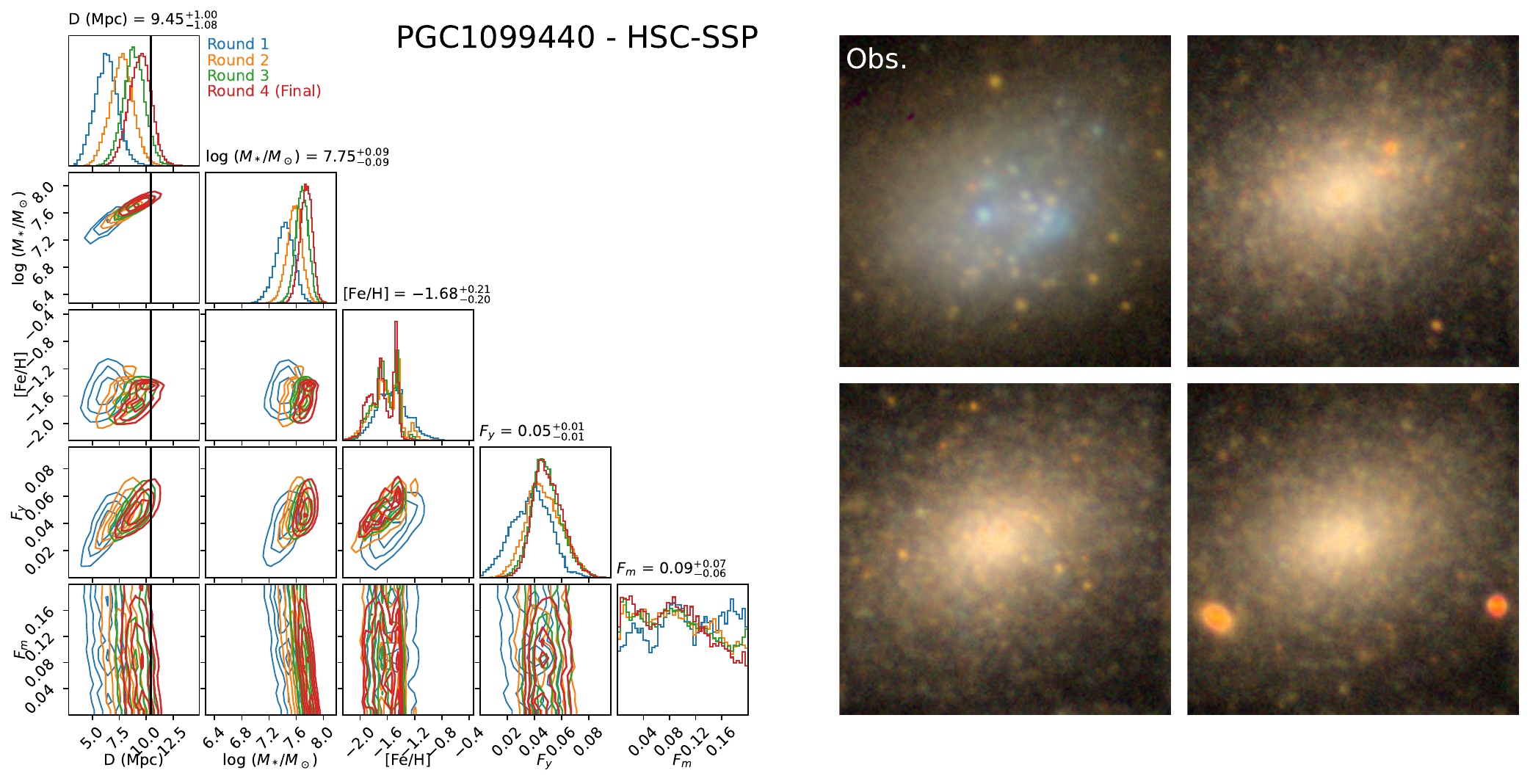}
\includegraphics[width = 0.2\textwidth]{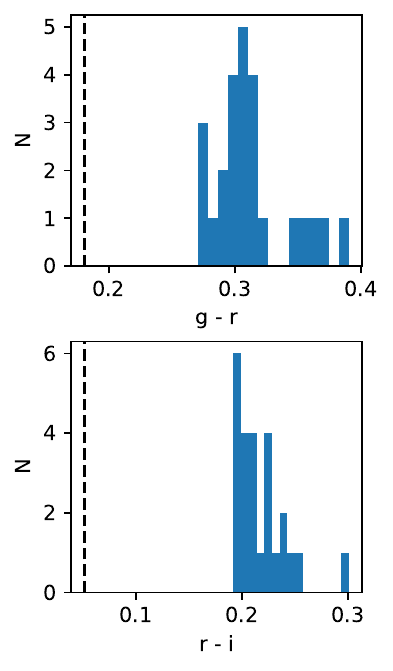}
\caption{Continuation of Figure~\ref{fig:corner_im_all_0}}
\label{fig:corner_im_all_2}
\end{figure*}
\begin{figure*}
\centering
\includegraphics[width = 0.75\textwidth]{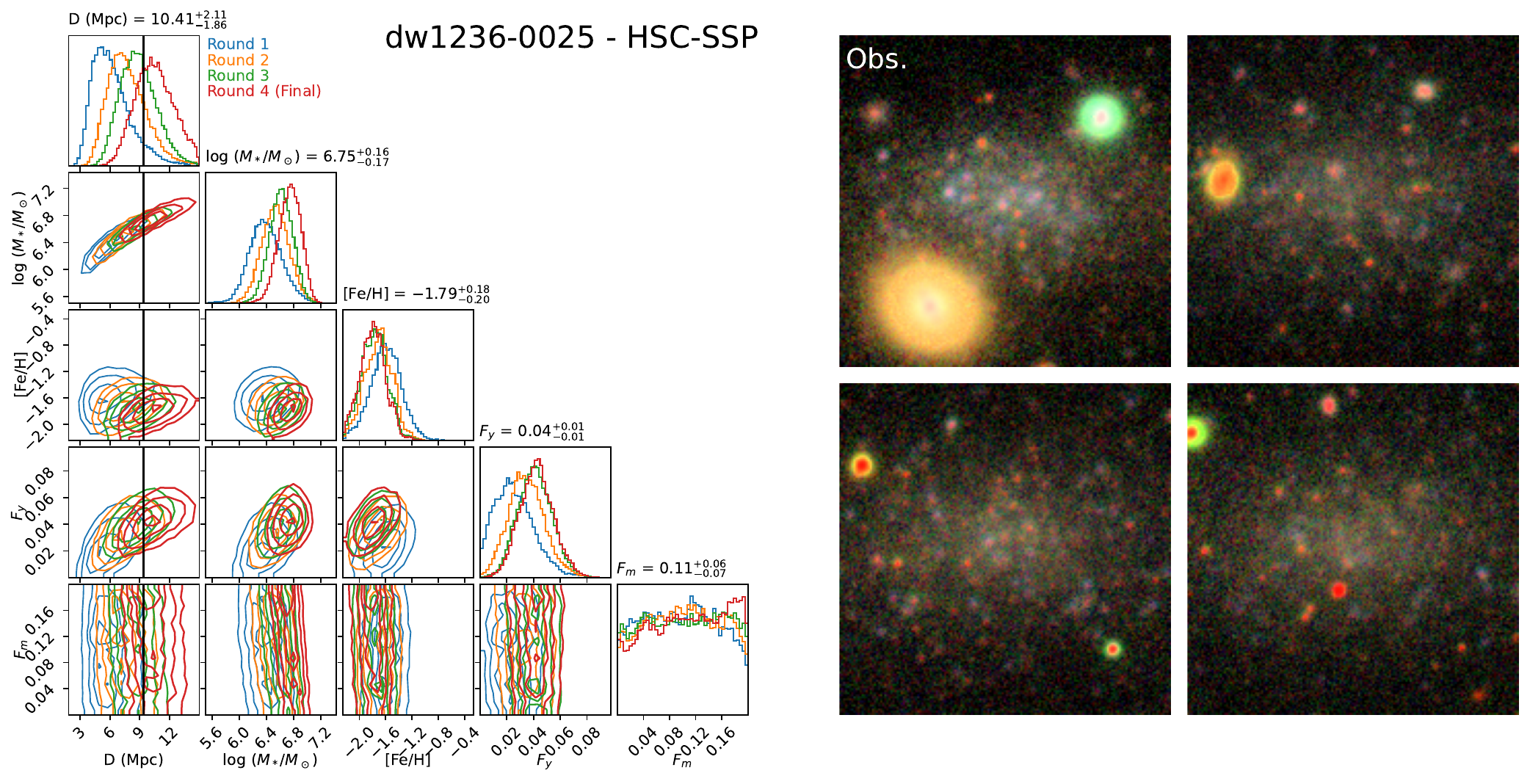}
\includegraphics[width = 0.2\textwidth]{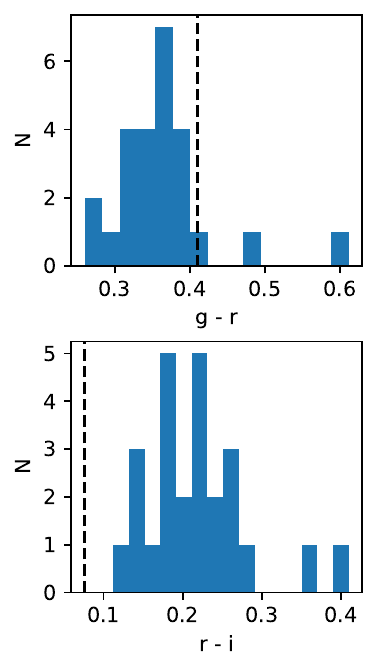}
\includegraphics[width = 0.75\textwidth]{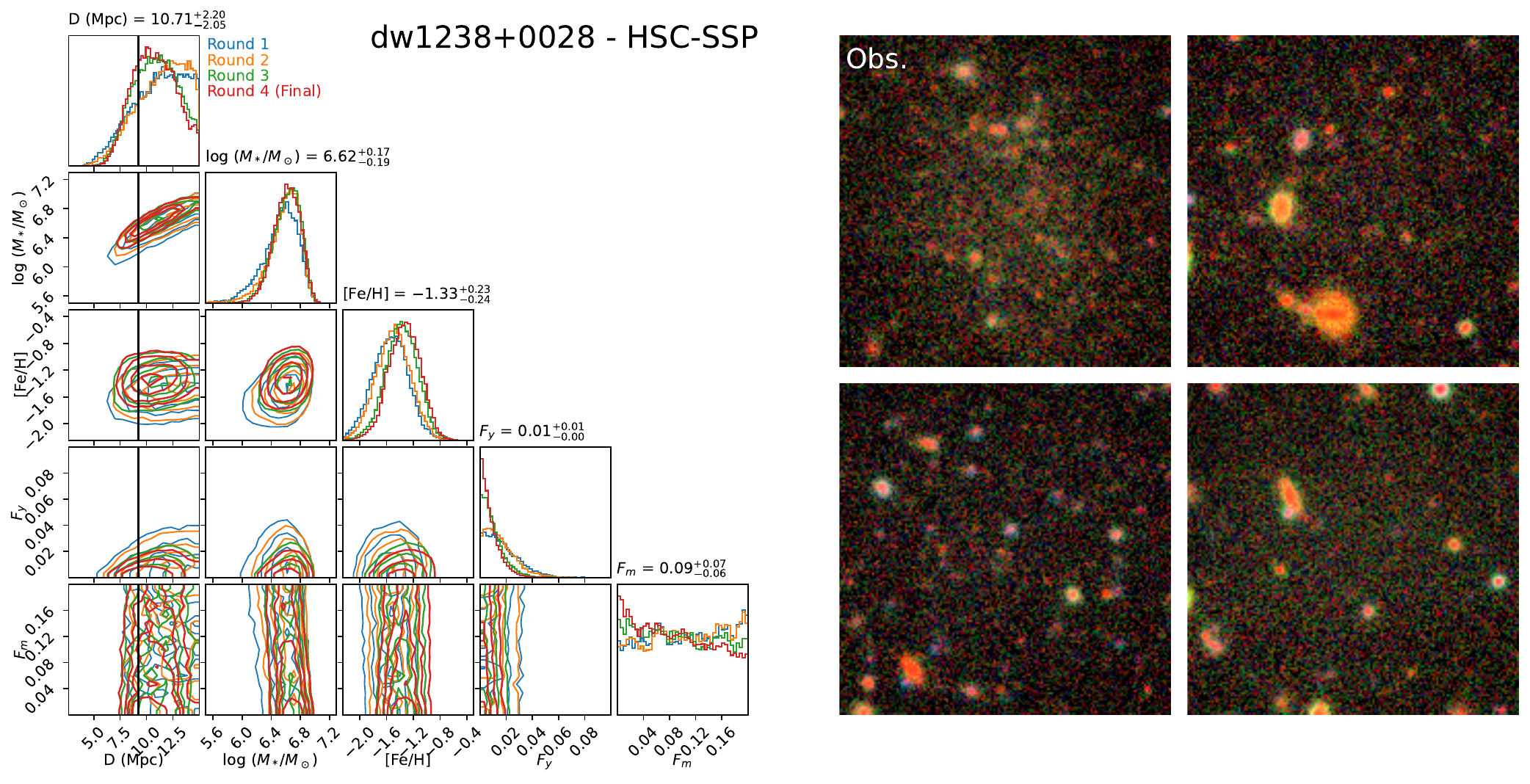}
\includegraphics[width = 0.2\textwidth]{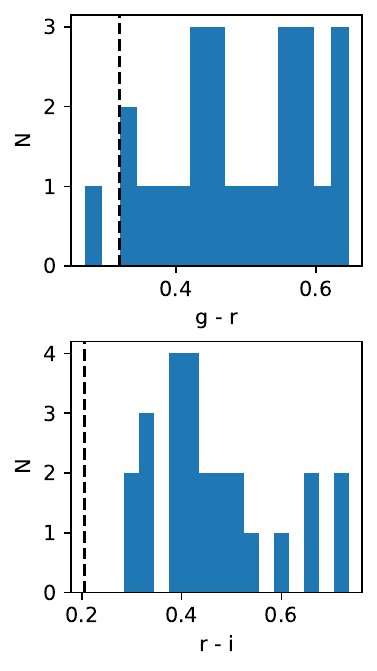}
\includegraphics[width = 0.75\textwidth]{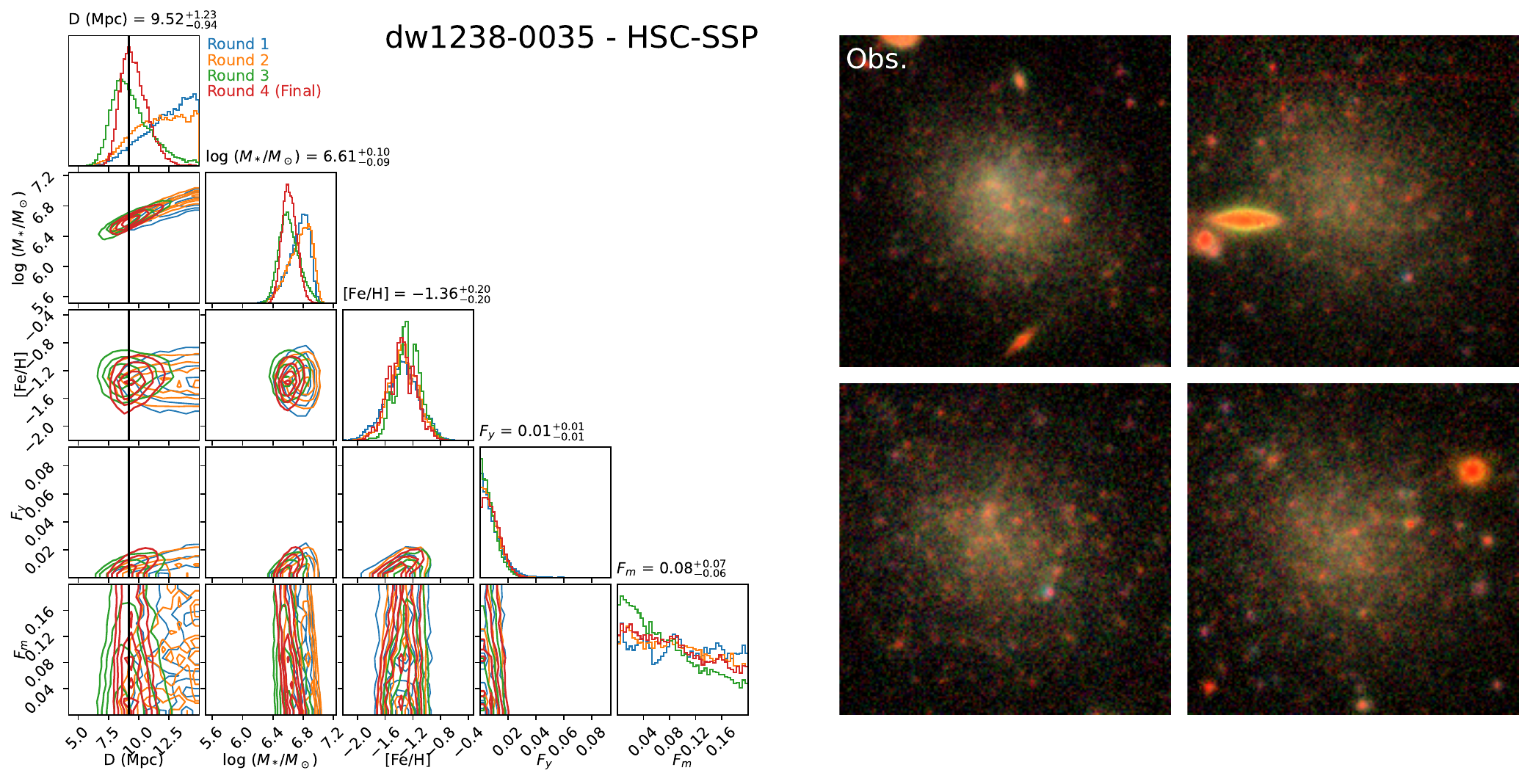}
\includegraphics[width = 0.2\textwidth]{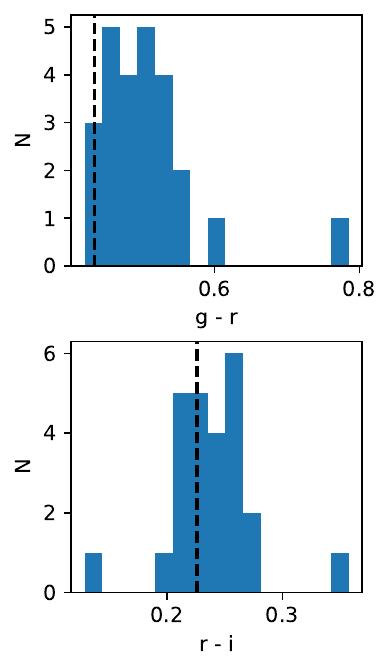}
\caption{Continuation of Figure~\ref{fig:corner_im_all_0}}
\label{fig:corner_im_all_3}
\end{figure*}
\begin{figure*}
\centering
\includegraphics[width = 0.75\textwidth]{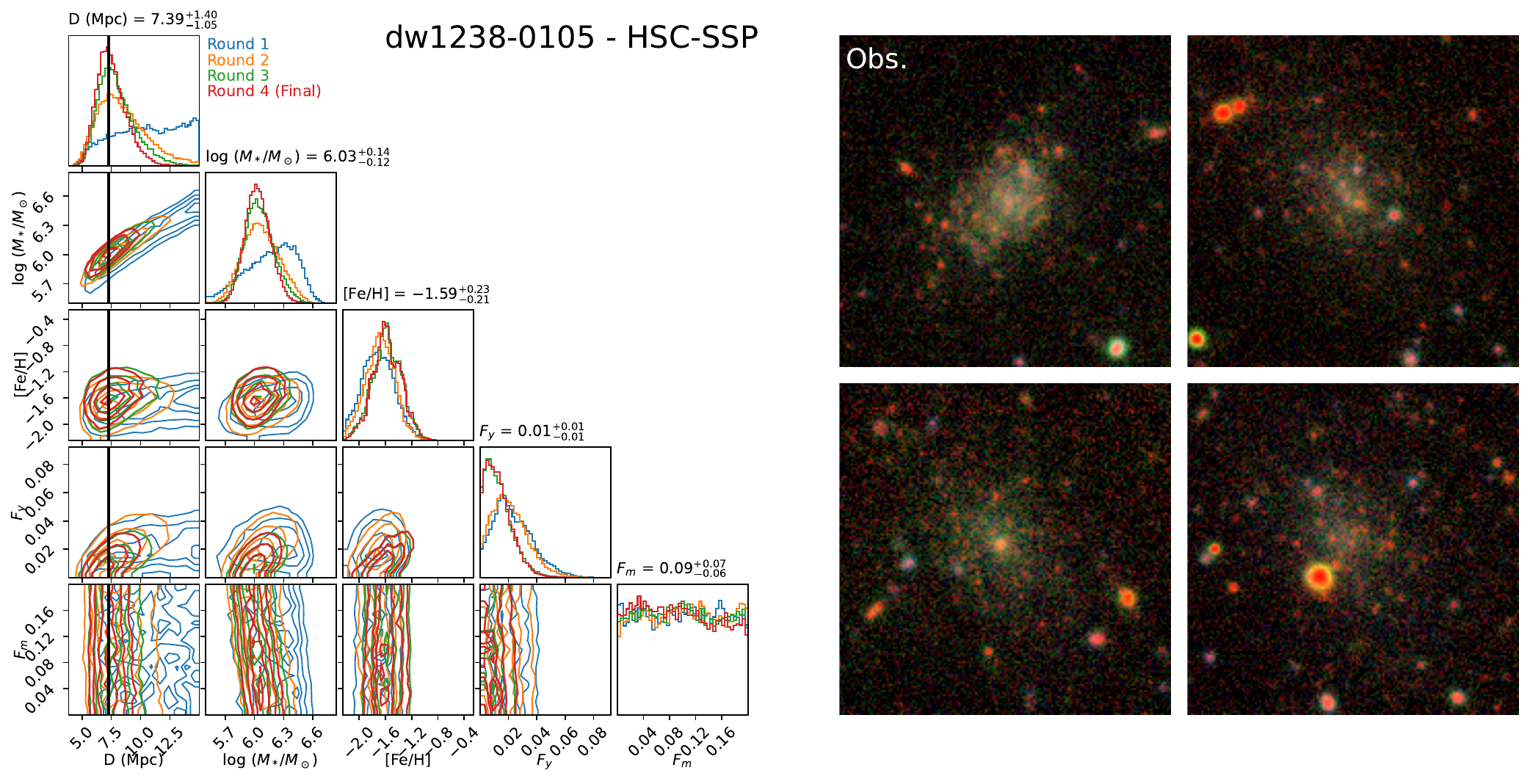}
\includegraphics[width = 0.2\textwidth]{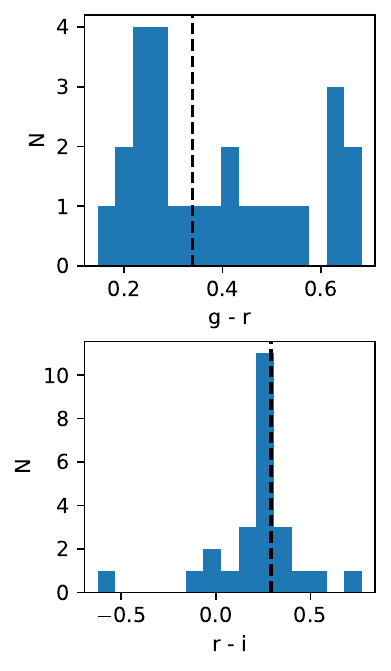}
\includegraphics[width = 0.75\textwidth]{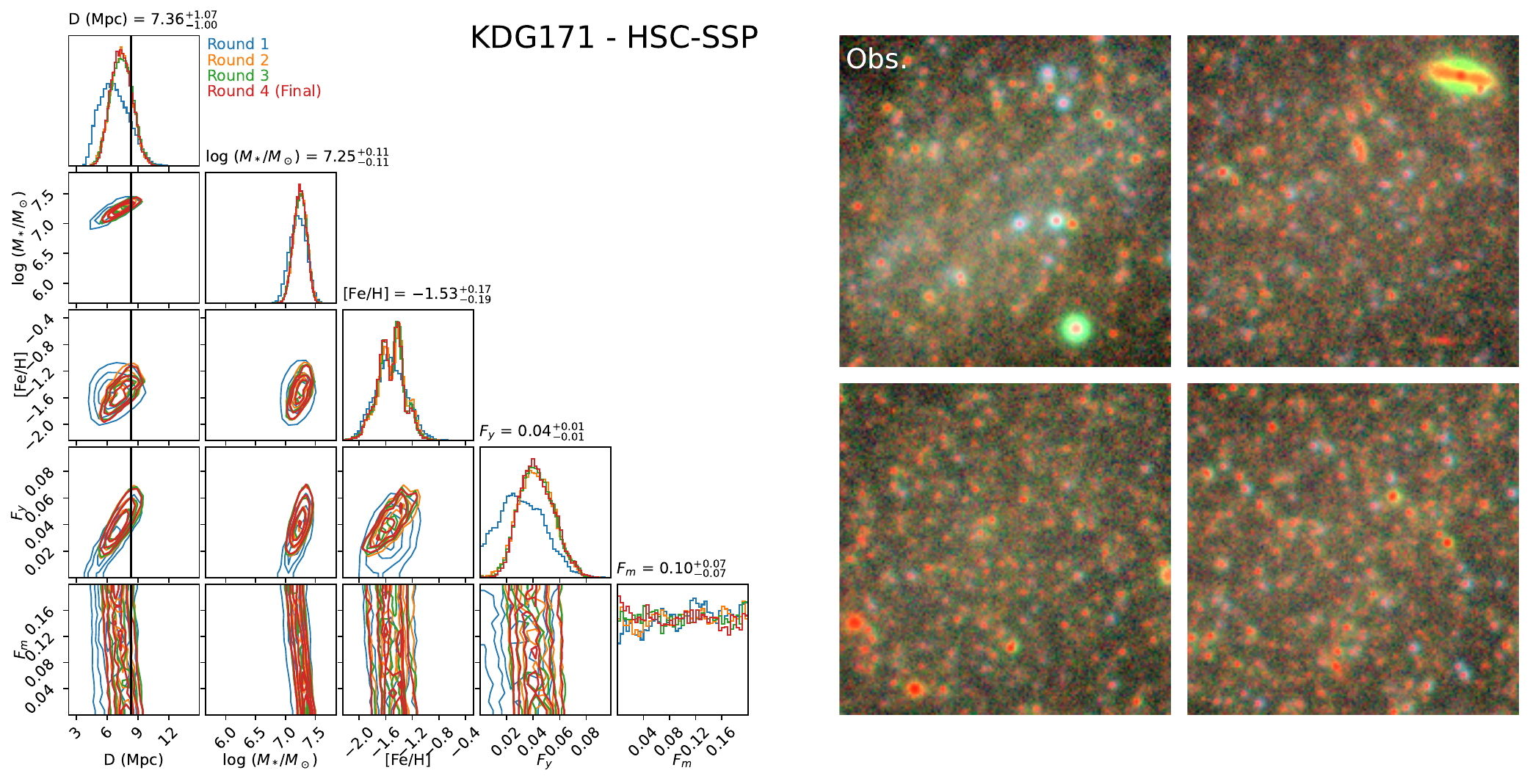}
\includegraphics[width = 0.2\textwidth]{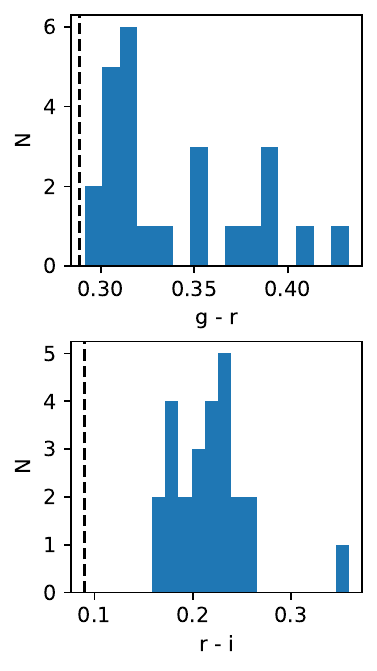}
\includegraphics[width = 0.75\textwidth]{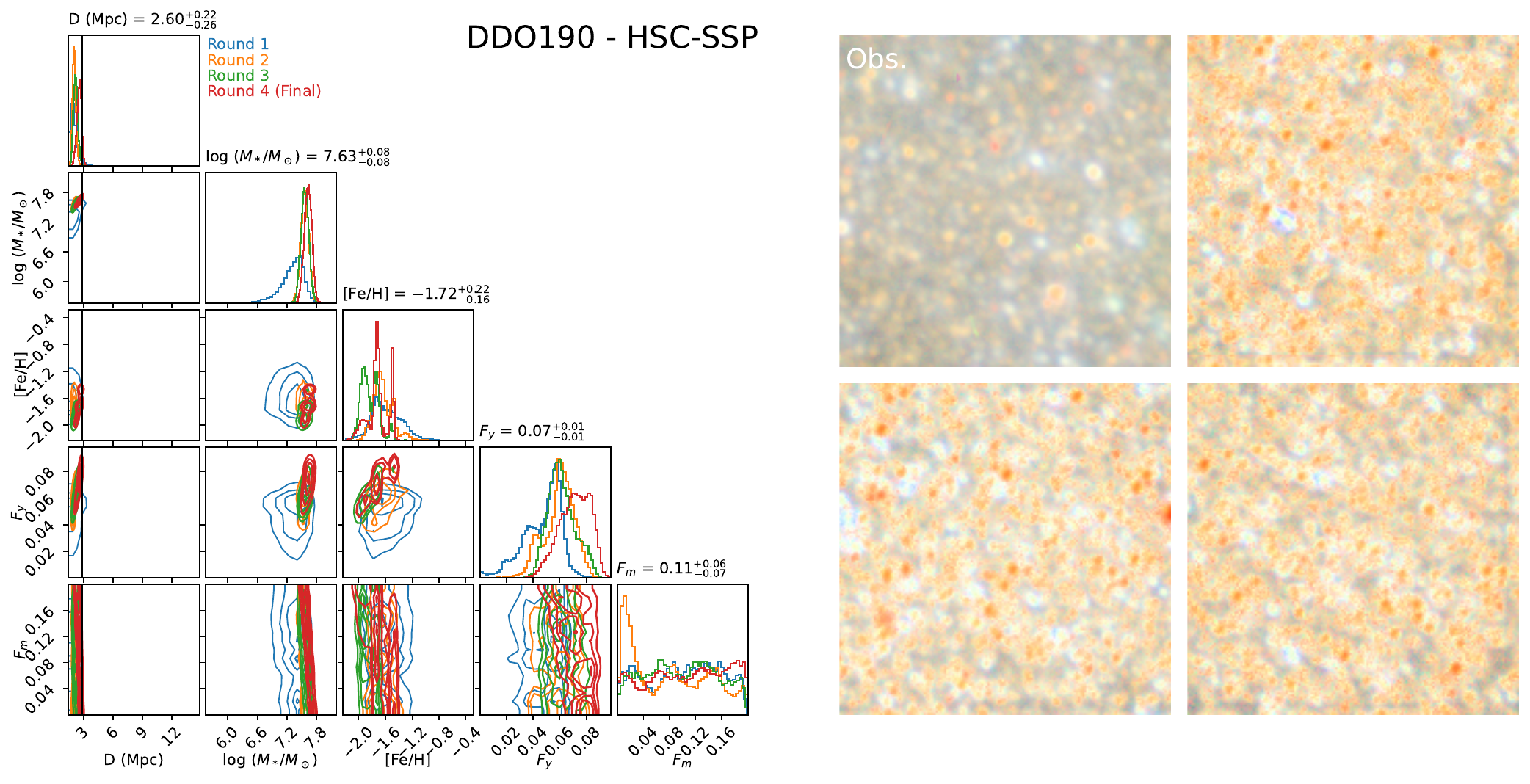}
\includegraphics[width = 0.2\textwidth]{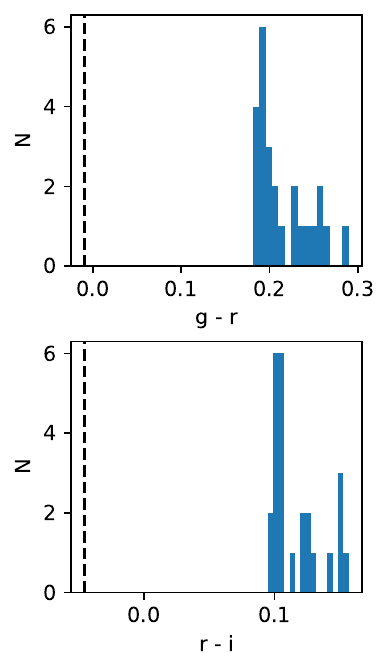}
\caption{Continuation of Figure~\ref{fig:corner_im_all_0}}
\label{fig:corner_im_all_4}
\end{figure*}

\section{Galaxies where \code{} distance differs from literature values}
\label{sec:outliers}
In this section we further analyze the galaxies where the \code{} inferred distance is discrepant from the measured literature distance. We focus on the five galaxies where the literature distance is more than the 95\% percentile removed from the Distance posterior, to search for systematic issues. A corner plot showing the full posterior distribution and posterior predictive images are shown, similar to Sec.~\ref{sec:examp_gal} are shown in Figure~\ref{fig:corner_im_outlier_0} and Figure~\ref{fig:corner_im_outlier_1}.

From the joint posterior distributions we find no obvious systematic issues. For the galaxies KK65 and KK182, where DECaLS data is used, the literature distance is consistent with the wide posterior from the first round of training but the later rounds narrow the posterior away from that distance. For the galaxies dw1232+0015 and CGCG 014-054, even after the initial round of training the distance discrepant with the literature value. In addition these five galaxies span a range of $Z$ and $F_y$, eliminating the possibility of a systematic issue in modelling for one of these parameters.

Further examination of the posterior predictive images gives no further clues. By eye there are no obvious trends among these four galaxies. The starkest difference is that the observed image CGCG 014-054 appears bluer than the simulated images. However the $F_y$ measured by silkscreen is only 0.04, meaning there is plenty of room within the prior to increase $F_y$ resulting in a bluer population. For KK62, dw1232+0015 and CGCG 014-054 the images appear to be more centrally concentrated compared to the observed image. This could indicate an issue with the Sersic fitting that is an input to \code{} that could bias the resulting inference. However, looking at the residuals from our Sersic morphology fits (described in section ~\ref{sec:data}), we find no significant issues.

Another possibility is that there are issues with the distances measured in the literature. For dw1232+0015, whose distance was measured using SBF in \citet{Carlsten2022}, the difference in distance would imply an offset in absolute SBF magnitude of 0.53 mag. This is only $ 1.6\sigma $ away from the calibration assuming a rms scatter of 0.32 mag, as reported in \citet{carlsten2019}. Taking into account these uncertainties, it is much less of an outlier. For TRGB measurements, there is a known issue where blends can produce a "phantom tip" to the red giant branch.\citep{Bailan2011,vandokkum2018} However in this case, this would imply the true distance is larger than reported, worsening the discrepancies. Given that this is only the initial implementation of \code{}, and it remains relatively untested for a large population of galaxies, we do not favor such explanations. Moving forward, comparing different distance measurements will help overcome systematic issues associated with each method to obtain robust results.

In summary, we find no conclusive answer to the question of what causes these discrepancies; there is nothing systematically different for these five galaxies. With such a small sample it is difficult to further examine possible causes. With a much larger sample of galaxies we will be able to better separate discrepant measurements as a function of galaxy properties, to investigate ways to improve \code{} and how to flag where it is likely to produce anomalous results.

\begin{figure*}
\centering
\includegraphics[width = 0.75\textwidth]{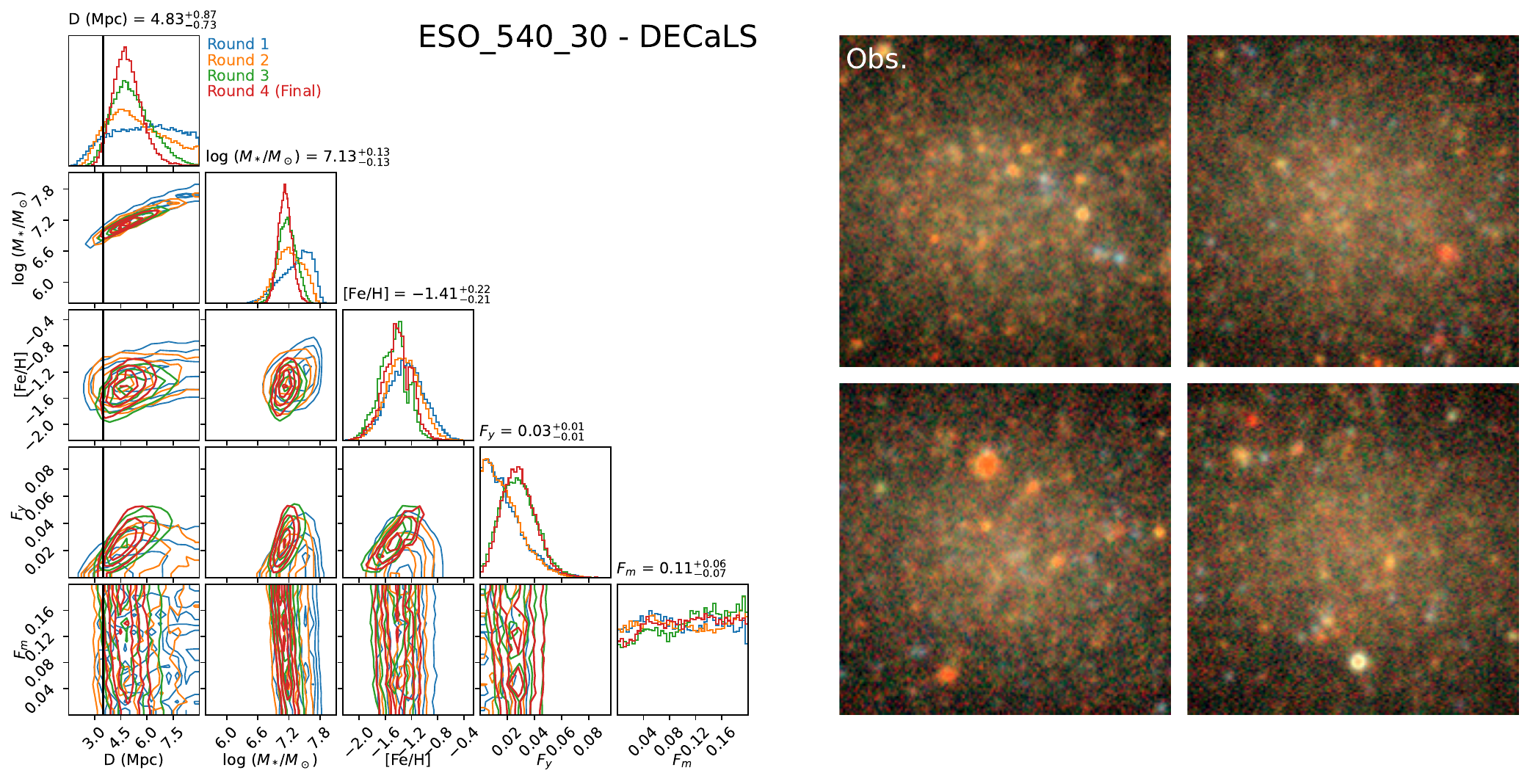}
\includegraphics[width = 0.2\textwidth]{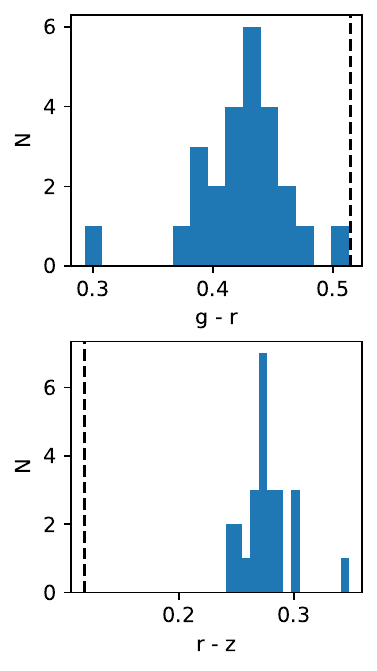}
\includegraphics[width = 0.75\textwidth]{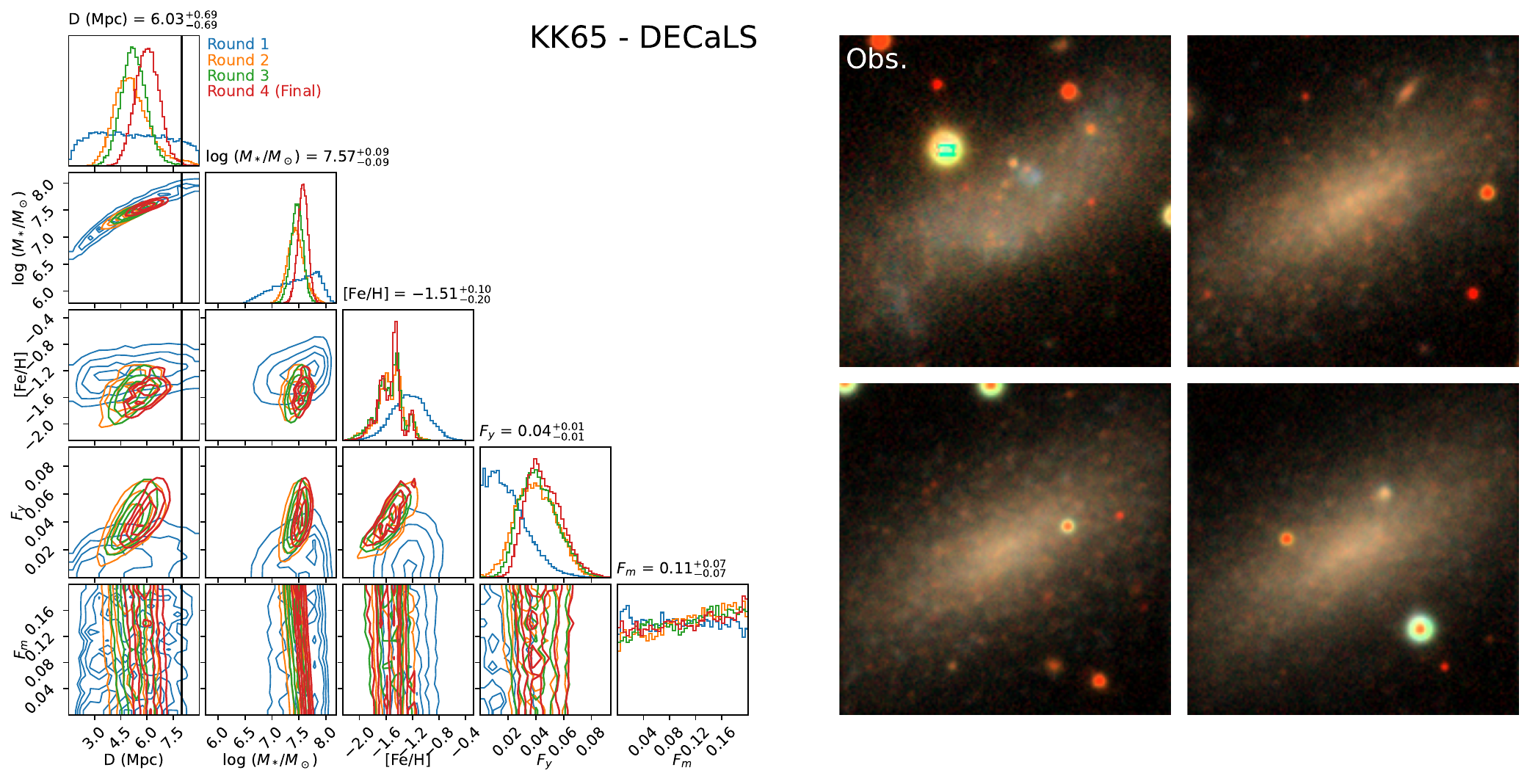}
\includegraphics[width = 0.2\textwidth]{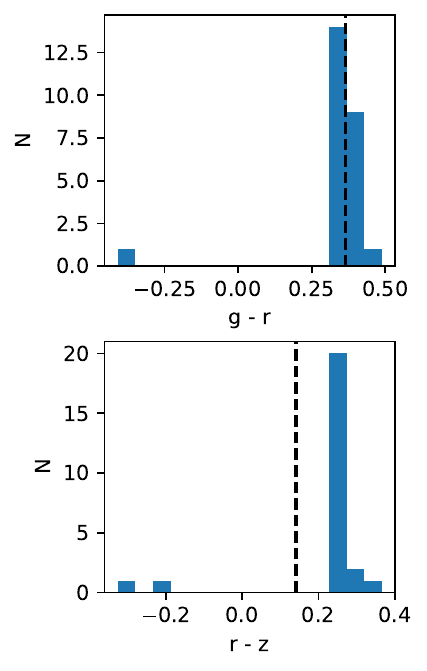}
\includegraphics[width = 0.75\textwidth]{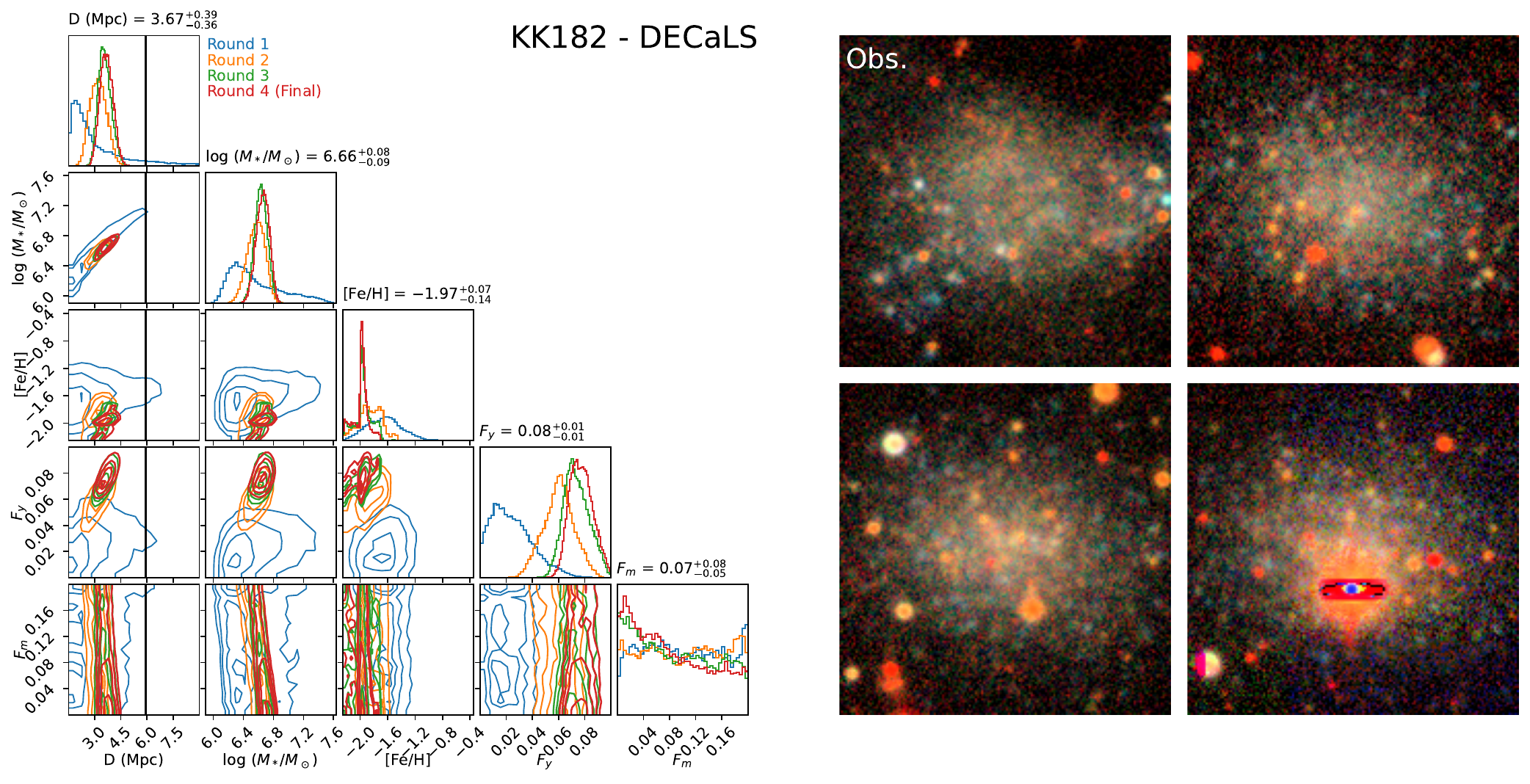}
\includegraphics[width = 0.2\textwidth]{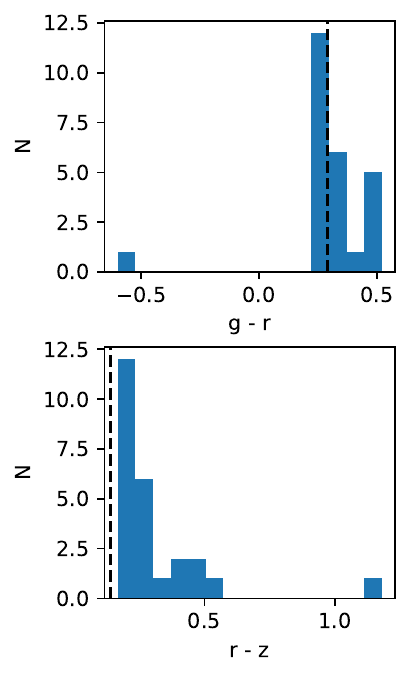}
\caption{Continuation of Figure~\ref{fig:corner_im_all_0} but focused on the galaxies for which the \code{} inferred distance does match the literature value. We choose these when the literature distances lie outside the 5\%-95\% percentile range of the posteriors produced by \code.}
\label{fig:corner_im_outlier_0}
\end{figure*}
\begin{figure*}
\centering
\includegraphics[width = 0.75\textwidth]{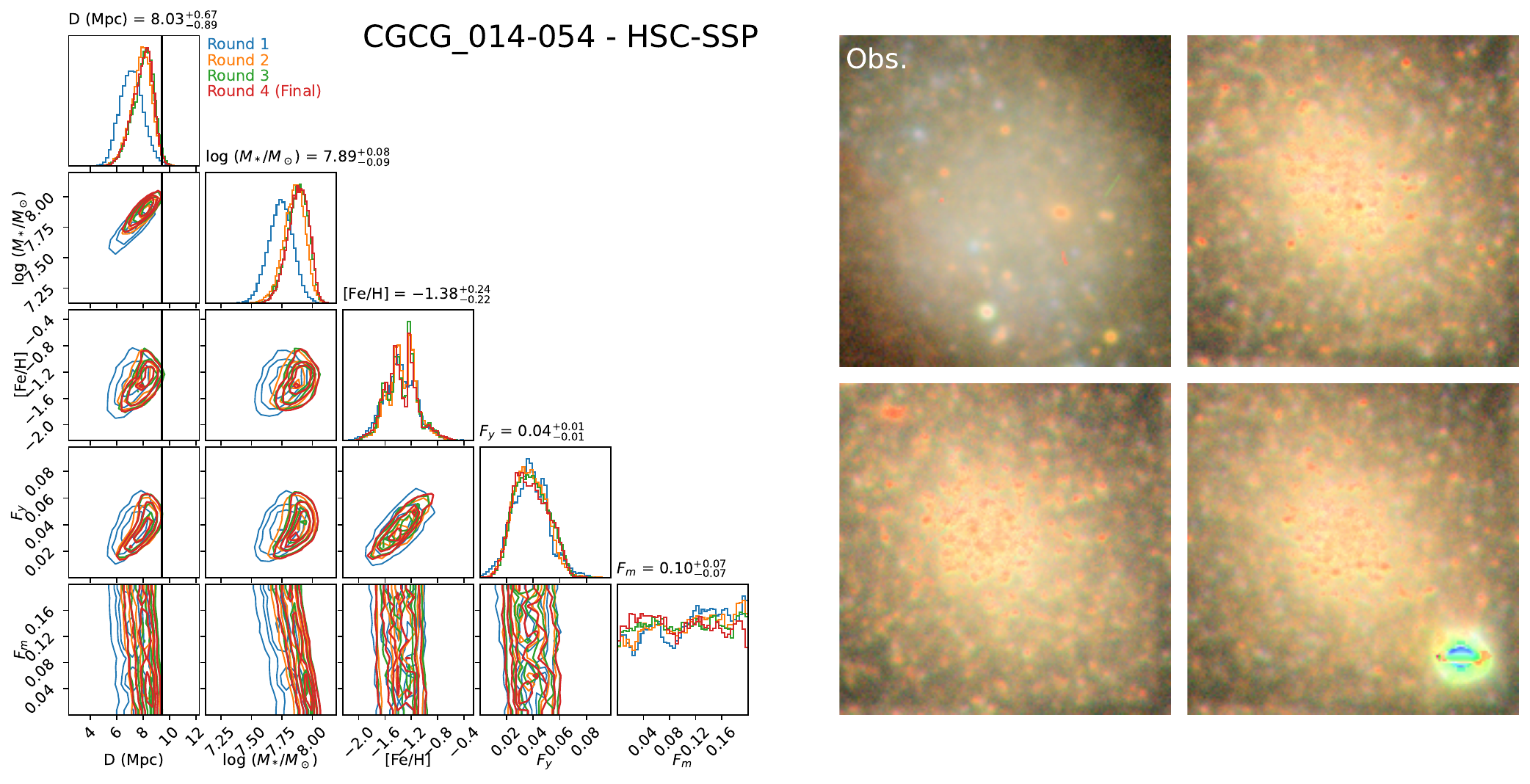}
\includegraphics[width = 0.2\textwidth]{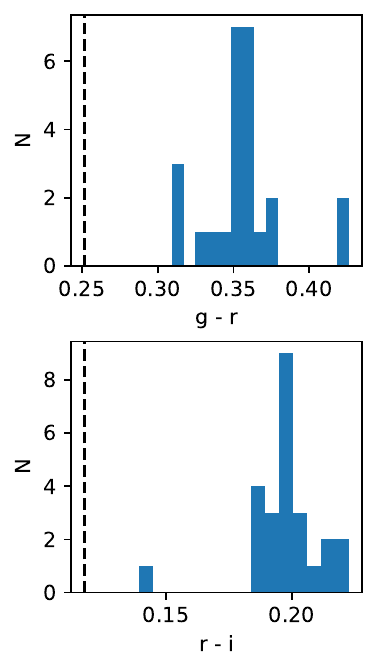}
\includegraphics[width = 0.75\textwidth]{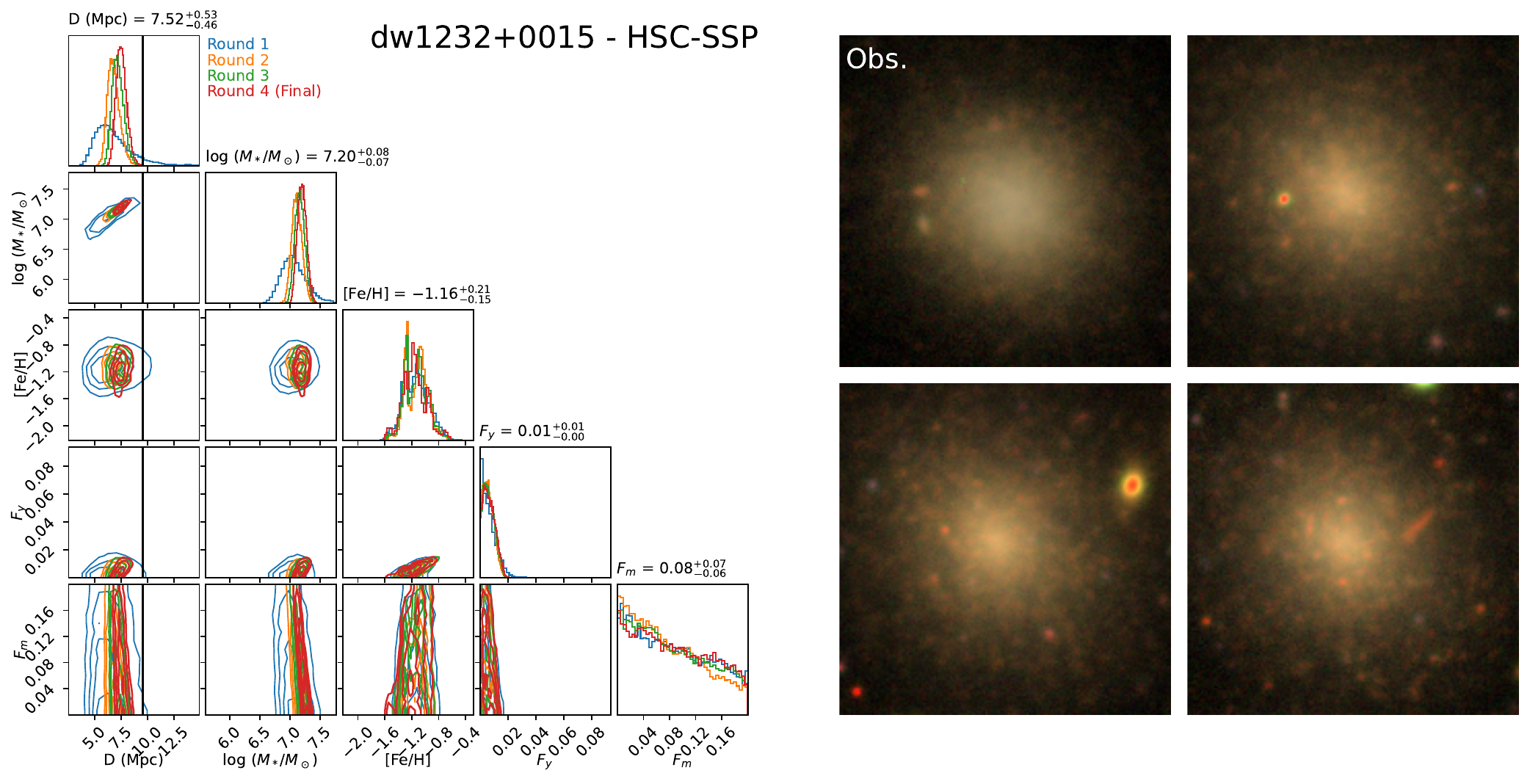}
\includegraphics[width = 0.2\textwidth]{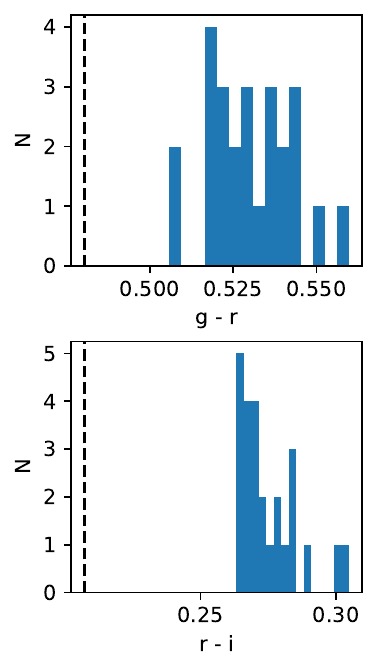}
\caption{Continuation of Figure~\ref{fig:corner_im_outlier_0}, focusing on two more galaxies with discrepant literature and \code{} distances.}
\label{fig:corner_im_outlier_1}
\end{figure*}

\bibliography{bib}{}

\begin{thebibliography}{}
\expandafter\ifx\csname natexlab\endcsname\relax\def\natexlab#1{#1}\fi
\providecommand{\url}[1]{\href{#1}{#1}}
\providecommand{\dodoi}[1]{doi:~\href{http://doi.org/#1}{\nolinkurl{#1}}}
\providecommand{\doeprint}[1]{\href{http://ascl.net/#1}{\nolinkurl{http://ascl.net/#1}}}
\providecommand{\doarXiv}[1]{\href{https://arxiv.org/abs/#1}{\nolinkurl{https://arxiv.org/abs/#1}}}

\bibitem[{{Aihara} {et~al.}(2018){Aihara}, {Arimoto}, {Armstrong}, {Arnouts}, {Bahcall}, {Bickerton}, {Bosch}, {Bundy}, {Capak}, {Chan}, {Chiba}, {Coupon}, {Egami}, {Enoki}, {Finet}, {Fujimori}, {Fujimoto}, {Furusawa}, {Furusawa}, {Goto}, {Goulding}, {Greco}, {Greene}, {Gunn}, {Hamana}, {Harikane}, {Hashimoto}, {Hattori}, {Hayashi}, {Hayashi}, {He{\l}miniak}, {Higuchi}, {Hikage}, {Ho}, {Hsieh}, {Huang}, {Huang}, {Ikeda}, {Imanishi}, {Inoue}, {Iwasawa}, {Iwata}, {Jaelani}, {Jian}, {Kamata}, {Karoji}, {Kashikawa}, {Katayama}, {Kawanomoto}, {Kayo}, {Koda}, {Koike}, {Kojima}, {Komiyama}, {Konno}, {Koshida}, {Koyama}, {Kusakabe}, {Leauthaud}, {Lee}, {Lin}, {Lin}, {Lupton}, {Mandelbaum}, {Matsuoka}, {Medezinski}, {Mineo}, {Miyama}, {Miyatake}, {Miyazaki}, {Momose}, {More}, {More}, {Moritani}, {Moriya}, {Morokuma}, {Mukae}, {Murata}, {Murayama}, {Nagao}, {Nakata}, {Niida}, {Niikura}, {Nishizawa}, {Obuchi}, {Oguri}, {Oishi}, {Okabe}, {Okamoto}, {Okura}, {Ono}, {Onodera}, {Onoue}, {Osato}, {Ouchi}, {Price}, {Pyo},
  {Sako}, {Sawicki}, {Shibuya}, {Shimasaku}, {Shimono}, {Shirasaki}, {Silverman}, {Simet}, {Speagle}, {Spergel}, {Strauss}, {Sugahara}, {Sugiyama}, {Suto}, {Suyu}, {Suzuki}, {Tait}, {Takada}, {Takata}, {Tamura}, {Tanaka}, {Tanaka}, {Tanaka}, {Tanaka}, {Terai}, {Terashima}, {Toba}, {Tominaga}, {Toshikawa}, {Turner}, {Uchida}, {Uchiyama}, {Umetsu}, {Uraguchi}, {Urata}, {Usuda}, {Utsumi}, {Wang}, {Wang}, {Wong}, {Yabe}, {Yamada}, {Yamanoi}, {Yasuda}, {Yeh}, {Yonehara}, \& {Yuma}}]{aihara2018}
{Aihara}, H., {Arimoto}, N., {Armstrong}, R., {et~al.} 2018, \pasj, 70, S4, \dodoi{10.1093/pasj/psx066}

\bibitem[{{Akeson} {et~al.}(2019){Akeson}, {Armus}, {Bachelet}, {Bailey}, {Bartusek}, {Bellini}, {Benford}, {Bennett}, {Bhattacharya}, {Bohlin}, {Boyer}, {Bozza}, {Bryden}, {Calchi Novati}, {Carpenter}, {Casertano}, {Choi}, {Content}, {Dayal}, {Dressler}, {Dor{\'e}}, {Fall}, {Fan}, {Fang}, {Filippenko}, {Finkelstein}, {Foley}, {Furlanetto}, {Kalirai}, {Gaudi}, {Gilbert}, {Girard}, {Grady}, {Greene}, {Guhathakurta}, {Heinrich}, {Hemmati}, {Hendel}, {Henderson}, {Henning}, {Hirata}, {Ho}, {Huff}, {Hutter}, {Jansen}, {Jha}, {Johnson}, {Jones}, {Kasdin}, {Kelly}, {Kirshner}, {Koekemoer}, {Kruk}, {Lewis}, {Macintosh}, {Madau}, {Malhotra}, {Mandel}, {Massara}, {Masters}, {McEnery}, {McQuinn}, {Melchior}, {Melton}, {Mennesson}, {Peeples}, {Penny}, {Perlmutter}, {Pisani}, {Plazas}, {Poleski}, {Postman}, {Ranc}, {Rauscher}, {Rest}, {Roberge}, {Robertson}, {Rodney}, {Rhoads}, {Rhodes}, {Ryan}, {Sahu}, {Sand}, {Scolnic}, {Seth}, {Shvartzvald}, {Siellez}, {Smith}, {Spergel}, {Stassun}, {Street}, {Strolger}, {Szalay},
  {Trauger}, {Troxel}, {Turnbull}, {van der Marel}, {von der Linden}, {Wang}, {Weinberg}, {Williams}, {Windhorst}, {Wollack}, {Wu}, {Yee}, \& {Zimmerman}}]{Akeson2019}
{Akeson}, R., {Armus}, L., {Bachelet}, E., {et~al.} 2019, arXiv e-prints, arXiv:1902.05569, \dodoi{10.48550/arXiv.1902.05569}

\bibitem[{Akiba {et~al.}(2019)Akiba, Sano, Yanase, Ohta, \& Koyama}]{akiba2019}
Akiba, T., Sano, S., Yanase, T., Ohta, T., \& Koyama, M. 2019, in Proceedings of the 25th ACM SIGKDD international conference on knowledge discovery \& data mining, 2623--2631

\bibitem[{Anand {et~al.}(2021)Anand, Rizzi, Tully, Shaya, Karachentsev, Makarov, Makarova, Wu, Dolphin, \& Kourkchi}]{Anand2021}
Anand, G.~S., Rizzi, L., Tully, R.~B., {et~al.} 2021, The Astronomical Journal, 162, 80, \dodoi{10.3847/1538-3881/ac0440}

\bibitem[{{Bailin} {et~al.}(2011){Bailin}, {Bell}, {Chappell}, {Radburn-Smith}, \& {de Jong}}]{Bailan2011}
{Bailin}, J., {Bell}, E.~F., {Chappell}, S.~N., {Radburn-Smith}, D.~J., \& {de Jong}, R.~S. 2011, \apj, 736, 24, \dodoi{10.1088/0004-637X/736/1/24}

\bibitem[{{Blakeslee} {et~al.}(2001){Blakeslee}, {Vazdekis}, \& {Ajhar}}]{Blakeslee2001}
{Blakeslee}, J.~P., {Vazdekis}, A., \& {Ajhar}, E.~A. 2001, \mnras, 320, 193, \dodoi{10.1046/j.1365-8711.2001.03937.x}

\bibitem[{{Blakeslee} {et~al.}(2010){Blakeslee}, {Cantiello}, {Mei}, {C{\^o}t{\'e}}, {Barber DeGraaff}, {Ferrarese}, {Jord{\'a}n}, {Peng}, {Tonry}, \& {Worthey}}]{blakeslee2010}
{Blakeslee}, J.~P., {Cantiello}, M., {Mei}, S., {et~al.} 2010, \apj, 724, 657, \dodoi{10.1088/0004-637X/724/1/657}

\bibitem[{{Bothun}(1986)}]{bouthon1986}
{Bothun}, G.~D. 1986, \aj, 91, 507, \dodoi{10.1086/114029}

\bibitem[{{Cantiello} {et~al.}(2018){Cantiello}, {Blakeslee}, {Ferrarese}, {C{\^o}t{\'e}}, {Roediger}, {Raimondo}, {Peng}, {Gwyn}, {Durrell}, \& {Cuillandre}}]{cantiello2018}
{Cantiello}, M., {Blakeslee}, J.~P., {Ferrarese}, L., {et~al.} 2018, \apj, 856, 126, \dodoi{10.3847/1538-4357/aab043}

\bibitem[{{Carleton} {et~al.}(2023){Carleton}, {Ellsworth-Bowers}, {Windhorst}, {Cohen}, {Conselice}, {Diego}, {Zitrin}, {Archer}, {McIntyre}, {Kamieneski}, {Jansen}, {Summers}, {D'Silva}, {Koekemoer}, {Coe}, {Driver}, {Frye}, {Grogin}, {Marshall}, {Nonino}, {Pirzkal}, {Robotham}, {Ryan}, {Ortiz}, {Tompkins}, {Willmer}, {Yan}, \& {Holwerda}}]{Carleton2023}
{Carleton}, T., {Ellsworth-Bowers}, T., {Windhorst}, R.~A., {et~al.} 2023, arXiv e-prints, arXiv:2309.16028, \dodoi{10.48550/arXiv.2309.16028}

\bibitem[{{Carlsten} {et~al.}(2019){Carlsten}, {Beaton}, {Greco}, \& {Greene}}]{carlsten2019}
{Carlsten}, S.~G., {Beaton}, R.~L., {Greco}, J.~P., \& {Greene}, J.~E. 2019, \apj, 879, 13, \dodoi{10.3847/1538-4357/ab22c1}

\bibitem[{{Carlsten} {et~al.}(2022){Carlsten}, {Greene}, {Beaton}, {Danieli}, \& {Greco}}]{Carlsten2022}
{Carlsten}, S.~G., {Greene}, J.~E., {Beaton}, R.~L., {Danieli}, S., \& {Greco}, J.~P. 2022, \apj, 933, 47, \dodoi{10.3847/1538-4357/ac6fd7}

\bibitem[{{Choi} {et~al.}(2016){Choi}, {Dotter}, {Conroy}, {Cantiello}, {Paxton}, \& {Johnson}}]{choi2016}
{Choi}, J., {Dotter}, A., {Conroy}, C., {et~al.} 2016, \apj, 823, 102, \dodoi{10.3847/0004-637X/823/2/102}

\bibitem[{{Cohen} {et~al.}(2018){Cohen}, {van Dokkum}, {Danieli}, {Romanowsky}, {Abraham}, {Merritt}, {Zhang}, {Mowla}, {Kruijssen}, {Conroy}, \& {Wasserman}}]{cohen2018}
{Cohen}, Y., {van Dokkum}, P., {Danieli}, S., {et~al.} 2018, \apj, 868, 96, \dodoi{10.3847/1538-4357/aae7c8}

\bibitem[{{Conroy} \& {van Dokkum}(2016)}]{conroy2016}
{Conroy}, C., \& {van Dokkum}, P.~G. 2016, \apj, 827, 9, \dodoi{10.3847/0004-637X/827/1/9}

\bibitem[{{Cook} {et~al.}(2020){Cook}, {Conroy}, \& {van Dokkum}}]{cook2020}
{Cook}, B.~A., {Conroy}, C., \& {van Dokkum}, P. 2020, \apj, 893, 160, \dodoi{10.3847/1538-4357/ab83ea}

\bibitem[{{Cook} {et~al.}(2019){Cook}, {Conroy}, {van Dokkum}, \& {Speagle}}]{cook2019}
{Cook}, B.~A., {Conroy}, C., {van Dokkum}, P., \& {Speagle}, J.~S. 2019, \apj, 876, 78, \dodoi{10.3847/1538-4357/ab16e5}

\bibitem[{Cook {et~al.}(2006)Cook, Gelman, \& Rubin}]{Cook2006}
Cook, S.~R., Gelman, A., \& Rubin, D.~B. 2006, Journal of Computational and Graphical Statistics, 15, 675

\bibitem[{{Cranmer} {et~al.}(2020){Cranmer}, {Brehmer}, \& {Louppe}}]{Cranmer2020}
{Cranmer}, K., {Brehmer}, J., \& {Louppe}, G. 2020, Proceedings of the National Academy of Science, 117, 30055, \dodoi{10.1073/pnas.1912789117}

\bibitem[{{Da Costa} {et~al.}(2010){Da Costa}, {Rejkuba}, {Jerjen}, \& {Grebel}}]{dacosta2010}
{Da Costa}, G.~S., {Rejkuba}, M., {Jerjen}, H., \& {Grebel}, E.~K. 2010, \apjl, 708, L121, \dodoi{10.1088/2041-8205/708/2/L121}

\bibitem[{{Danieli} {et~al.}(2020){Danieli}, {van Dokkum}, {Abraham}, {Conroy}, {Dolphin}, \& {Romanowsky}}]{Danieli2020}
{Danieli}, S., {van Dokkum}, P., {Abraham}, R., {et~al.} 2020, \apjl, 895, L4, \dodoi{10.3847/2041-8213/ab8dc4}

\bibitem[{Deistler {et~al.}(2022)Deistler, Goncalves, \& Macke}]{deistler2022}
Deistler, M., Goncalves, P.~J., \& Macke, J.~H. 2022, Advances in Neural Information Processing Systems, 35, 23135

\bibitem[{{Dey} {et~al.}(2019){Dey}, {Schlegel}, {Lang}, {Blum}, {Burleigh}, {Fan}, {Findlay}, {Finkbeiner}, {Herrera}, {Juneau}, {Landriau}, {Levi}, {McGreer}, {Meisner}, {Myers}, {Moustakas}, {Nugent}, {Patej}, {Schlafly}, {Walker}, {Valdes}, {Weaver}, {Y{\`e}che}, {Zou}, {Zhou}, {Abareshi}, {Abbott}, {Abolfathi}, {Aguilera}, {Alam}, {Allen}, {Alvarez}, {Annis}, {Ansarinejad}, {Aubert}, {Beechert}, {Bell}, {BenZvi}, {Beutler}, {Bielby}, {Bolton}, {Brice{\~n}o}, {Buckley-Geer}, {Butler}, {Calamida}, {Carlberg}, {Carter}, {Casas}, {Castander}, {Choi}, {Comparat}, {Cukanovaite}, {Delubac}, {DeVries}, {Dey}, {Dhungana}, {Dickinson}, {Ding}, {Donaldson}, {Duan}, {Duckworth}, {Eftekharzadeh}, {Eisenstein}, {Etourneau}, {Fagrelius}, {Farihi}, {Fitzpatrick}, {Font-Ribera}, {Fulmer}, {G{\"a}nsicke}, {Gaztanaga}, {George}, {Gerdes}, {Gontcho}, {Gorgoni}, {Green}, {Guy}, {Harmer}, {Hernandez}, {Honscheid}, {Huang}, {James}, {Jannuzi}, {Jiang}, {Joyce}, {Karcher}, {Karkar}, {Kehoe}, {Kneib}, {Kueter-Young}, {Lan},
  {Lauer}, {Le Guillou}, {Le Van Suu}, {Lee}, {Lesser}, {Perreault Levasseur}, {Li}, {Mann}, {Marshall}, {Mart{\'\i}nez-V{\'a}zquez}, {Martini}, {du Mas des Bourboux}, {McManus}, {Meier}, {M{\'e}nard}, {Metcalfe}, {Mu{\~n}oz-Guti{\'e}rrez}, {Najita}, {Napier}, {Narayan}, {Newman}, {Nie}, {Nord}, {Norman}, {Olsen}, {Paat}, {Palanque-Delabrouille}, {Peng}, {Poppett}, {Poremba}, {Prakash}, {Rabinowitz}, {Raichoor}, {Rezaie}, {Robertson}, {Roe}, {Ross}, {Ross}, {Rudnick}, {Safonova}, {Saha}, {S{\'a}nchez}, {Savary}, {Schweiker}, {Scott}, {Seo}, {Shan}, {Silva}, {Slepian}, {Soto}, {Sprayberry}, {Staten}, {Stillman}, {Stupak}, {Summers}, {Sien Tie}, {Tirado}, {Vargas-Maga{\~n}a}, {Vivas}, {Wechsler}, {Williams}, {Yang}, {Yang}, {Yapici}, {Zaritsky}, {Zenteno}, {Zhang}, {Zhang}, {Zhou}, \& {Zhou}}]{dey2019}
{Dey}, A., {Schlegel}, D.~J., {Lang}, D., {et~al.} 2019, \aj, 157, 168, \dodoi{10.3847/1538-3881/ab089d}

\bibitem[{{Dotter}(2016)}]{dotter2016}
{Dotter}, A. 2016, \apjs, 222, 8, \dodoi{10.3847/0067-0049/222/1/8}

\bibitem[{Durkan {et~al.}(2019)Durkan, Bekasov, Murray, \& Papamakarios}]{durkan2019}
Durkan, C., Bekasov, A., Murray, I., \& Papamakarios, G. 2019, Advances in neural information processing systems, 32

\bibitem[{{Euclid Collaboration} {et~al.}(2022){Euclid Collaboration}, {Scaramella}, {Amiaux}, {Mellier}, {Burigana}, {Carvalho}, {Cuillandre}, {Da Silva}, {Derosa}, {Dinis}, {Maiorano}, {Maris}, {Tereno}, {Laureijs}, {Boenke}, {Buenadicha}, {Dupac}, {Gaspar Venancio}, {G{\'o}mez-{\'A}lvarez}, {Hoar}, {Lorenzo Alvarez}, {Racca}, {Saavedra-Criado}, {Schwartz}, {Vavrek}, {Schirmer}, {Aussel}, {Azzollini}, {Cardone}, {Cropper}, {Ealet}, {Garilli}, {Gillard}, {Granett}, {Guzzo}, {Hoekstra}, {Jahnke}, {Kitching}, {Maciaszek}, {Meneghetti}, {Miller}, {Nakajima}, {Niemi}, {Pasian}, {Percival}, {Pottinger}, {Sauvage}, {Scodeggio}, {Wachter}, {Zacchei}, {Aghanim}, {Amara}, {Auphan}, {Auricchio}, {Awan}, {Balestra}, {Bender}, {Bodendorf}, {Bonino}, {Branchini}, {Brau-Nogue}, {Brescia}, {Candini}, {Capobianco}, {Carbone}, {Carlberg}, {Carretero}, {Casas}, {Castander}, {Castellano}, {Cavuoti}, {Cimatti}, {Cledassou}, {Congedo}, {Conselice}, {Conversi}, {Copin}, {Corcione}, {Costille}, {Courbin}, {Degaudenzi}, {Douspis},
  {Dubath}, {Duncan}, {Dusini}, {Farrens}, {Ferriol}, {Fosalba}, {Fourmanoit}, {Frailis}, {Franceschi}, {Franzetti}, {Fumana}, {Gillis}, {Giocoli}, {Grazian}, {Grupp}, {Haugan}, {Holmes}, {Hormuth}, {Hudelot}, {Kermiche}, {Kiessling}, {Kilbinger}, {Kohley}, {Kubik}, {K{\"u}mmel}, {Kunz}, {Kurki-Suonio}, {Lahav}, {Ligori}, {Lilje}, {Lloro}, {Mansutti}, {Marggraf}, {Markovic}, {Marulli}, {Massey}, {Maurogordato}, {Melchior}, {Merlin}, {Meylan}, {Mohr}, {Moresco}, {Morin}, {Moscardini}, {Munari}, {Nichol}, {Padilla}, {Paltani}, {Peacock}, {Pedersen}, {Pettorino}, {Pires}, {Poncet}, {Popa}, {Pozzetti}, {Raison}, {Rebolo}, {Rhodes}, {Rix}, {Roncarelli}, {Rossetti}, {Saglia}, {Schneider}, {Schrabback}, {Secroun}, {Seidel}, {Serrano}, {Sirignano}, {Sirri}, {Skottfelt}, {Stanco}, {Starck}, {Tallada-Cresp{\'\i}}, {Tavagnacco}, {Taylor}, {Teplitz}, {Toledo-Moreo}, {Torradeflot}, {Trifoglio}, {Valentijn}, {Valenziano}, {Verdoes Kleijn}, {Wang}, {Welikala}, {Weller}, {Wetzstein}, {Zamorani}, {Zoubian}, {Andreon},
  {Baldi}, {Bardelli}, {Boucaud}, {Camera}, {Di Ferdinando}, {Fabbian}, {Farinelli}, {Galeotta}, {Graci{\'a}-Carpio}, {Maino}, {Medinaceli}, {Mei}, {Neissner}, {Polenta}, {Renzi}, {Romelli}, {Rosset}, {Sureau}, {Tenti}, {Vassallo}, {Zucca}, {Baccigalupi}, {Balaguera-Antol{\'\i}nez}, {Battaglia}, {Biviano}, {Borgani}, {Bozzo}, {Cabanac}, {Cappi}, {Casas}, {Castignani}, {Colodro-Conde}, {Coupon}, {Courtois}, {Cuby}, {de la Torre}, {Desai}, {Dole}, {Fabricius}, {Farina}, {Ferreira}, {Finelli}, {Flose-Reimberg}, {Fotopoulou}, {Ganga}, {Gozaliasl}, {Hook}, {Keihanen}, {Kirkpatrick}, {Liebing}, {Lindholm}, {Mainetti}, {Martinelli}, {Martinet}, {Maturi}, {McCracken}, {Metcalf}, {Morgante}, {Nightingale}, {Nucita}, {Patrizii}, {Potter}, {Riccio}, {S{\'a}nchez}, {Sapone}, {Schewtschenko}, {Schultheis}, {Scottez}, {Teyssier}, {Tutusaus}, {Valiviita}, {Viel}, {Vriend}, \& {Whittaker}}]{Euclid2022}
{Euclid Collaboration}, {Scaramella}, R., {Amiaux}, J., {et~al.} 2022, \aap, 662, A112, \dodoi{10.1051/0004-6361/202141938}

\bibitem[{{Foster} {et~al.}(2024){Foster}, {Taylor}, \& {Blakeslee}}]{foster2024}
{Foster}, L.~M., {Taylor}, J.~E., \& {Blakeslee}, J.~P. 2024, \mnras, 527, 1656, \dodoi{10.1093/mnras/stad3235}

\bibitem[{{Freedman} {et~al.}(2020){Freedman}, {Madore}, {Hoyt}, {Jang}, {Beaton}, {Lee}, {Monson}, {Neeley}, \& {Rich}}]{Freedman2020}
{Freedman}, W.~L., {Madore}, B.~F., {Hoyt}, T., {et~al.} 2020, \apj, 891, 57, \dodoi{10.3847/1538-4357/ab7339}

\bibitem[{{Geha} {et~al.}(2013){Geha}, {Brown}, {Tumlinson}, {Kalirai}, {Simon}, {Kirby}, {VandenBerg}, {Mu{\~n}oz}, {Avila}, {Guhathakurta}, \& {Ferguson}}]{geha2013}
{Geha}, M., {Brown}, T.~M., {Tumlinson}, J., {et~al.} 2013, \apj, 771, 29, \dodoi{10.1088/0004-637X/771/1/29}

\bibitem[{{Greco} \& {Danieli}(2022)}]{greco2022}
{Greco}, J.~P., \& {Danieli}, S. 2022, \apj, 941, 26, \dodoi{10.3847/1538-4357/ac75b7}

\bibitem[{{Greco} {et~al.}(2021){Greco}, {van Dokkum}, {Danieli}, {Carlsten}, \& {Conroy}}]{greco2021}
{Greco}, J.~P., {van Dokkum}, P., {Danieli}, S., {Carlsten}, S.~G., \& {Conroy}, C. 2021, \apj, 908, 24, \dodoi{10.3847/1538-4357/abd030}

\bibitem[{{Greco} {et~al.}(2018){Greco}, {Greene}, {Strauss}, {Macarthur}, {Flowers}, {Goulding}, {Huang}, {Kim}, {Komiyama}, {Leauthaud}, {Leisman}, {Lupton}, {Sif{\'o}n}, \& {Wang}}]{greco2018}
{Greco}, J.~P., {Greene}, J.~E., {Strauss}, M.~A., {et~al.} 2018, \apj, 857, 104, \dodoi{10.3847/1538-4357/aab842}

\bibitem[{Greenberg {et~al.}(2019)Greenberg, Nonnenmacher, \& Macke}]{greenberg2019}
Greenberg, D., Nonnenmacher, M., \& Macke, J. 2019, in International Conference on Machine Learning, PMLR, 2404--2414

\bibitem[{{Hahn} \& {Melchior}(2022)}]{hahn2022}
{Hahn}, C., \& {Melchior}, P. 2022, \apj, 938, 11, \dodoi{10.3847/1538-4357/ac7b84}

\bibitem[{He {et~al.}(2016)He, Zhang, Ren, \& Sun}]{he2016}
He, K., Zhang, X., Ren, S., \& Sun, J. 2016, in Proceedings of the IEEE conference on computer vision and pattern recognition, 770--778

\bibitem[{Hermans {et~al.}(2022)Hermans, Delaunoy, Rozet, Wehenkel, \& Louppe}]{Hermans2022}
Hermans, J., Delaunoy, A., Rozet, F., Wehenkel, A., \& Louppe, G. 2022, Transactions on Machine Learning Research.
\newblock \url{https://openreview.net/pdf?id=LHAbHkt6Aq}

\bibitem[{{Hidalgo} {et~al.}(2018){Hidalgo}, {Pietrinferni}, {Cassisi}, {Salaris}, {Mucciarelli}, {Savino}, {Aparicio}, {Silva Aguirre}, \& {Verma}}]{Hidalgo2018}
{Hidalgo}, S.~L., {Pietrinferni}, A., {Cassisi}, S., {et~al.} 2018, \apj, 856, 125, \dodoi{10.3847/1538-4357/aab158}

\bibitem[{{Iglesias-Navarro} {et~al.}(2024){Iglesias-Navarro}, {Huertas-Company}, {Mart{\'\i}n-Navarro}, {Knapen}, \& {Pernet}}]{iglesias-navarro2024}
{Iglesias-Navarro}, P., {Huertas-Company}, M., {Mart{\'\i}n-Navarro}, I., {Knapen}, J.~H., \& {Pernet}, E. 2024, arXiv e-prints, arXiv:2406.18661, \dodoi{10.48550/arXiv.2406.18661}

\bibitem[{{Ivezi{\'c}} {et~al.}(2019){Ivezi{\'c}}, {Kahn}, {Tyson}, {Abel}, {Acosta}, {Allsman}, {Alonso}, {AlSayyad}, {Anderson}, {Andrew}, {Angel}, {Angeli}, {Ansari}, {Antilogus}, {Araujo}, {Armstrong}, {Arndt}, {Astier}, {Aubourg}, {Auza}, {Axelrod}, {Bard}, {Barr}, {Barrau}, {Bartlett}, {Bauer}, {Bauman}, {Baumont}, {Bechtol}, {Bechtol}, {Becker}, {Becla}, {Beldica}, {Bellavia}, {Bianco}, {Biswas}, {Blanc}, {Blazek}, {Blandford}, {Bloom}, {Bogart}, {Bond}, {Booth}, {Borgland}, {Borne}, {Bosch}, {Boutigny}, {Brackett}, {Bradshaw}, {Brandt}, {Brown}, {Bullock}, {Burchat}, {Burke}, {Cagnoli}, {Calabrese}, {Callahan}, {Callen}, {Carlin}, {Carlson}, {Chandrasekharan}, {Charles-Emerson}, {Chesley}, {Cheu}, {Chiang}, {Chiang}, {Chirino}, {Chow}, {Ciardi}, {Claver}, {Cohen-Tanugi}, {Cockrum}, {Coles}, {Connolly}, {Cook}, {Cooray}, {Covey}, {Cribbs}, {Cui}, {Cutri}, {Daly}, {Daniel}, {Daruich}, {Daubard}, {Daues}, {Dawson}, {Delgado}, {Dellapenna}, {de Peyster}, {de Val-Borro}, {Digel}, {Doherty}, {Dubois},
  {Dubois-Felsmann}, {Durech}, {Economou}, {Eifler}, {Eracleous}, {Emmons}, {Fausti Neto}, {Ferguson}, {Figueroa}, {Fisher-Levine}, {Focke}, {Foss}, {Frank}, {Freemon}, {Gangler}, {Gawiser}, {Geary}, {Gee}, {Geha}, {Gessner}, {Gibson}, {Gilmore}, {Glanzman}, {Glick}, {Goldina}, {Goldstein}, {Goodenow}, {Graham}, {Gressler}, {Gris}, {Guy}, {Guyonnet}, {Haller}, {Harris}, {Hascall}, {Haupt}, {Hernandez}, {Herrmann}, {Hileman}, {Hoblitt}, {Hodgson}, {Hogan}, {Howard}, {Huang}, {Huffer}, {Ingraham}, {Innes}, {Jacoby}, {Jain}, {Jammes}, {Jee}, {Jenness}, {Jernigan}, {Jevremovi{\'c}}, {Johns}, {Johnson}, {Johnson}, {Jones}, {Juramy-Gilles}, {Juri{\'c}}, {Kalirai}, {Kallivayalil}, {Kalmbach}, {Kantor}, {Karst}, {Kasliwal}, {Kelly}, {Kessler}, {Kinnison}, {Kirkby}, {Knox}, {Kotov}, {Krabbendam}, {Krughoff}, {Kub{\'a}nek}, {Kuczewski}, {Kulkarni}, {Ku}, {Kurita}, {Lage}, {Lambert}, {Lange}, {Langton}, {Le Guillou}, {Levine}, {Liang}, {Lim}, {Lintott}, {Long}, {Lopez}, {Lotz}, {Lupton}, {Lust}, {MacArthur}, {Mahabal},
  {Mandelbaum}, {Markiewicz}, {Marsh}, {Marshall}, {Marshall}, {May}, {McKercher}, {McQueen}, {Meyers}, {Migliore}, {Miller}, {Mills}, {Miraval}, {Moeyens}, {Moolekamp}, {Monet}, {Moniez}, {Monkewitz}, {Montgomery}, {Morrison}, {Mueller}, {Muller}, {Mu{\~n}oz Arancibia}, {Neill}, {Newbry}, {Nief}, {Nomerotski}, {Nordby}, {O'Connor}, {Oliver}, {Olivier}, {Olsen}, {O'Mullane}, {Ortiz}, {Osier}, {Owen}, {Pain}, {Palecek}, {Parejko}, {Parsons}, {Pease}, {Peterson}, {Peterson}, {Petravick}, {Libby Petrick}, {Petry}, {Pierfederici}, {Pietrowicz}, {Pike}, {Pinto}, {Plante}, {Plate}, {Plutchak}, {Price}, {Prouza}, {Radeka}, {Rajagopal}, {Rasmussen}, {Regnault}, {Reil}, {Reiss}, {Reuter}, {Ridgway}, {Riot}, {Ritz}, {Robinson}, {Roby}, {Roodman}, {Rosing}, {Roucelle}, {Rumore}, {Russo}, {Saha}, {Sassolas}, {Schalk}, {Schellart}, {Schindler}, {Schmidt}, {Schneider}, {Schneider}, {Schoening}, {Schumacher}, {Schwamb}, {Sebag}, {Selvy}, {Sembroski}, {Seppala}, {Serio}, {Serrano}, {Shaw}, {Shipsey}, {Sick}, {Silvestri},
  {Slater}, {Smith}, {Smith}, {Sobhani}, {Soldahl}, {Storrie-Lombardi}, {Stover}, {Strauss}, {Street}, {Stubbs}, {Sullivan}, {Sweeney}, {Swinbank}, {Szalay}, {Takacs}, {Tether}, {Thaler}, {Thayer}, {Thomas}, {Thornton}, {Thukral}, {Tice}, {Trilling}, {Turri}, {Van Berg}, {Vanden Berk}, {Vetter}, {Virieux}, {Vucina}, {Wahl}, {Walkowicz}, {Walsh}, {Walter}, {Wang}, {Wang}, {Warner}, {Wiecha}, {Willman}, {Winters}, {Wittman}, {Wolff}, {Wood-Vasey}, {Wu}, {Xin}, {Yoachim}, \& {Zhan}}]{ivezic2019}
{Ivezi{\'c}}, {\v{Z}}., {Kahn}, S.~M., {Tyson}, J.~A., {et~al.} 2019, \apj, 873, 111, \dodoi{10.3847/1538-4357/ab042c}

\bibitem[{{Jensen} {et~al.}(2003){Jensen}, {Tonry}, {Barris}, {Thompson}, {Liu}, {Rieke}, {Ajhar}, \& {Blakeslee}}]{Jensen2003}
{Jensen}, J.~B., {Tonry}, J.~L., {Barris}, B.~J., {et~al.} 2003, \apj, 583, 712, \dodoi{10.1086/345430}

\bibitem[{{Jensen} {et~al.}(2021){Jensen}, {Blakeslee}, {Ma}, {Milne}, {Brown}, {Cantiello}, {Garnavich}, {Greene}, {Lucey}, {Phan}, {Tully}, \& {Wood}}]{Jensen2021}
{Jensen}, J.~B., {Blakeslee}, J.~P., {Ma}, C.-P., {et~al.} 2021, \apjs, 255, 21, \dodoi{10.3847/1538-4365/ac01e7}

\bibitem[{{Jerjen} {et~al.}(2001){Jerjen}, {Rekola}, {Takalo}, {Coleman}, \& {Valtonen}}]{jerjen2001}
{Jerjen}, H., {Rekola}, R., {Takalo}, L., {Coleman}, M., \& {Valtonen}, M. 2001, \aap, 380, 90, \dodoi{10.1051/0004-6361:20011408}

\bibitem[{{Karachentsev} {et~al.}(2013){Karachentsev}, {Makarov}, \& {Kaisina}}]{karachentsev2013}
{Karachentsev}, I.~D., {Makarov}, D.~I., \& {Kaisina}, E.~I. 2013, \aj, 145, 101, \dodoi{10.1088/0004-6256/145/4/101}

\bibitem[{{Karachentsev} {et~al.}(2022){Karachentsev}, {Makarova}, {Anand}, \& {Tully}}]{Karachentsev2022}
{Karachentsev}, I.~D., {Makarova}, L.~N., {Anand}, G.~S., \& {Tully}, R.~B. 2022, \aj, 163, 234, \dodoi{10.3847/1538-3881/ac5ab5}

\bibitem[{{Karachentsev} {et~al.}(2003){Karachentsev}, {Grebel}, {Sharina}, {Dolphin}, {Geisler}, {Guhathakurta}, {Hodge}, {Karachentseva}, {Sarajedini}, \& {Seitzer}}]{karachentsev2003}
{Karachentsev}, I.~D., {Grebel}, E.~K., {Sharina}, M.~E., {et~al.} 2003, \aap, 404, 93, \dodoi{10.1051/0004-6361:20030170}

\bibitem[{{Karachentseva} \& {Karachentsev}(1998)}]{Karachentseva1998}
{Karachentseva}, V.~E., \& {Karachentsev}, I.~D. 1998, \aaps, 127, 409, \dodoi{10.1051/aas:1998109}

\bibitem[{{Khullar} {et~al.}(2022){Khullar}, {Nord}, {{\'C}iprijanovi{\'c}}, {Poh}, \& {Xu}}]{Khullar2022}
{Khullar}, G., {Nord}, B., {{\'C}iprijanovi{\'c}}, A., {Poh}, J., \& {Xu}, F. 2022, Machine Learning: Science and Technology, 3, 04LT04, \dodoi{10.1088/2632-2153/ac98f4}

\bibitem[{{Kirby} {et~al.}(2013){Kirby}, {Cohen}, {Guhathakurta}, {Cheng}, {Bullock}, \& {Gallazzi}}]{kirby2013}
{Kirby}, E.~N., {Cohen}, J.~G., {Guhathakurta}, P., {et~al.} 2013, \apj, 779, 102, \dodoi{10.1088/0004-637X/779/2/102}

\bibitem[{Kobyzev {et~al.}(2020)Kobyzev, Prince, \& Brubaker}]{kobyzev2020}
Kobyzev, I., Prince, S.~J., \& Brubaker, M.~A. 2020, IEEE transactions on pattern analysis and machine intelligence, 43, 3964

\bibitem[{Kroupa(2001)}]{kroupa2001}
Kroupa, P. 2001, Monthly Notices of the Royal Astronomical Society, 322, 231

\bibitem[{{Lanyon-Foster} {et~al.}(2007){Lanyon-Foster}, {Conselice}, \& {Merrifield}}]{lanyonfoster2007}
{Lanyon-Foster}, M.~M., {Conselice}, C.~J., \& {Merrifield}, M.~R. 2007, \mnras, 380, 571, \dodoi{10.1111/j.1365-2966.2007.12132.x}

\bibitem[{{Lee} {et~al.}(2018){Lee}, {Pak}, {Lee}, \& {Oh}}]{lee2018}
{Lee}, J.~H., {Pak}, M., {Lee}, H.-R., \& {Oh}, S. 2018, \apj, 857, 102, \dodoi{10.3847/1538-4357/aab892}

\bibitem[{{Lee} {et~al.}(1993){Lee}, {Freedman}, \& {Madore}}]{lee1993}
{Lee}, M.~G., {Freedman}, W.~L., \& {Madore}, B.~F. 1993, \apj, 417, 553, \dodoi{10.1086/173334}

\bibitem[{Loshchilov \& Hutter(2018)}]{loshchilov2018}
Loshchilov, I., \& Hutter, F. 2018, in International Conference on Learning Representations

\bibitem[{Lueckmann {et~al.}(2021)Lueckmann, Boelts, Greenberg, Goncalves, \& Macke}]{Lueckmann2021}
Lueckmann, J.-M., Boelts, J., Greenberg, D., Goncalves, P., \& Macke, J. 2021, in Proceedings of Machine Learning Research, Vol. 130, Proceedings of The 24th International Conference on Artificial Intelligence and Statistics, ed. A.~Banerjee \& K.~Fukumizu (PMLR), 343--351.
\newblock \url{https://proceedings.mlr.press/v130/lueckmann21a.html}

\bibitem[{maintainers \& contributors(2016)}]{torchvision2016}
maintainers, T., \& contributors. 2016, TorchVision: PyTorch's Computer Vision library, \url{https://github.com/pytorch/vision},  GitHub

\bibitem[{{Marigo} {et~al.}(2017){Marigo}, {Girardi}, {Bressan}, {Rosenfield}, {Aringer}, {Chen}, {Dussin}, {Nanni}, {Pastorelli}, {Rodrigues}, {Trabucchi}, {Bladh}, {Dalcanton}, {Groenewegen}, {Montalb{\'a}n}, \& {Wood}}]{Margio2017}
{Marigo}, P., {Girardi}, L., {Bressan}, A., {et~al.} 2017, \apj, 835, 77, \dodoi{10.3847/1538-4357/835/1/77}

\bibitem[{{Marleau} {et~al.}(2024){Marleau}, {Cuillandre}, {Cantiello}, {Carollo}, {Duc}, {Habas}, {Hunt}, {Jablonka}, {Mirabile}, {Mondelin}, {Poulain}, {Saifollahi}, {S{\'a}nchez-Janssen}, {Sola}, {Urbano}, {Z{\"o}ller}, {Bolzonella}, {Lan{\c{c}}on}, {Laureijs}, {Marchal}, {Schirmer}, {Stone}, {Boselli}, {Ferr{\'e}-Mateu}, {Hatch}, {Kluge}, {Montes}, {Sorce}, {Tortora}, {Venhola}, {Golden-Marx}, {Aghanim}, {Amara}, {Andreon}, {Auricchio}, {Baldi}, {Balestra}, {Bardelli}, {Battaglia}, {Bender}, {Bodendorf}, {Branchini}, {Brescia}, {Brinchmann}, {Camera}, {Candini}, {Capobianco}, {Carbone}, {Carretero}, {Casas}, {Castellano}, {Cavuoti}, {Cimatti}, {Congedo}, {Conselice}, {Conversi}, {Copin}, {Courbin}, {Courtois}, {Cropper}, {Da Silva}, {Degaudenzi}, {Di Giorgio}, {Dinis}, {Douspis}, {Duncan}, {Dupac}, {Dusini}, {Ealet}, {Farina}, {Farrens}, {Ferriol}, {Fosalba}, {Fotopoulou}, {Frailis}, {Franceschi}, {Fumana}, {Galeotta}, {Garilli}, {Gillard}, {Gillis}, {Giocoli}, {G{\'o}mez-Alvarez}, {Grazian}, {Grupp},
  {Guzzo}, {Hailey}, {Haugan}, {Hoar}, {Hoekstra}, {Holmes}, {Hook}, {Hormuth}, {Hornstrup}, {Hu}, {Hudelot}, {Jahnke}, {Jhabvala}, {Keih{\"a}nen}, {Kermiche}, {Kiessling}, {Kitching}, {Kohley}, {Kubik}, {Kuijken}, {K{\"u}mmel}, {Kunz}, {Kurki-Suonio}, {Lahav}, {Le Mignant}, {Ligori}, {Lilje}, {Lindholm}, {Lloro}, {Maino}, {Maiorano}, {Mansutti}, {Marggraf}, {Markovic}, {Martinet}, {Marulli}, {Massey}, {Maurogordato}, {McCracken}, {Medinaceli}, {Mei}, {Mellier}, {Meneghetti}, {Merlin}, {Meylan}, {Moresco}, {Moscardini}, {Munari}, {Nakajima}, {Nichol}, {Niemi}, {Padilla}, {Paltani}, {Pasian}, {Pedersen}, {Percival}, {Pettorino}, {Pires}, {Polenta}, {Poncet}, {Popa}, {Pozzetti}, {Raison}, {Rebolo}, {Refregier}, {Renzi}, {Rhodes}, {Riccio}, {Rix}, {Romelli}, {Roncarelli}, {Rossetti}, {Saglia}, {Sapone}, {Scaramella}, {Schneider}, {Secroun}, {Seidel}, {Seiffert}, {Serrano}, {Sirignano}, {Sirri}, {Stanco}, {Tallada-Cresp{\'\i}}, {Taylor}, {Teplitz}, {Tereno}, {Toledo-Moreo}, {Tsyganov}, {Tutusaus}, {Valentijn},
  {Valenziano}, {Vassallo}, {Verdoes Kleijn}, {Veropalumbo}, {Wang}, {Weller}, {Williams}, {Zamorani}, {Zucca}, {Baccigalupi}, {Biviano}, {Burigana}, {De Lucia}, {George}, {Scottez}, {Viel}, {Simon}, {Mora}, {Mart{\'\i}n-Fleitas}, \& {Scott}}]{Marleau2024}
{Marleau}, F.~R., {Cuillandre}, J.~C., {Cantiello}, M., {et~al.} 2024, arXiv e-prints, arXiv:2405.13502, \dodoi{10.48550/arXiv.2405.13502}

\bibitem[{{McQuinn} {et~al.}(2014){McQuinn}, {Cannon}, {Dolphin}, {Skillman}, {Salzer}, {Haynes}, {Adams}, {Cave}, {Elson}, {Giovanelli}, {Ott}, \& {Saintonge}}]{McQuinn2014}
{McQuinn}, K. B.~W., {Cannon}, J.~M., {Dolphin}, A.~E., {et~al.} 2014, \apj, 785, 3, \dodoi{10.1088/0004-637X/785/1/3}

\bibitem[{{McQuinn} {et~al.}(2024){McQuinn}, {B. Newman}, {Savino}, {Dolphin}, {Weisz}, {Williams}, {Boyer}, {Cohen}, {Correnti}, {Cole}, {Geha}, {Gennaro}, {Kallivayalil}, {Sandstrom}, {Skillman}, {Anderson}, {Bolatto}, {Boylan-Kolchin}, {Garling}, {Gilbert}, {Girardi}, {Kalirai}, {Mazzi}, {Pastorelli}, {Richstein}, \& {Warfield}}]{mcquinn2024}
{McQuinn}, K. B.~W., {B. Newman}, M.~J., {Savino}, A., {et~al.} 2024, \apj, 961, 16, \dodoi{10.3847/1538-4357/ad1105}

\bibitem[{Mentz {et~al.}(2016)Mentz, La~Barbera, Peletier, Falc{\'o}n-Barroso, Lisker, van~de Ven, Loubser, Hilker, S{\'a}nchez-Janssen, Napolitano, {et~al.}}]{mentz2016}
Mentz, J., La~Barbera, F., Peletier, R., {et~al.} 2016, Monthly Notices of the Royal Astronomical Society, 463, 2819

\bibitem[{{Moresco} {et~al.}(2022){Moresco}, {Amati}, {Amendola}, {Birrer}, {Blakeslee}, {Cantiello}, {Cimatti}, {Darling}, {Della Valle}, {Fishbach}, {Grillo}, {Hamaus}, {Holz}, {Izzo}, {Jimenez}, {Lusso}, {Meneghetti}, {Piedipalumbo}, {Pisani}, {Pourtsidou}, {Pozzetti}, {Quartin}, {Risaliti}, {Rosati}, \& {Verde}}]{Moresco2022}
{Moresco}, M., {Amati}, L., {Amendola}, L., {et~al.} 2022, Living Reviews in Relativity, 25, 6, \dodoi{10.1007/s41114-022-00040-z}

\bibitem[{{Mutlu-Pakdil} {et~al.}(2021){Mutlu-Pakdil}, {Sand}, {Crnojevi{\'c}}, {Drlica-Wagner}, {Caldwell}, {Guhathakurta}, {Seth}, {Simon}, {Strader}, \& {Toloba}}]{mutlupakdil2021}
{Mutlu-Pakdil}, B., {Sand}, D.~J., {Crnojevi{\'c}}, D., {et~al.} 2021, \apj, 918, 88, \dodoi{10.3847/1538-4357/ac0db8}

\bibitem[{{Newman} {et~al.}(2024{\natexlab{a}}){Newman}, {McQuinn}, {Skillman}, {Boyer}, {Cohen}, {Dolphin}, \& {Telford}}]{Newmann2024a}
{Newman}, M. J.~B., {McQuinn}, K. B.~W., {Skillman}, E.~D., {et~al.} 2024{\natexlab{a}}, \apj, 966, 175, \dodoi{10.3847/1538-4357/ad306d}

\bibitem[{{Newman} {et~al.}(2024{\natexlab{b}}){Newman}, {McQuinn}, {Skillman}, {Boyer}, {Cohen}, {Dolphin}, \& {Telford}}]{Newmann2024b}
---. 2024{\natexlab{b}}, \apj, 975, 195, \dodoi{10.3847/1538-4357/ad79f8}

\bibitem[{{Olsen} {et~al.}(2003){Olsen}, {Blum}, \& {Rigaut}}]{olsen2003}
{Olsen}, K. A.~G., {Blum}, R.~D., \& {Rigaut}, F. 2003, \aj, 126, 452, \dodoi{10.1086/375648}

\bibitem[{Papamakarios \& Murray(2016)}]{papamakarios2016}
Papamakarios, G., \& Murray, I. 2016, in Advances in Neural Information Processing Systems, ed. D.~Lee, M.~Sugiyama, U.~Luxburg, I.~Guyon, \& R.~Garnett, Vol.~29 (Curran Associates, Inc.).
\newblock \url{https://proceedings.neurips.cc/paper_files/paper/2016/file/6aca97005c68f1206823815f66102863-Paper.pdf}

\bibitem[{Papamakarios {et~al.}(2021)Papamakarios, Nalisnick, Rezende, Mohamed, \& Lakshminarayanan}]{papamakarios2021}
Papamakarios, G., Nalisnick, E., Rezende, D.~J., Mohamed, S., \& Lakshminarayanan, B. 2021, The Journal of Machine Learning Research, 22, 2617

\bibitem[{Papamakarios {et~al.}(2017)Papamakarios, Pavlakou, \& Murray}]{papamakarios2017}
Papamakarios, G., Pavlakou, T., \& Murray, I. 2017, Advances in neural information processing systems, 30

\bibitem[{Pasha \& Miller(2023)}]{Pasha2023}
Pasha, I., \& Miller, T.~B. 2023, Journal of Open Source Software, 8, 5703, \dodoi{10.21105/joss.05703}

\bibitem[{{Plummer}(1911)}]{Plummer1911}
{Plummer}, H.~C. 1911, \mnras, 71, 460, \dodoi{10.1093/mnras/71.5.460}

\bibitem[{{Raimondo}(2009)}]{Raimondo2009}
{Raimondo}, G. 2009, \apj, 700, 1247, \dodoi{10.1088/0004-637X/700/2/1247}

\bibitem[{{Raimondo} {et~al.}(2005){Raimondo}, {Brocato}, {Cantiello}, \& {Capaccioli}}]{Raimondo2005}
{Raimondo}, G., {Brocato}, E., {Cantiello}, M., \& {Capaccioli}, M. 2005, \aj, 130, 2625, \dodoi{10.1086/497591}

\bibitem[{{Sakai} {et~al.}(1996){Sakai}, {Madore}, \& {Freedman}}]{sakai1996}
{Sakai}, S., {Madore}, B.~F., \& {Freedman}, W.~L. 1996, \apj, 461, 713, \dodoi{10.1086/177096}

\bibitem[{Selvaraju {et~al.}(2020)Selvaraju, Cogswell, Das, Vedantam, Parikh, \& Batra}]{selvaraju2020}
Selvaraju, R.~R., Cogswell, M., Das, A., {et~al.} 2020, International journal of computer vision, 128, 336

\bibitem[{{Sersic}(1968)}]{sersic1968}
{Sersic}, J.~L. 1968, {Atlas de Galaxias Australes}

\bibitem[{Sharma(2017)}]{sharma2017}
Sharma, S. 2017, Annual Review of Astronomy and Astrophysics, 55, 213

\bibitem[{{Shen} {et~al.}(2021){Shen}, {Danieli}, {van Dokkum}, {Abraham}, {Brodie}, {Conroy}, {Dolphin}, {Romanowsky}, {Kruijssen}, \& {Dutta Chowdhury}}]{Shen2021}
{Shen}, Z., {Danieli}, S., {van Dokkum}, P., {et~al.} 2021, \apjl, 914, L12, \dodoi{10.3847/2041-8213/ac0335}

\bibitem[{{Talts} {et~al.}(2018){Talts}, {Betancourt}, {Simpson}, {Vehtari}, \& {Gelman}}]{talts2018}
{Talts}, S., {Betancourt}, M., {Simpson}, D., {Vehtari}, A., \& {Gelman}, A. 2018, arXiv e-prints, arXiv:1804.06788, \dodoi{10.48550/arXiv.1804.06788}

\bibitem[{Tejero-Cantero {et~al.}(2020)Tejero-Cantero, Boelts, Deistler, Lueckmann, Durkan, Gonçalves, Greenberg, \& Macke}]{tejero-cantero2020}
Tejero-Cantero, A., Boelts, J., Deistler, M., {et~al.} 2020, Journal of Open Source Software, 5, 2505, \dodoi{10.21105/joss.02505}

\bibitem[{{Tonry} \& {Schneider}(1988)}]{tonry1988}
{Tonry}, J., \& {Schneider}, D.~P. 1988, \aj, 96, 807, \dodoi{10.1086/114847}

\bibitem[{{Tonry} {et~al.}(1997){Tonry}, {Blakeslee}, {Ajhar}, \& {Dressler}}]{tonry1997}
{Tonry}, J.~L., {Blakeslee}, J.~P., {Ajhar}, E.~A., \& {Dressler}, A. 1997, \apj, 475, 399, \dodoi{10.1086/303576}

\bibitem[{{Tonry} {et~al.}(2001){Tonry}, {Dressler}, {Blakeslee}, {Ajhar}, {Fletcher}, {Luppino}, {Metzger}, \& {Moore}}]{tonry2001}
{Tonry}, J.~L., {Dressler}, A., {Blakeslee}, J.~P., {et~al.} 2001, \apj, 546, 681, \dodoi{10.1086/318301}

\bibitem[{{Tully} {et~al.}(2013){Tully}, {Courtois}, {Dolphin}, {Fisher}, {H{\'e}raudeau}, {Jacobs}, {Karachentsev}, {Makarov}, {Makarova}, {Mitronova}, {Rizzi}, {Shaya}, {Sorce}, \& {Wu}}]{Tully2013}
{Tully}, R.~B., {Courtois}, H.~M., {Dolphin}, A.~E., {et~al.} 2013, \aj, 146, 86, \dodoi{10.1088/0004-6256/146/4/86}

\bibitem[{{van Dokkum} {et~al.}(2018){van Dokkum}, {Danieli}, {Cohen}, {Romanowsky}, \& {Conroy}}]{vandokkum2018}
{van Dokkum}, P., {Danieli}, S., {Cohen}, Y., {Romanowsky}, A.~J., \& {Conroy}, C. 2018, \apjl, 864, L18, \dodoi{10.3847/2041-8213/aada4d}

\bibitem[{Villar(2022)}]{villar2022}
Villar, V.~A. 2022, in {36th Conference on Neural Information Processing Systems}: {Workshop on Machine Learning and the Physical Sciences}.
\newblock \doarXiv{2211.04480}

\bibitem[{{Wang} {et~al.}(2023){Wang}, {Leja}, {Villar}, \& {Speagle}}]{wang2023}
{Wang}, B., {Leja}, J., {Villar}, V.~A., \& {Speagle}, J.~S. 2023, \apjl, 952, L10, \dodoi{10.3847/2041-8213/ace361}

\bibitem[{{Weisz} {et~al.}(2011){Weisz}, {Dalcanton}, {Williams}, {Gilbert}, {Skillman}, {Seth}, {Dolphin}, {McQuinn}, {Gogarten}, {Holtzman}, {Rosema}, {Cole}, {Karachentsev}, \& {Zaritsky}}]{weisz2011}
{Weisz}, D.~R., {Dalcanton}, J.~J., {Williams}, B.~F., {et~al.} 2011, \apj, 739, 5, \dodoi{10.1088/0004-637X/739/1/5}

\bibitem[{Yan {et~al.}(2020)Yan, Jerabkova, \& Kroupa}]{yan2020}
Yan, Z., Jerabkova, T., \& Kroupa, P. 2020, Astronomy \& Astrophysics, 637, A68

\bibitem[{{Zaritsky} {et~al.}(2019){Zaritsky}, {Donnerstein}, {Dey}, {Kadowaki}, {Zhang}, {Karunakaran}, {Mart{\'\i}nez-Delgado}, {Rahman}, \& {Spekkens}}]{zaritsky2019}
{Zaritsky}, D., {Donnerstein}, R., {Dey}, A., {et~al.} 2019, \apjs, 240, 1, \dodoi{10.3847/1538-4365/aaefe9}

\bibitem[{{Zhang} {et~al.}(2023){Zhang}, {Jayasinghe}, \& {Bloom}}]{Zhang2023}
{Zhang}, K., {Jayasinghe}, T., \& {Bloom}, J. 2023, in Machine Learning for Astrophysics, 39, \dodoi{10.48550/arXiv.2312.05687}

\end{thebibliography}
\bibliographystyle{aasjournal}

\end{document}